\documentclass[aps,pra,twocolumn,showpacs,
preprintnumbers,longbibliography]
{revtex4-1}

\usepackage{graphicx}
\usepackage{dcolumn}
\usepackage{bm}
\usepackage{amsmath}
\usepackage{amssymb}
\usepackage{amsthm}
\usepackage{amsfonts}
\usepackage{enumerate}
\usepackage{latexsym}
\usepackage[T1]{fontenc}
\usepackage{lmodern}
\usepackage[colorlinks=true,urlcolor=blue,citecolor=blue,linkcolor=blue]{hyperref}

\begin{document}

\title{Band geometry, Berry curvature and superfluid weight}

\author{Long Liang}
\author{Tuomas I. Vanhala}
\author{Sebastiano Peotta}
\author{Topi Siro}
\affiliation{COMP Centre of Excellence, Department of Applied Physics, Aalto University, Helsinki, Finland}

\author{Ari Harju}
\email{ari.harju@aalto.fi}
\affiliation{COMP Centre of Excellence, Department of Applied Physics, Aalto University, Helsinki, Finland}
\author{P\"{a}ivi T\"{o}rm\"{a}}
\email{paivi.torma@aalto.fi}
\affiliation{COMP Centre of Excellence, Department of Applied Physics, Aalto University, Helsinki, Finland}

\begin{abstract}

We present a theory of the superfluid weight in multiband attractive Hubbard models  within the Bardeen-Cooper-Schrieffer (BCS)  mean field framework. We show how to separate the geometric contribution to the superfluid weight from the conventional one, and that the geometric contribution is associated with the interband matrix elements of the current operator. 
Our theory can be applied to systems with or without time reversal symmetry. In both cases the geometric superfluid weight can be related to the quantum metric of the corresponding noninteracting systems.
 This leads to a lower bound on the superfluid weight given by the absolute value of the Berry curvature.  We apply our theory to the attractive Kane-Mele-Hubbard  and Haldane-Hubbard models, which can be realized in ultracold atom gases. Quantitative comparisons are made to  state of the art dynamical mean-field
theory and exact diagonalization results.
\end{abstract}
\maketitle

\section{introduction and main results}
 
A manifold of quantum states possess a natural geometric structure given by the quantum geometric tensor~\cite{Berry:1989ai}, whose imaginary part is the Berry curvature and real part is the quantum metric \cite{Provost1980}. The integral of Berry curvature over a surface in parameter spaces gives the Berry phase \cite{Berry84}, which measures the phase change of a quantum state along the boundary of the surface. Related to the phase change, the amplitude change  is characterized by the quantum metric.

The Berry curvature provides a coherent understanding  of basic phenomena such as electron transport, polarization, and orbital magnetisation \cite{XCN2010,resta1994,resta2010}. For systems with discrete translational invariance, the integral of the Berry curvature over the Brillouin zone gives the Chern number, which can be used to characterize topological phases. The quantum metric has found applications in the theory of quantum entanglement and quantum information \cite{GQS} and can be used to detect quantum phase transitions~\cite{QPT_rev}.

Recently, there has been great interest in models with flat or quasi-flat bands. A band is called quasi-flat if the ratio between the bandwidth and the energy gap to neighbouring bands (the flatness ratio) is much smaller than unity. A quasi-flat band with nonzero Chern number may support various fractional quantum Hall states \cite{SarmaFlatBand, WenFlatBand,fqh2011,ShengFlatBand_boson,doi:10.1142/S021797921330017X,roy2014,Roy2015}, and it has been shown that flat bands can enhance the superconducting transition temperature because of the high density of states \cite{Volovik_Flatband1,Volovik_Flatband2,FlatBandTc,PhysRevB.93.214505}. The importance of the quantum geometric tensor, especially in flat-band systems, has been revealed in connection with lattice fractional quantum Hall states \cite{roy2014,Roy2015}, flat band superfluidity \cite{PT, LiebLattice}, and orbital magnetic susceptibility \cite{GYN2015,PRFM2016}.
  
Nonzero superfluid weight is a defining property of superconductors and leads to the Meissner effect and dissipationless transport. It sets  the phase fluctuation energy scale that plays a significant role in high temperature superconductors  \cite{Emery1995}. In Ref.~\cite{PT}, two of us pointed out that the superfluid weight is related to the quantum metric and bounded from below by the Chern number in the isolated flat band limit. 
However, a flat band with zero Chern number can also support a large superfluid weight \cite{LiebLattice}, which is consistent with, but can not be explained by the Chern number bound.
The theory developed in \cite{PT} depends on time reversal symmetry (TRS) and the relation between the superfluid weight and the quantum metric is obtained only in the specific case of the flat band limit.

In this article, we develop a general theory of the superfluid weight in multiband attractive Hubbard models in the framework of linear response theory. The superfluid weight defined through linear response is equivalent to the one defined in terms of the thermodynamic potentials, but it offers several advantages. For example, it is the starting point for investigating beyond-mean-field effects through many-body perturbation theory and facilitates the derivation of useful sum rules~\cite{BaymStAndrews,StringariBook}. 

We show here how to separate in general the geometric contribution to the superfluid weight from the conventional one. 
The linear response approach clarifies the origin of the geometric effect. 
The conventional contribution is associated with the diagonal (intraband) matrix elements of the current operator that are derivatives of the band dispersions. Thus, the conventional contribution vanishes in the flat band limit.  The geometric contribution, however, is associated with the {\it off-diagonal (interband) matrix elements of the current operator} and can be nonzero even for a flat band.

In the presence of TRS and for uniform pairing, we find a novel form for the geometric contribution $D^s_{\mathrm{geom}}$, where the quantum metric appears explicitly both in the case of an isolated, not necessarily flat, band and in two-band systems. A Bloch band is called an isolated band if it is separated from other bands by large enough band gaps. In the isolated band limit (see Sec.~\ref{Sec2_a} for details), we obtain
\begin{eqnarray}\label{Eq:geom}
D^s_{\mathrm{geom},\mu\nu}&=&2\Delta^2\sum_{\mathbf{k}}\frac{\tanh{(\beta E_{\mathbf{k}}/2)}}{E_{\mathbf{k}}}g_{\mu\nu}(\mathbf{k}),
\end{eqnarray}
where $\mu,\nu = x, y, z$ are spatial indices, $\Delta$ is the pairing order parameter, $\beta=1/T$ is the inverse temperature (the Boltzmann constant is taken to be 1 throughout this article), $E_{\mathbf{k}}\geq 0$ is the BCS theory quasiparticle excitation energy,  and $g_{\mu\nu}(\mathbf{k})$ is the quantum metric of the isolated Bloch band. A similar expression for two-band systems is given by Eq. (\ref{Eq:DS_TRS_2bands}) below. 
Furthermore, the quantum metric also appears in the isolated band limit of two-band systems without TRS, see Eq. (\ref{Eq:SW_geom_isolated_HH}) below.
These are important generalizations of the previous results, where the quantum metric has been related to the superfluid weight only in the isolated flat band limit with TRS \cite{PT,LiebLattice}, and show that the quantum metric affects the superfluid properties in a broad class of systems. 

The quantum metric is non-negative everywhere in the Brillouin zone and allows us to derive a bound on  the superfluid weight using the Berry curvature $B_{\mu\nu}(\mathbf{k})$. For two-dimensional isotropic systems, 
\begin{eqnarray}\label{Eq:geom_bound}
D^s_{\mathrm{geom}}\ge 2\Delta^2\sum_{\mathbf{k}}\frac{\tanh{(\beta E_{\mathbf{k}}/2)}}{E_{\mathbf{k}}} | B_{xy}(\mathbf{k})|.
\end{eqnarray}
Importantly, this bound depends on the absolute value of the Berry curvature. Therefore the geometric contribution is nonzero for any nonzero Berry curvature. Eq. (\ref{Eq:geom_bound}) provides a much stronger lower bound than the Chern number bound \cite{PT} since a nonzero Berry curvature can still integrate to zero. 
This explains why bands with zero Chern number, such as the Lieb lattice flat band \cite{LiebLattice}, can still have a nonzero superfluid weight.

\begin{figure} [t]
	\includegraphics[width=\columnwidth]{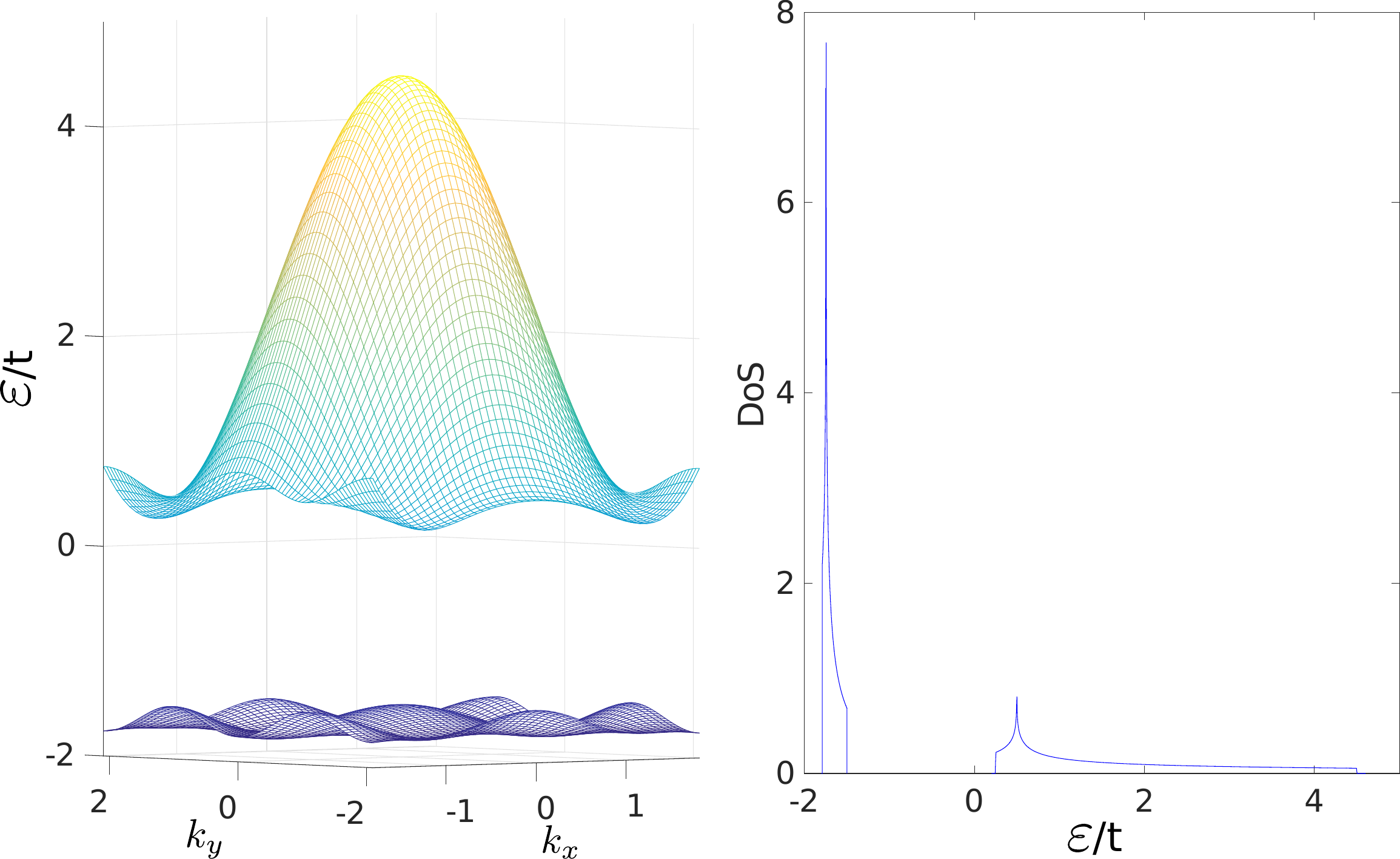}
	\caption{Band structure and density of states (DoS) of the non-interacting Haldane model for $\cos(\phi)=t/(4t')=3\sqrt{3/43}$. The bandwidth of the lower band is about $0.29t$ and the gap between upper and lower band is $1.75t$. The quasi-flat lower band has much larger density of states than the dispersive upper band.} \label{Fig:noninteracting}
\end{figure}

We apply our general theory of the superfluid weight for multiband systems to
the attractive Kane-Mele-Hubbard (KMH) model and the spinful Haldane-Hubbard (HH)  model, which are paradigmatic models of interacting topological systems and also of great interest in current ultracold atomic gas experiments \cite{Jotzu2014,Flaschner1091}. The KMH model has TRS, while in the spinful HH model TRS is broken but $SU(2)$ spin symmetry is present.  The Haldane model \cite{Haldane1988} is a representative model of Chern insulators. The Kane-Mele model is a time reversal symmetric generalization of the Haldane model and is a representative model of $\mathbb{Z}_2$ topological insulators \cite{KM2005_1}. The band energies of the noninteracting part of both the Kane-Mele-Hubbard model and the Haldane-Hubbard model are shown in Fig.~\ref{Fig:noninteracting} and the corresponding lattice in Fig.~\ref{Fig:Lattice}. The nearest-neighbour hopping $t$ and the complex next-nearest-neighbour hopping $t'e^{i\phi}$  are chosen to minimize the  flatness ratio~\cite{fqh2011}. Consequently the lower band is quasi-flat while the higher one is highly dispersive.  

\begin{figure}
\includegraphics[width=\columnwidth] {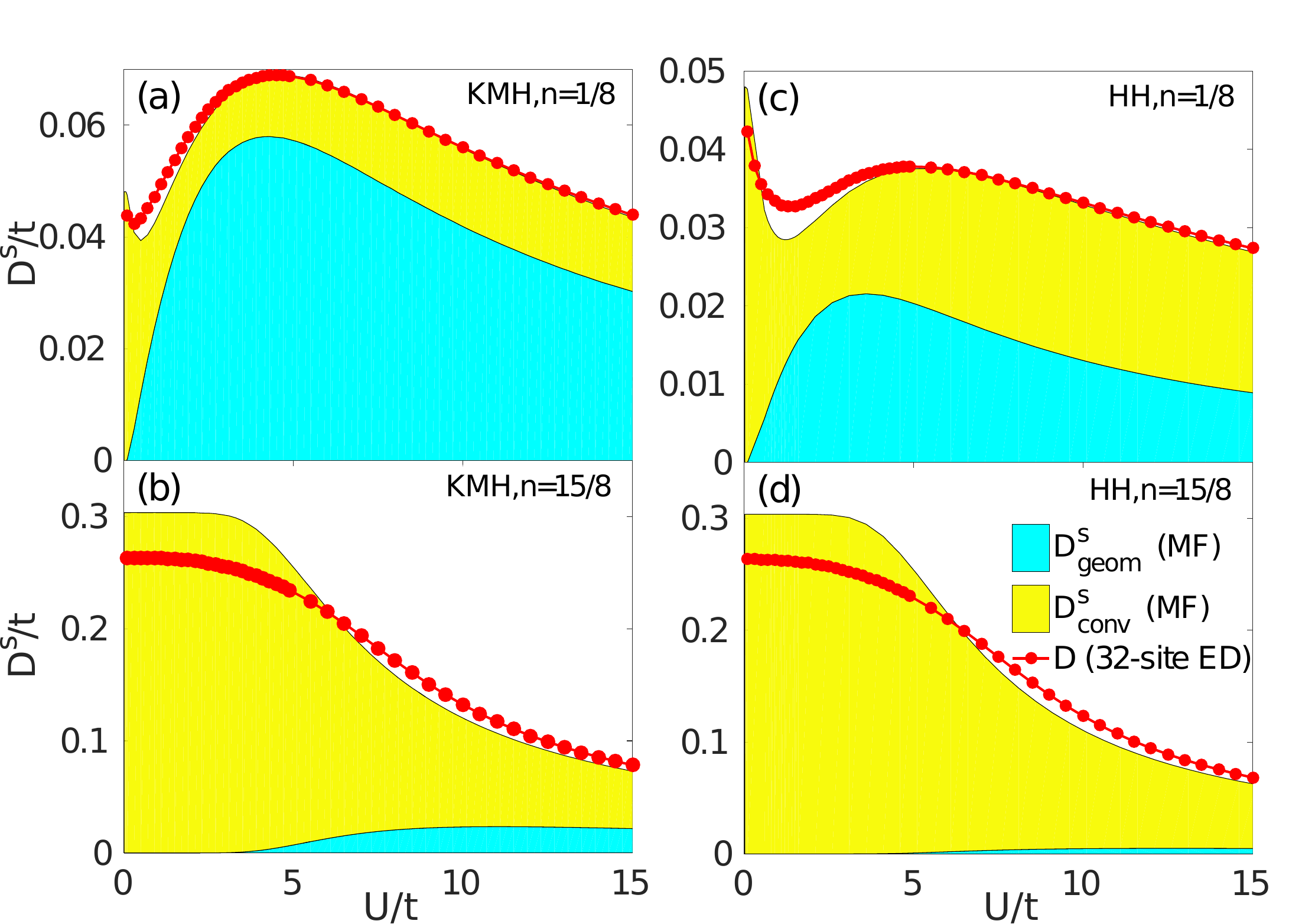} 
\caption{
Zero-temperature superfluid weight for the KMH  model [(a)-(b)] and the HH model [(c)-(d)] at filling  $n=1/8$ [(a) and (c), flat band], $15/8$ [(b) and (d), dispersive band]. Geometric contributions and conventional contributions are obtained using the theory developed in Section~\ref{Sec2}. The geometric contribution is large and dominant or comparable to the conventional one for the flat bands. Solid curves are the Drude weight $D$ obtained from exact diagonalization (ED) on a 32-site cluster.}\label{Fig:SW_Zero}
\end{figure}

Our mean-field results for the critical temperature are consistent with previous theoretical predictions \cite{Volovik_Flatband1, Volovik_Flatband2, FlatBandTc}. For the flat band, the critical temperature is proportional to the Hubbard interaction, while for the dispersive band it is exponentially small in the weak coupling limit. The superfluid weight obtained within  mean-field theory is shown in Fig. \ref{Fig:SW_Zero}. For the quasi-flat bands [Fig. \ref{Fig:SW_Zero} (a) and (c)] the geometric contribution is comparable to, or even larger than the conventional one, while for the strongly dispersive bands the conventional contribution dominates [Fig. \ref{Fig:SW_Zero} (b) and (d)].

It is not immediately clear whether BCS theory is a good approximation especially in the highly degenerate flat band case. 
In a recent work, two of us have shown that the BCS wave function is the exact ground state of the attractive Hubbard interaction term projected on the flat band subspace if TRS is present and the uniform pairing condition [see Eq.~\eqref{Eq:UPC} below] is satisfied~\cite{2016arXiv160800976T}.
The validity of BCS theory in the strong coupling limit can be justified using perturbation theory (see Sec.~\ref{Sec3:ValidityBCS}). Furthermore, we employ dynamical mean-field theory (DMFT) to calculate the order parameter and the superfluid weight and find good agreement with mean-field results. DMFT goes beyond static mean-field theory by including local fluctuations. However, non-local fluctuations are not included and the method might be biased by the choice of order parameters. We thus perform also exact diagonalization (ED) calculations to get the Drude weight (red dots in Fig.~\ref{Fig:SW_Zero}) for a finite system. The Drude weight in the bulk limit is equivalent to the superfluid weight for a gapped system \cite{swz}.  ED gives unbiased results which are in good agreement with the mean-field results.

The rest of this article is organized as follows. In Sec. \ref{Sec2}, we present the derivation of the superfluid weight within the BCS theory and show how to separate the geometric contribution from the conventional one. We discuss the cases with and without TRS in Sec. \ref{Sec2_a} and Sec. \ref{Sec2_b}, respectively. In Sec. \ref{Sec3} we apply our theory to the attractive KMH model and HH model and compare the mean-field results with DMFT and ED results. Finally, conclusions and future prospects are presented in Sec. \ref{Sec4}.

\section{superfluid weight in multiband attractive Hubbard models}\label{Sec2}
We start from the lattice Hamiltonian
\begin{eqnarray}
H&=&-\sum_{i\alpha,j\beta,\sigma}t^{\sigma}_{i\alpha,j\beta}c^{\dag}_{i\alpha\sigma}c_{j\beta \sigma}\nonumber\\
&&-U\sum_{i\alpha} n_{i\alpha\uparrow}n_{i\alpha\downarrow}-\mu\sum_{i\alpha\sigma} n_{i\alpha\sigma}, 
\end{eqnarray}
where $c^\dag_{i\alpha\sigma}$ ($c_{i\alpha\sigma}$) is the creation (annihilation) operator  labelled by  orbital $i\alpha$ ($i$ is the unit cell label and $\alpha$ the sublattice label) and spin $\sigma$, and $n_{i\alpha\sigma}=c^\dag_{i\alpha\sigma}c_{i\alpha\sigma}$ is the particle number operator. The first term in the Hamiltonian is the kinetic energy, which can be spin dependent. Note that we only consider the case of kinetic Hamiltonians which commute with the spin operator along the $z$-axis. The second term is the Hubbard interaction which we assume to be attractive, i.e., $U>0$. The filling $n\equiv (N_\uparrow+N_\downarrow)/N_\mathrm{sites}$ is controlled by the chemical potential $\mu$. 

The superfluid weight can be obtained using linear response theory~\cite{BaymStAndrews}, which relates the response functions to the correlation functions  evaluated on the  ground state. In our case the BCS ground state is a good approximation of the true ground state, therefore the use of linear response theory is legitimate. 
 To calculate the current-current response function, we introduce a slowly varying vector potential $\mathbf{A}$ by the Peierls substitution, so that the hopping $t_{ij}$ is modified by the phase factor $e^{-i\mathbf{A}\cdot (\mathbf{r}_{j}-\mathbf{r}_{i})}$. Expanding the phase factors up to $A^2$ order,  we get
$H(\mathbf{A})
=H+j^p_{\mu}A_{\mu}+T_{\mu\nu}A_{\mu}A_{\nu}/2$, where $j^p_{\mu}=i[x_{\mu},H], ~T_{\mu\nu}=i[x_{\mu},j^p_{\nu}]$ and  $x_{\mu}=\sum_{i\alpha} r_{i,\mu}c^{\dag}_{i\alpha\sigma}c_{i\alpha\sigma}$ is the position operator. Here $j^p_{\mu}$ is the paramagnetic current operator while $T_{\mu\nu}A_{\nu}$ is the diamagnetic current operator. From linear response theory we find that the current density is $ j_{\mu}(\mathbf{q},\omega)=K_{\mu\nu}(\mathbf{q},\omega)A_{\nu}(\mathbf{q},\omega)$, where $K_{\mu\nu}$ is the current-current response function, 
\begin{eqnarray}
\begin{split}
K_{\mu\nu}&(\mathbf{q},\omega)=\langle T_{\mu\nu} \rangle\\
&-i\int^\infty_0 \mathrm{d}t e^{i(\omega +i0^+)t}\langle [j^p_{\mu}(\mathbf{q},t),j^p_{\nu}(-\mathbf{q},0)]\rangle\,.
\end{split}
\end{eqnarray}
The superfluid weight is defined through the static Meissner effect by taking the proper zero momentum limit of the transverse component of the current-current response function. At mean-field level one can use the result~\cite{swz}
\begin{eqnarray}
D^s_{\mu\nu}=K_{\mu\nu}(\mathbf{q}\to 0,\omega=0).
\end{eqnarray} 
This is equivalent
to the  definition in terms of the thermodynamic potentials, see Appendix \ref{appendix:equivalence}. 
  We calculate $K_{\mu\nu}(\mathbf{q},\omega)$ within the BCS framework by decoupling the Hubbard interaction as
  \begin{eqnarray}\label{Eq:deouple}
 -U\sum_{i\alpha} n_{i\alpha\uparrow}n_{i\alpha\downarrow}\approx \sum_{i\alpha} (\Delta_{i\alpha} c^{\dag}_{i\alpha\uparrow}c^{\dag}_{i\alpha\downarrow}+\mathrm{H.c.}),
 \end{eqnarray}
 where the order parameter $\Delta_{i\alpha}=-U\langle c_{i\alpha\downarrow}c_{i\alpha\uparrow} \rangle $ should be determined self-consistently. We  consider mean-field solutions that preserve the translational symmetry. Then the mean-field Hamiltonian reads $H_{\mathrm{MF}}=\sum_{\mathbf{k}}\Psi^\dag_{\mathbf{k}}\mathcal{H}(\mathbf{k})\Psi_{\mathbf{k}}$. The Nambu field is $\Psi_{\mathbf{k}}=(c_{\alpha\mathbf{k}\uparrow}, c^{\dag}_{\alpha-\mathbf{k}\uparrow})^{T}$ with $\alpha=1,2,\cdots,M$ denoting the orbital. The Bogoliubov-de Gennes (BdG)  Hamiltonian reads
 \begin{eqnarray}\label{Eq:BdgHamiltonian}
 \mathcal{H}(\mathbf{k})=\left[  \begin{array}{cc}
  \mathcal{H}_{\uparrow}(\mathbf{k})-\mu & \bm \Delta \\ 
 \bm \Delta^\dag & -\mathcal{H}^{\ast}_{\downarrow}(-\mathbf{k})+\mu 
 \end{array} \right],
 \end{eqnarray}
where the $M$ by $M$ matrix $\mathcal{H}_{\sigma}(\mathbf{k})$ is the Fourier transform of the hopping terms and $\bm \Delta=\mathrm{diag}(\Delta_1,\Delta_2,\cdots,\Delta_M)$ are  momentum independent order parameters in orbital space. 

The diamagnetic and paramagnetic current operators are given respectively by 
\begin{eqnarray}
T_{\mu\nu}=\sum_{\mathbf{k},\sigma}c^\dag_{\mathbf{k}\sigma}\partial_{\mu}\partial_{\nu}\mathcal{H}_\sigma(\mathbf{k})c_{\mathbf{k}\sigma},
\end{eqnarray}
and
\begin{eqnarray}
j^p_{\mu}(\mathbf{q})=\sum_{\mathbf{k},\sigma}c^\dag_{\mathbf{k}\sigma}\partial_{\mu}\mathcal{H}_\sigma(\mathbf{k+q}/2)c_{\mathbf{k+q}\sigma}.
\end{eqnarray}
It is convenient to calculate the response function in imaginary time using the Matsubara formalism. Within the BCS mean-field theory, we obtain
\begin{eqnarray}
K_{\mu\nu}(\mathbf{q},i\omega_n)&=&\frac{1}{\beta}\sum_{\mathbf{k}}\sum_{\Omega_m} \mathrm{Tr}\bigg[\partial_\mu \partial_\nu \mathcal{H}(\mathbf{k}) G(i\Omega_m,\mathbf{k})\nonumber\\
&&+
 G(i\Omega_m,\mathbf{k})\partial_{\nu} \mathcal{H}(\mathbf{k+q/2})\gamma^z\\
&&\times G(i\omega_n+i\Omega_m,\mathbf{k+q})\partial_{\mu}\mathcal{H}(\mathbf{k+q/2})\gamma^z\bigg],\nonumber 
\end{eqnarray}
where $\omega_n=2\pi n/\beta$, $\Omega_m=2\pi (m+1/2)/\beta$ are bosonic and fermionic Matsubara frequencies and $\partial_\mu\equiv \partial_{k_\mu}$ is the derivative with respect to the momentum $k_\mu$. For simplicity, the volume (area in two dimensions) is taken to be 1. Here $\gamma^z=\tau^z\otimes I_{M\times M}$ and $\tau^i$ are Pauli matrices acting in the particle-hole space, and $I_{M\times M}$ is the $M$ by $M$ identity matrix. Furthermore, 
 \begin{eqnarray}
  G(i\omega_n,\mathbf{k})=\frac{1}{i\omega_n-\mathcal{H}(\mathbf{k})}=\sum^{2M}_{j=1}\frac{|\psi_j(\mathbf{k}) \rangle\langle \psi_j(\mathbf{k})|}{i\omega_n-E_{j,\mathbf{k}}},
 \end{eqnarray}
 is the Green's function and $|\psi_j(\mathbf{k}) \rangle$ is the $j$-th eigenvector of the BdG Hamiltonian with eigenvalue $E_{j,\mathbf{k}}$.  Hereafter the $\mathbf{k}$-dependence of quantities will be omitted with some exceptions. Performing the Matsubara frequency summation and taking the $\omega=0$, $\mathbf{q}\to 0$ limit,  we get the superfluid weight
\begin{eqnarray}\label{Eq:SW}
D^s_{\mu\nu}&=&\sum_{\mathbf{k},i,j}\frac{n(E_{j})-n(E_i)}{E_i-E_j}\bigg(\langle\psi_i |\partial_{\mu}\mathcal{H} |\psi_j \rangle \langle\psi_j |\partial_{\nu}\mathcal{H} |\psi_i \rangle \nonumber\\
&&-\langle\psi_i |\partial_{\mu}\mathcal{H} \gamma^z|\psi_j \rangle \langle\psi_j |\gamma^z\partial_{\nu}\mathcal{H} |\psi_i \rangle\bigg),
\end{eqnarray}
where $n(E_i)=1/(e^{\beta E_i}+1)$ is the Fermi-Dirac distribution and the prefactor should be understood as $-\partial_E n(E)$ when $i=j$ or $E_i$ and $E_j$ are degenerate.   The first term in the parenthesis is the diamagnetic term and the second term is the paramagnetic term. For a single-band system, it is well known that the paramagnetic term vanishes at zero temperature, and the nonzero  diamagnetic term  leads to the Meissner effect. However, for a multiband system the paramagnetic term remains finite even at zero temperature.

For our purposes it is convenient  to write  Eq.~\eqref{Eq:SW} in terms of the matrix elements of the current operator.
To this end, we expand the BdG wave functions in terms of the Bloch functions $|m\rangle_\sigma$: $|\psi_i\rangle=\sum^M_{m=1}\big(w_{+,im}|m\rangle_{\uparrow}\otimes|+\rangle+w_{-,im}|m^\ast_-\rangle_{\downarrow}\otimes|-\rangle\big)$, where $|m\rangle_{\uparrow}$ is the eigenvector of $\mathcal{H}_\uparrow(\mathbf{k})$ with eigenvalue $\varepsilon_{\uparrow,m,\mathbf{k}}$, $|m^\ast_-\rangle_{\downarrow}$ is the eigenvector of $\mathcal{H}^\ast_\downarrow(-\mathbf{k})$ with eigenvalue $\varepsilon_{\downarrow,m,-\mathbf{k}}$, and  $|\pm\rangle$ is the eigenvector of $\tau^z$ with eigenvalue $\pm 1$. Since the order parameters and the chemical potential are momentum independent, the derivative of the BdG Hamiltonian can be written as  $\partial_{\mu}\mathcal{H}(\mathbf{k})=P_+\partial_{\mu}\mathcal{H}_\uparrow(\mathbf{k})-P_-\partial_{\mu}\mathcal{H}^\ast_\downarrow(-\mathbf{k})$. Here $P_+$ ($P_-$) is the projection operator onto the particle (hole) space. Inserting these expressions into Eq. (\ref{Eq:SW}) leads to
\begin{eqnarray}\label{Eq:eq_main}
D^s_{\mu\nu}&=&\sum_{\mathbf{k}}\sum_{m,n,p,q}C^{mn}_{pq}[j_{\mu,\uparrow}(\mathbf{k})]_{mn} [j_{\nu,\downarrow}(-\mathbf{k})]_{qp},    
\end{eqnarray}
where
\begin{eqnarray}
\scalebox{1}[1]{$C^{mn}_{pq}=2\sum_{i,j}\frac{n(E_i)-n(E_j)}{E_j-E_i}w^\ast_{+,im}w_{+,jn}w^\ast_{-,jp}w_{-,iq}$}.
\end{eqnarray}
The matrix element of the current operator is
\begin{eqnarray}\label{eq:current_operator}
[j_{\mu,\sigma}(\mathbf{k})]_{mn}&=&{}_{\sigma}\langle m| \partial_\mu \mathcal{H}_{\sigma}(\mathbf{k})|n\rangle_{\sigma}\\
&=&\partial_\mu\varepsilon_{\sigma,m}\delta_{mn}+(\varepsilon_{\sigma,m}-\varepsilon_{\sigma,n}){}_{\sigma}\langle\partial_{\mu}m|n\rangle_{\sigma}.\nonumber 
\end{eqnarray} 
where $\delta_{mn}$ is the Kronecker delta function.  The diagonal matrix elements of the current operator are given by the derivatives of the band dispersions and are therefore zero in the flat band limit. Therefore we separate the superfluid weight into two terms,
 \begin{eqnarray}\label{Eq:eq_split}
 D^s_{\mu\nu}&=&D^s_{\mathrm{conv},\mu\nu}+D^s_{\mathrm{geom},\mu\nu},
 \end{eqnarray}
where the geometric contribution $D^s_{\mathrm{geom},\mu\nu}$ is defined as the terms that depends only on the off-diagonal elements of the current operator,
\begin{eqnarray}\label{Eq:eq_geom_general}
	D^s_{\mathrm{geom},\mu\nu}&=&\sum_{\mathbf{k}}
	\sum_{\substack{
			m\ne n \\
			p\ne q
		}} 
	C^{mn}_{pq}[j_{\mu,\uparrow}(\mathbf{k})]_{mn} [j_{\nu,\downarrow}(-\mathbf{k})]_{qp}, 
\end{eqnarray}    
and the terms containing diagonal elements of the current operator are defined to be the conventional superfluid weight $D^s_{\mathrm{conv},\mu\nu}$. 

Our results, Eqs. (\ref{Eq:SW})-(\ref{Eq:eq_geom_general}), are quite general and can be applied to various systems provided that the BCS approximation is good. Although our theory is developed for the simplest intra-orbital Hubbard interaction, it can straightforwardly be generalized to inter-orbital interactions, in which case one needs to consider inter-orbital pairings.  Here, we focus on the simplest Hubbard interaction which admits simple mean-field solutions. We will show that in the presence of extra symmetries, Eq. (\ref{Eq:eq_main}) can be further simplified to the point that the geometric contribution can be written solely in terms of the quantum metric in the isolated band approximation. In the following we discuss two cases that can be applied to the KMH model and HH model. 

Naively one may think that in the isolated band limit the only relevant terms in Eq.~\eqref{Eq:SW} are the ones involving only the quasiparticle wave functions $|\psi_i\rangle$ adiabatically connected  to the isolated band when $\Delta_\alpha \to 0$. This is wrong since the off-diagonal matrix elements of the current operator \eqref{eq:current_operator} scale with the energy gap and for this reason the isolated flat band limit of Eq.~\eqref{Eq:SW} must be taken with care. We will show below that all terms in Eq.~\eqref{Eq:SW} can provide a nonzero contribution in the isolated flat-band limit and this leads precisely to the geometric term of the superfluid weight. Moreover in the case of broken TRS we will show that it is necessary to calculate the quasiparticle eigenstates $|\psi_i\rangle$ adiabatically connected to the isolated band up to first order in the order parameters $\Delta_\alpha$, before taking the isolated band limit. This first order correction also generates off-diagonal matrix elements of the current operator. Details are in Appendix~\ref{app:derivation_HH}. We note that this is an effect of interactions and such complications do not arise in the noninteracting limit ($U = \Delta_\alpha = 0$).
The observation that the isolated band limit of Eq.~\eqref{Eq:SW} is rather subtle is crucial for the present work. A similar situation is encountered in Quantum Hall systems where it is found that the current operator is purely off-diagonal, namely the matrix elements of the current operator between states in the same Landau level are vanishing~\cite{Moon:1995}. 

\subsection{Time reversal symmetric and uniform pairing case}\label{Sec2_a}

In the presence of TRS, the kinetic energy for spin up and spin down particles are related:  $\mathcal{H}_{\uparrow}(\mathbf{k})=\mathcal{H}^{\ast}_{\downarrow}(-\mathbf{k})$. 
We further assume that the order parameter is uniform in orbital space, namely the matrix $\bm \Delta=\Delta I_{M\times M}$ is proportional to the identity. In this case we can choose a gauge such that $\Delta$ is real. This is equivalent to the uniform pairing condition introduced in Ref.~\cite{2016arXiv160800976T}. As we mentioned before, the order parameter should be determined self-consistently and whether the uniform pairing ansatz is good or not depends on the specific problem. However, the uniform pairing state already captures a lot of interesting physical systems ~\cite{PT,2016arXiv160800976T}. For the KMH and HH models studied here, this assumption is fulfilled because of inversion symmetry.

An important consequence of uniform pairing together with time reversal symmetry is the absence of interband pairing. This means that when  the diagonal blocks of the BdG Hamiltonian in Eq.~\eqref{Eq:BdgHamiltonian} are diagonalized by going from orbital to band space, the off-diagonal blocks retain their diagonal form. In fact it is easy to see that the off-diagonal blocks transform as $\bm \Delta(\mathbf{k})=\mathcal{G}^\dag(\mathbf{k}) \bm \Delta \mathcal{G}(\mathbf{k})=\bm \Delta$, where $\mathcal{G}(\mathbf{k})$ is the unitary matrix that diagonalizes $\mathcal{H}_{\uparrow}(\mathbf{k})$ and whose matrix elements are given by the Bloch functions $\mathcal{G}(\mathbf{k})_{\alpha m} = \left\langle\alpha | m_{\mathbf{k}}\right\rangle$.
Then the BdG Hamiltonian in band space takes the simple form $\mathcal{H}(\mathbf{k})=\sum^{M}_{m=1} [(\varepsilon_m-\mu)\tau^z+\Delta\tau^x]\otimes |m\rangle\langle m|$. 
It is straightforward to write down the eigenfunctions and the eigenvalues. The eigenvalues appear in pairs: $E^{\pm}_m=\pm E_m=\pm\sqrt{(\varepsilon_m-\mu)^2+\Delta^2}$. The corresponding eigenfunctions are 
$|\psi^+_m\rangle=(u_m|+\rangle+v_m|-\rangle)\otimes|m\rangle$ and 
$|\psi^-_m\rangle=(-v_m|+\rangle+u_m|-\rangle)\otimes|m\rangle$, 
where 
\begin{eqnarray}
u_m=\frac{1}{\sqrt{2}}\sqrt{1+\frac{\varepsilon_m-\mu}{E_m}},  v_m=\frac{1}{\sqrt{2}}\sqrt{1-\frac{\varepsilon_m-\mu}{E_m}}. 
\end{eqnarray}
Substituting these into Eq. (\ref{Eq:SW}) and using the definitions Eq. (\ref{Eq:eq_split}) and Eq. (\ref{Eq:eq_geom_general}), we get 
\begin{eqnarray}\label{Eq:DS_TRS_conv}
D^s_{\mathrm{conv},\mu\nu}&=&\sum_{\mathbf{k},m}\left[-\frac{\beta}{2\cosh^2{(\beta E_m/2)}}+\frac{\tanh{(\beta E_m/2)}}{E_m}\right]\nonumber\\
&&\times\frac{\Delta^2}{E^2_m}\partial_\mu \varepsilon_m  \partial_\nu \varepsilon_m, 
\end{eqnarray}
and 

\begin{eqnarray}\label{Eq:DS_TRS_geom}
D^s_{\mathrm{geom},\mu\nu}&=&\sum_{\mathbf{k},m\ne n}\left[\frac{\tanh{(\beta E_m/2)}}{E_m}-\frac{\tanh{(\beta E_n/2)}}{E_n}\right]\\
&&\times \frac{\Delta^2(\varepsilon_n-\varepsilon_m)}{\varepsilon_n+\varepsilon_m-2\mu}\bigg(\langle  \partial_\mu  m|  n \rangle  \langle n|\partial_\nu  m\rangle+\mathrm{H.c.}\bigg ).\nonumber 
\end{eqnarray}
 The conventional term contains only diagonal elements of the current operator $\propto \partial_\mu\varepsilon_m$  and has exactly the same  form as the superfluid weight for a single-band system summed over all bands.  In the flat band limit, the conventional term is negligible. The geometric term comes from the off-diagonal part of the current operator and depends on the derivatives of Bloch wave functions. The geometric term is reduced to the quantum metric in the isolated band limit and in two-band systems, as will be shown below.
 
First, let us discuss the isolated band limit.
Suppose the chemical potential  lies within an isolated band $\bar{m}$, 
 then we can perform a large band gap expansion as done in Appendix \ref{appendix:corrections}. 
At zero order in the expansion in inverse powers of the band gap, the superfluid weight is determined solely by the properties of the isolated band and the geometric contribution takes a simple form 
\begin{eqnarray}\label{Eq:DS_TRS_geom_isolated}
D^s_{\mathrm{geom},\mu\nu}&=&2\Delta^2\sum_{\mathbf{k}}\frac{\tanh{(\beta E_{\bar{m}}/2)}}{E_{\bar{m}}}\nonumber\\
&&\times[\langle  \partial_\mu  \bar{m}| ( 1-|\bar{m} \rangle  \langle \bar{m}|)|\partial_\nu  \bar{m}\rangle+\mathrm{H.c.}]\nonumber\\
&=&2\Delta^2\sum_{\mathbf{k}}\frac{\tanh{(\beta E_{\bar{m}}/2)}}{E_{\bar{m}}}g^{\bar{m}}_{\mu\nu},
\end{eqnarray}
where  $g^{\bar{m}}_{\mu\nu}$ is the quantum metric that defines a distance in Hilbert space
\begin{eqnarray}
\mathrm{d}s^2&\equiv& 1- |\langle \bar{m}(\mathbf{k}) | \bar{m}(\mathbf{k}+\mathrm{d}\mathbf{k})\rangle|^2\nonumber\\
&=&\frac{1}{2}g^{\bar{m}}_{\mu\nu}\mathrm{d}k_{\mu}\mathrm{d}k_{\nu}+O(k^3).
\end{eqnarray}
The quantum metric can be expressed in a compact form as the real part of the quantum geometric tensor $\mathcal{R}^{\bar{m}}_{\mu\nu}$:
\begin{eqnarray} \mathcal{R}^{\bar{m}}_{\mu\nu}=2\mathrm{Tr}[ P_{\bar{m}}\partial_{\mu}P_{\bar{m}}\partial_{\nu}P_{\bar{m}}],
\end{eqnarray}
 where $P_{\bar{m}}=| \bar{m}\rangle\langle \bar{m}|$ is the projection operator onto band $\bar{m}$.    The quantum geometric tensor is gauge invariant since the projection operator is gauge invariant. Our result can be easily generalized to a set of degenerate isolated bands. In that case, $P$ is the projection operator onto those degenerate bands. 
 
The result, Eq. (\ref{Eq:DS_TRS_geom_isolated}), is  surprising because if we had started from a single-band effective model, we would have obtained only the conventional term. However, here we show that the multiband effects can be written solely in terms of the quantum metric of the isolated Bloch band. 
It is possible to provide a general lower bound on the geometric contribution. 
The quantum generic tensor $\mathcal{R}_{\mu\nu}=g_{\mu\nu}+iB_{\mu\nu}$ is positive semidefinite \cite{Provost1980}. It can be shown \cite{roy2014} that $\det g_{\mu\nu}\ge |B_{\mu\nu}|^2$ and $\mathrm{Tr} \mathcal{R}_{\mu\nu}=\mathrm{Tr}  g_{\mu\nu}\ge 2|B_{\mu\nu}|$. 
For two-dimensional isotropic systems $D^s_{\mathrm{geom},xx}=D^s_{\mathrm{geom},yy}\equiv D^s_{\mathrm{geom}}$ and 
\begin{eqnarray}\label{Eq:geom_bound_2}
D^s_{\mathrm{geom}}\ge 2\Delta^2\sum_{\mathbf{k}}\frac{\tanh{(\beta E_{\bar{m}}/2)}}{E_{\bar{m}}} | B^{\bar{m}}_{xy}|.
\end{eqnarray}
This bound can be straightforwardly generalized to three-dimensional systems.
It is worth mentioning that the quantum metric can be nonzero even if the Berry curvature vanishes~\cite{PRFM2016}.  Generally speaking, in a multiband (i.e. multiorbital) system the Bloch wave functions at adjacent momenta correspond to different linear combinations of the orbital states, and thus the modulus of the overlap of Bloch wave functions is less than unity, resulting in a nonzero quantum metric.

For two-band systems, the noninteracting Hamiltonian can be written as  $\mathcal{H}_{\uparrow}(\mathbf{k})=h_0(\mathbf{k}) I+\mathbf{h}(\mathbf{k})\cdot\bm{ \sigma}$ where $\sigma^i$ are 2 by 2 Pauli matrices and $I$ is the identity matrix. The two Bloch bands are denoted by $|\pm\rangle$ with the band energy $\varepsilon_{\pm}=h_0\pm |\mathbf{h}|$. The corresponding quasiparticle energies are $E_{\pm}=\sqrt{(\varepsilon_{\pm}-\mu)^2+\Delta^2}$. The geometric contribution, Eq. (\ref{Eq:DS_TRS_geom}), can be simplified as

\begin{eqnarray}\label{Eq:DS_TRS_2bands}
D^s_{\mathrm{geom},\mu\nu}
&=&
2\Delta^2\sum_{\mathbf{k},s=\pm}\frac{\tanh{(\beta E_s/2)}}{sE_s}
\frac{|\mathbf{h}|}{\mu-h_0}g_{\mu\nu}.
\end{eqnarray}
The quantum metric for the two bands is the same for the two bands,
$g_{\mu\nu}=\partial_\mu \hat{\mathbf{h}}\cdot\partial_\nu \hat{\mathbf{h}}/2$, while the Berry curvature is $B_{\pm,\mu\nu}=\pm\hat{\mathbf{h}}\cdot (\partial_\mu \hat{\mathbf{h}} \times \partial_\nu\hat{\mathbf{h}})/2$, where $\hat{\mathbf{h}}=\mathbf{h}/|\mathbf{h}|$ is a unit vector. 
 In the two band case the equality $\det g_{\mu\nu}=B^2_{\pm,\mu\nu}$ holds. However, we emphasize that, even in this case, the quantum metric can be nonzero even if the Berry curvature is zero.

As a direct application of Eq. (\ref{Eq:DS_TRS_2bands}), we study the superfluid weight in superconducting graphene. The low energy properties of graphene are governed by the Dirac equation. The superconducting properties of Dirac particles are currently a subject of theoretical and experimental investigations and our theory might provide new insights in this problem. Using Eqs. (\ref{Eq:DS_TRS_conv}) and  (\ref{Eq:DS_TRS_2bands}), we find for the superfluid weight of a graphene-like material (see Appendix \ref{appendix:Dirac} for details)

\begin{eqnarray}
D^s=\frac{1}{\pi}\bigg(\sqrt{\Delta^2+\mu^2}+\frac{\Delta^2}{|\mu|}\ln\frac{|\mu|+\sqrt{\Delta^2+\mu^2}}{|\Delta|}\bigg).
\end{eqnarray}
The first term is the conventional contribution and the second term is the geometric contribution. The same expression is obtained in Ref.~\cite{SuperfluidGraphene} using a different approach. The important role of interband effects was also emphasised in Ref.~\cite{2015JPSJ84h4704M}. Our approach has the advantage of providing a simple derivation  and at the same time it reveals the deep connection with the geometric properties of the manifold of Bloch states and allows to obtain the bound in Eq.~\eqref{Eq:geom_bound_2}.

Before closing this section, it is worth mentioning that
 Eqs. (\ref{Eq:DS_TRS_conv})-(\ref{Eq:DS_TRS_geom_isolated}) are  direct consequences of the simple wave functions $|\psi^\pm_m\rangle$.
 In the Lieb lattice studied in Ref.~\cite{LiebLattice}, the order parameters are orbital dependent, but the flat band states are supported only on two sublattices where the order parameter is uniform by symmetry. Also in this more general case the result~\eqref{Eq:DS_TRS_geom_isolated} in the isolated flat band limit is valid, while Eq.~\eqref{Eq:DS_TRS_geom} is not.

\subsection{Case with $SU(2)$ spin and inversion symmetries}\label{Sec2_b}

In this section we consider the case with spin $SU(2)$ symmetry, which includes, e.g., the HH model. The conventional spin $SU(2)$ rotational symmetry implies that $\mathcal{H}_{\uparrow}(\mathbf{k})=\mathcal{H}_{\downarrow}(\mathbf{k})$. We further assume that there is inversion symmetry so that the Bloch Hamiltonians for opposite momenta are related by a unitary transformation $R$, $\mathcal{H}_{\uparrow}(-\mathbf{k})=R\mathcal{H}_{\uparrow}(\mathbf{k})R^\dag$, and the order parameters are invariant under inversion. Suppose $|n(\mathbf{k})\rangle_{\uparrow}$ is an eigenvector of $\mathcal{H}_{\uparrow}(\mathbf{k})$, then $|n(-\mathbf{k})\rangle_{\uparrow}=R |n(\mathbf{k})\rangle_{\uparrow}$ is an eigenvector of $\mathcal{H}_{\uparrow}(-\mathbf{k})$, therefore
 the matrix elements of the spin-down current operator and the spin-up current operator are related as
\begin{equation}
\begin{split}
&[j_{\nu,\downarrow}(-\mathbf{k})]_{mn}=-_{\downarrow}\langle m(-\mathbf{k}) |\partial_\nu \mathcal{H}_{\downarrow}(-\mathbf{k})|n(-\mathbf{k})\rangle_{\downarrow}\\
&=-_{\uparrow}\langle m(-\mathbf{k}) |\partial_\nu \mathcal{H}_{\uparrow}(-\mathbf{k})|n(-\mathbf{k})\rangle_{\uparrow}\\
&=-_{\uparrow}\langle m(\mathbf{k}) |R^\dag R  \partial_\nu \mathcal{H}_{\uparrow}(\mathbf{k}) R^\dag R|n(\mathbf{k})\rangle_{\uparrow}=-[j_{\nu,\uparrow}(\mathbf{k})]_{mn}\,.
\end{split}
\end{equation} 
Particle-hole symmetry of the BdG Hamiltonian and inversion symmetry imply that for each $\mathbf{k}$ the BdG Hamiltonian has eigenvalues which always appear in pairs with opposite signs $\pm E_{i,\mathbf{k}}$. The eigenfunction for the positive energy $+E_{i,\mathbf{k}}$ can be written as $|\psi^+_i\rangle=\sum_n u_{in}|n\rangle\otimes|+\rangle+v_{in}R^\ast|n^\ast\rangle \otimes |-\rangle$ and the corresponding negative energy state is $|\psi^-_i\rangle=\sum_n -v^\ast_{in}|n\rangle \otimes|+\rangle+u^\ast_{in}R^\ast|n^\ast\rangle\otimes|-\rangle$. Using these expressions, it is straightforward to write the superfluid weight in terms of $u$, $v$, and the current operator. 

To further investigate the geometric contribution of the superfluid weight in systems without TRS, we study a two-band model, i.e., the HH model in detail. We find that, in the isolated band limit, $D^s_{\mathrm{geom}}$ is also related to the quantum metric (see Appendix \ref{app:derivation_HH}),  
 \begin{eqnarray}\label{Eq:SW_geom_isolated_HH}
 D^s_{\mathrm{geom},\mu\nu}
 &\approx & \Delta^2 \sum_{\mathbf{k}}\frac{ h^2_z(E_-+E_0-|\mathbf{h}|)^2}{2E^2_0E^3_-}g_{\mu\nu},
 \end{eqnarray}
 where $E_\pm=\sqrt{(h_0-\mu)^2+\Delta^2+|\mathbf{h}|^2\pm2|\mathbf{h}| E_0}$ are the energies of the higher ($+$) and lower ($-$) quasiparticle bands and  $E_0=\sqrt{(h_0-\mu)^2+\Delta^2 \hat{h}^2_z}$ is the gap between the two branches (see Appendix \ref{app:derivation_HH}).

\begin{figure}
\includegraphics[width=\columnwidth]{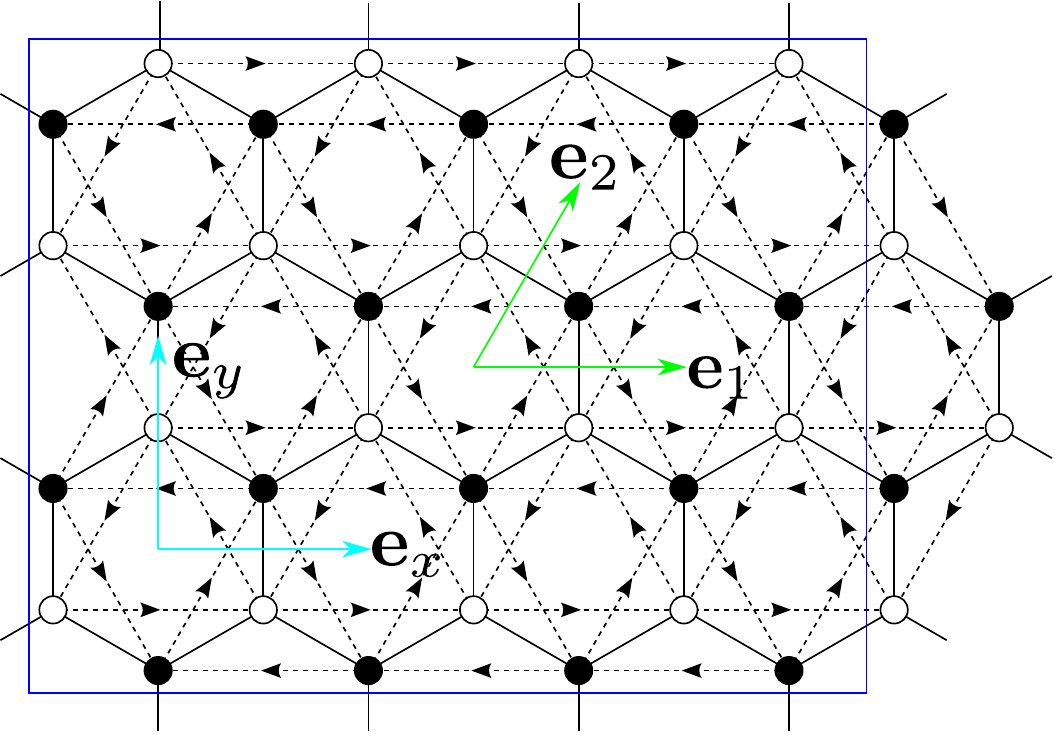}
\caption{A patch of the Haldane/Kane-Mele model.  The models consist of a honeycomb lattice with nearest- and next-nearest-neighbour (NNN)  hoppings.  Here $\mathbf{e}_1, ~\mathbf{e}_2$ are primitive vectors of the honeycomb lattice while $\mathbf{e}_x, ~\mathbf{e}_y$ are an orthogonal basis. Black and white dots are $A$ and $B$ sublattice sites. 
The  arrows  show  the  direction  of  positive
phase winding for the complex NNN hoppings for the spin up fermions.
The  rectangle shows the 32-site cluster used in exact diagonalization. This cluster has $C_6$ symmetry. }\label{Fig:Lattice}
\end{figure}

\section{Superfluid weight in the Kane-Mele-Hubbard  and Haldane-Hubbard models}\label{Sec3}

In this section, we apply our theory to the attractive KMH and HH models. The KMH model has TRS while the HH model does not, corresponding to the two cases discussed in Sec. \ref{Sec2_a} and Sec. \ref{Sec2_b}, respectively.

The interplay between band topology and Hubbard interactions, both repulsive and attractive, has been studied extensively and various phases have been predicted \cite{PhysRevB.82.075106,Kou_HH2011,Kou_HH2012,HH_attractive_kou,Kou_2014,HH_attractive_kou_2,Fiete2014,Ding_2015,PRB.91.134414,PRB.91.161107,PRB.93.115110,PRB.93.075131,PRL.116.137202,151205118,151103833Z,C1,PhysRevB.94.035109}. Investigation of possible exotic phases goes beyond the scope of this work.  Instead, we focus on the attractive interaction and search for the simplest BCS mean-field solutions. 

We write the Hamiltonian as
\begin{eqnarray}
H=H_{\mathrm{kin},\uparrow}+H_{\mathrm{kin},\downarrow}+H_{\mathrm{int}}-\mu N.
\end{eqnarray} 
 For the HH model $H_{\mathrm{kin},\uparrow}=H_{\mathrm{kin},\downarrow}$, while for the KMH model $H_{\mathrm{kin},\uparrow}$ and $H_{\mathrm{kin},\downarrow}$ are related by TRS. The kinetic energy is 
\begin{eqnarray}\label{Eq:HHNeg}
H_{\mathrm{kin},\sigma}=-t\sum_{\langle i,j\rangle}c^{\dag}_{i\sigma}c_{j\sigma}-t'\sum_{\langle\langle i,j\rangle\rangle}e^{i\phi^{\sigma}_{ij}}c^{\dag}_{i\sigma}c_{j\sigma},
\end{eqnarray}
where $c^{\dag}_{i\sigma}$ ($c_{i\sigma}$)  creates (annihilates) a spin-$\sigma$ fermion on site $i$. Here $\langle ij\rangle$ and $\langle\langle ij\rangle\rangle$ denote the nearest and next-to-nearest bonds on the honeycomb lattice. The directed phase $\phi^\sigma_{ij}$ represents the magnetic fields felt by spin-$\sigma$ fermions, see Fig. \ref{Fig:Lattice}. In the HH model, the complex hoppings are the same for spin-up and spin-down particles while they are complex conjugates ($\phi_{ij}^\uparrow = -\phi_{ij}^\downarrow$) in the KMH model.

The parameters $t$, $t'$, and $\phi_{ij}$ can be tuned in cold atom experiments \cite{Jotzu2014}. In this article, we take $t=1$ as the energy unit and set $\cos(\phi)=t/(4t')=3\sqrt{3/43}$. Under these parameters, the lower band ($n<1$) is quasi-flat with large flatness ratio and the upper band ($n>1$) is strongly dispersive  \cite{fqh2011}, see Fig. \ref{Fig:noninteracting}. This allows us to study the flat band limit as well as a dispersive band by tuning the filling. The onsite energy difference is set to be zero, so there is inversion symmetry, with $R=\sigma^x$, meaning that the Hamiltonian is invariant under the interchange of $A$ and $B$ sublattices (black and white dots in Fig. \ref{Fig:Lattice}). The Hamiltonians have 6-fold rotational symmetry that can be used to simplify computations. Using this symmetry, we show in  Appendix \ref{appendix:C6} that the superfluid weight  tensor is proportional to the identity, $D^s_{\mu\nu}=D^s\delta_{\mu\nu}$. This property is also checked numerically.
We decouple the attractive Hubbard interaction as in Eq. (\ref{Eq:deouple}) and the orbital index $\alpha$=$A, B$ denotes the sublattice.
We search for mean-field solutions that preserve the inversion symmetry, i.e., $\Delta_A=\Delta_B$.

\subsection{Validity of the BCS approximation}
\label{Sec3:ValidityBCS}

Our theory of superfluid weight is based on the BCS approximation, so before calculating the superfluid weight,  we first discuss the validity of this approximation.

For a flat band with TRS, it has been shown \cite{2016arXiv160800976T} that in the isolated flat band limit, the BCS wave function is an exact ground state provided that the uniform pairing condition is satisfied. For the KMH model, this condition means 
\begin{eqnarray}\label{Eq:UPC}
n_\alpha=\frac{V_c}{(2\pi)^2}\int_{\mathrm{B.Z.}}\mathrm{d}\mathbf{k}\,|\left\langle\alpha |m(\mathbf{k})\right\rangle|^2=\frac{1}{2},
\end{eqnarray}
where 
$V_c$ is the volume of the unit cell, $|m(\mathbf{k})\rangle$ is the Bloch function, and $|\alpha\rangle$ the wave function corresponding to a lattice site. For the KMH and HH models Eq. (\ref{Eq:UPC}) is a result of inversion symmetry. For the dispersive band, 
because of the existence of the well-defined Fermi surface, the BCS  approximation is reliable  in the weak coupling limit, as the BCS instability is an intrinsic instability of a Fermi liquid \cite{Shankar_RMP_1994}. 

The validity of the BCS wave function in the $U\to \infty$ limit has been investigated at both zero \cite{Leggett1980} and finite \cite{BCSBEC1985} temperature in the study of the BCS-BEC crossover \cite{RevModPhys.62.113}. Here we shall argue that, by mapping the Hubbard model to effective spin models \cite{PhysRevA.79.033620}, the 
BCS approximation is  good in the strong coupling limit for our systems. The starting point is the atomic limit, where each site is empty $|0\rangle $ or doubly occupied $c^{\dag}_\uparrow c^\dag_\downarrow |0\rangle$. These two states form an $SU(2)$ representation and can be viewed as a local `spin'. Turning on a weak tunnelling $t^\sigma_{ij}$, we can project the hopping terms to the local spin degrees of freedom and get an effective spin Hamiltonian. Up to the second order, it reads
\begin{eqnarray}
H_{\mathrm{eff}}&=&-\sum_{i,j}\frac{2t^{\uparrow }_{ij} t^{\downarrow}_{ ij}}{U}T^+_i T^-_j +\mathrm{H.c.}\nonumber\\ 
&+& \sum_{i,j}\frac{2(|t^{\uparrow}_{ ij}|^2+|t^{\downarrow}_{ ij}|^2)}{U}T^z_i T^z_j -\mu \sum_i T^z_i, 
\end{eqnarray}
where $T^+_i=c^{\dag}_{i\uparrow} c^\dag_{i\downarrow}$, $T^-_i=c_{i\downarrow} c_{i\uparrow} $ and  $T^z_i=(n_{i\uparrow}+n_{i\downarrow}-1)/2$. The chemical potential becomes the magnetic field and the magnetization $\langle T^z_i\rangle$ is determined by the filling.
Treating the pseudo-spin as a classical vector, that is, $\langle T^z_{i}\rangle=(n-1)/2$ and $\langle T^+_i\rangle \langle T^-_i \rangle+\langle T^z_{i}\rangle^2=1/4$, we find $\langle T^+_i \rangle = \langle T^-_i \rangle=\sqrt{2n-n^2}/2$. The order parameter in the large $U$ limit calculated using the original Hubbard model is $\Delta=U\langle T^+_i \rangle=U\sqrt{2n-n^2}/2$, which coincides with the classical spin approximation. Thus the BCS theory in the large $U$ limit is the classical approximation of the effective spin model, which should be  good  when the frustration is weak \cite{PhysRev.83.1260}. 
For our models, fluctuations are mainly induced by the $t'$ terms: if $t'$ vanishes, the effective model becomes the anti-ferromagnetic Heisenberg model on the honeycomb lattice and the classical approximation is good for this model \cite{Fouet2001}. 
For the parameters studied in this article, $t'^2/t^2\approx 0.1$ is small, so we expect the BCS approximation to be good in the large $U$ limit. It is possible that the classical approximation breaks down for other choices of the parameters, but this is beyond the scope of this article.

\begin{figure}[h]
	\includegraphics[width=\columnwidth]{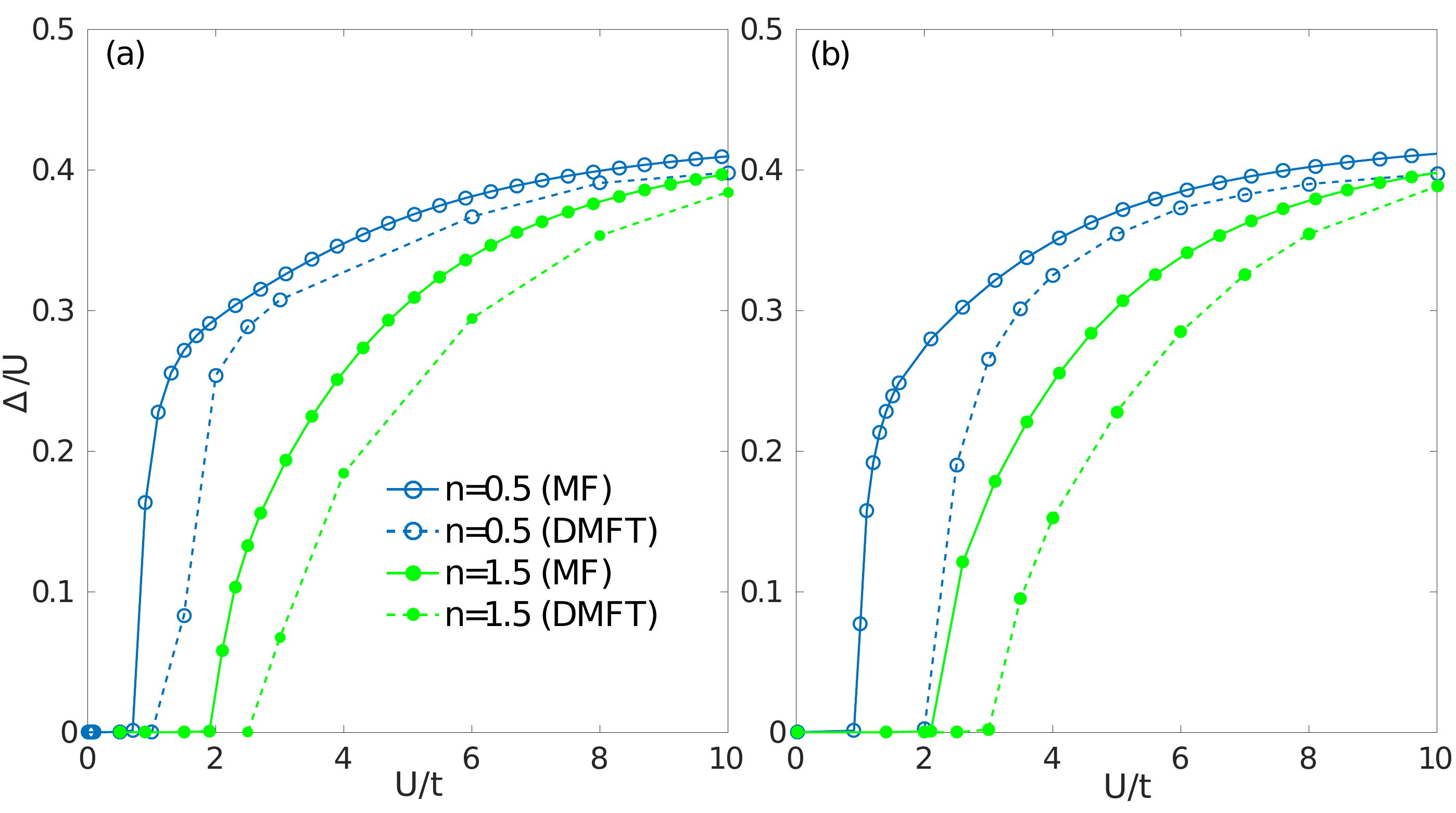} 
	\caption{Order parameters as functions of $U$ for different fillings obtained with mean-field BCS theory and DMFT at temperature $T=0.1t$. (a), KMH model. (b), HH model. DMFT captures the local fluctuations and therefore predicts higher critical values of the interaction. Away from the phase transition, the agreement between the BCS and DMFT  confirms the validity of the BCS approximation.  }\label{Fig:DMFT_MF_Delta}%
\end{figure} 
To further confirm the validity of the mean-field theory, 
we employ DMFT to calculate the order parameters, see Appendix \ref{appendix:DMFT}. 
 In Fig. \ref{Fig:DMFT_MF_Delta} we plot the order parameters as functions of the interaction strength $U$ for fillings chosen to coincide with the middle point of the flat band and the dispersive band, in the noninteracting limit. Because of the finite temperature $T=0.1t$, there is a critical interaction strength $U$ below which the superfluid is destroyed by thermal fluctuations. The local quantum and thermal fluctuations included in DMFT increase the critical $U$ compared to the BCS results. However, away from the phase transition the BCS results agree well with DMFT signalling that local fluctuations do not play an important role deep within the ordered state. 
 
 As the DMFT might be biased by the choice of order parameters and the lack of long range correlation effects, we also apply the unbiased ED method to calculate the Drude weight, which is equivalent to the superfluid weight in the bulk limit for a gapped system \cite{swz}. ED is performed   on a 32-site cluster, which preserves the $C_6$ rotational symmetry. Since the cluster is large, we can only calculate at fillings $n=1/8$ and $n=15/8$, corresponding to the bottom of the flat band and the top of the dispersive band, respectively. As shown in Fig. \ref{Fig:SW_Zero}, the Drude weight obtained from ED is in quantitative agreement with the mean-field superfluid weight.

\subsection{Superfluid weight}

\subsubsection{Zero temperature results}

Fig. \ref{Fig:SW_Zero} presents our main results for the zero temperature superfluid
weight. An important feature shown in Fig. \ref{Fig:SW_Zero} (a) and (c) is that,
although the lower band is not strictly flat, resulting in a finite
conventional contribution in the noninteracting limit, the geometric
contribution is still important: it is comparable to, or even larger
than the conventional one. For weak interactions, the geometric
contribution increases linearly with $U$, and reaches a maximum around
$U/t=4$. By contrast, the conventional contribution first decreases as $U$
becomes larger than the bandwidth, then increases as $U$ becomes
large enough to induce pairing in the other band, and finally
decreases again with increasing $U$. Together these effects produce the
peculiar nonmonotonic behaviour of the total superfluid weight. For
the dispersive bands [Fig. \ref{Fig:SW_Zero} (b) and (d)], the conventional
contribution dominates and, as a result, the superfluid weight is
roughly constant for weak interactions and decreases monotonically
with increasing $U$.

Fig. \ref{Fig:SW_Zero} also shows that $D^s$ from mean-field theory is in quantitative agreement with the ED results, which further confirms the validity of our theory. 
To limit the basis size of the ED calculation on the 32-site cluster, we use very low and high filling fractions. However, we emphasize that the qualitative behavior of the superfluid weight  depends only on  whether the band is flat or not. To confirm this, we also compare mean-field results to ED on smaller clusters at fillings $1/3$ and $5/3$ in appendix \ref{appendix:ED}, and find good agreement also in those cases.

\subsubsection{Finite temperature results}

\begin{figure}[h]
 \includegraphics[width=\columnwidth] {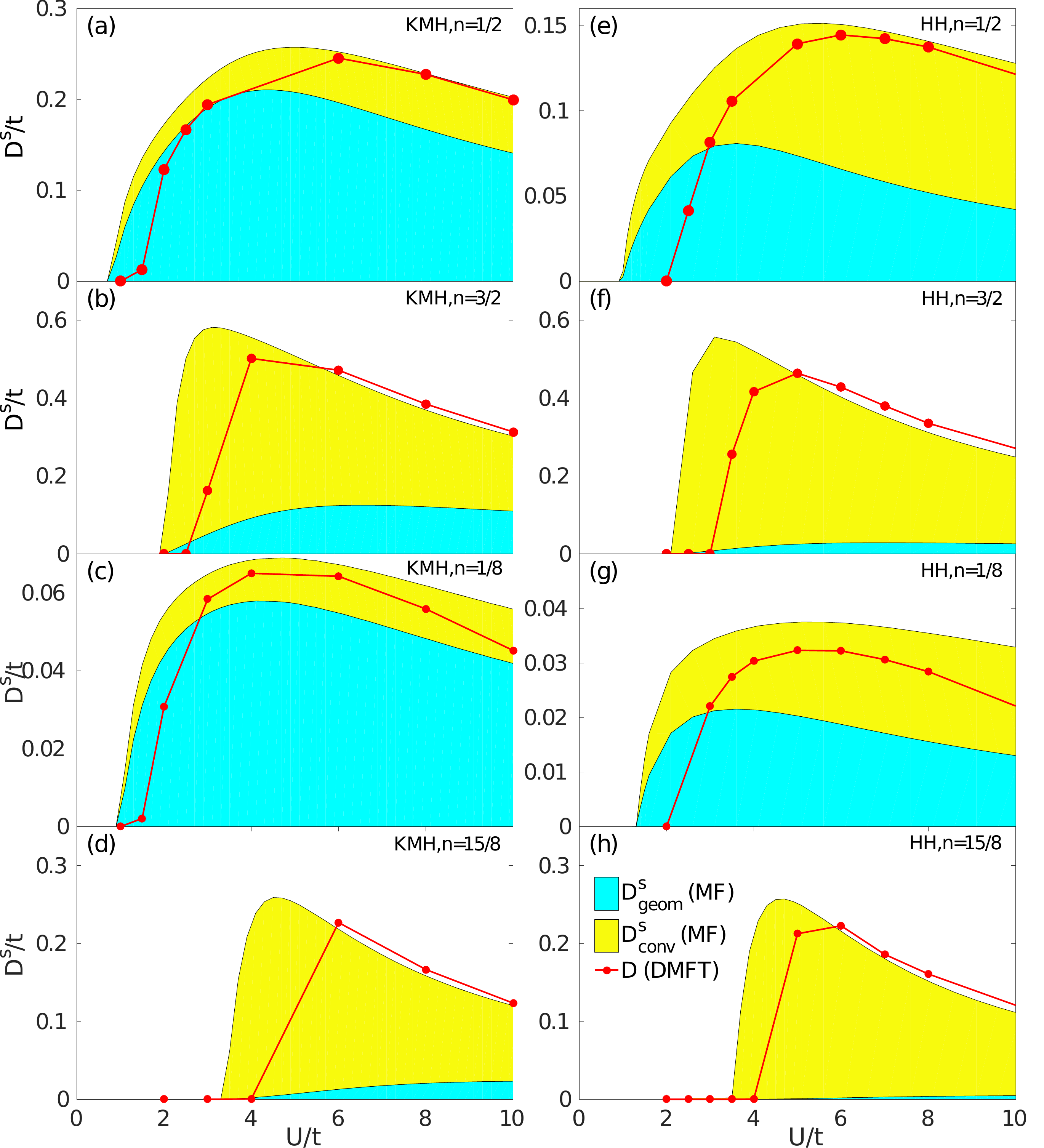}
 	\caption{
Finite-temperature superfluid weight for the KMH  model [(a)-(d)] and the HH model [(e)-(h)] at different fillings. The temperature is $T=0.1t$. Different from the zero temperature case,  both the flat and dispersive bands supefluid weight show nonmonotonic behavior. However, a remarkable similarity between the finite and zero temperature cases is that,  
the geometric contribution is  important for the flat bands.  Away from the phase transition the mean-field results agree well with DMFT results. 
 		 }\label{Fig:SW_FiniteT}
 \end{figure}

 In Fig. \ref{Fig:SW_FiniteT} we plot the finite temperature superfluid weight as a function of the interaction. 
  The DMFT results are obtained by calculating the system's current response to a small vector potential, see Appendix \ref{appendix:DMFT} for details. Away from the phase transition DMFT agrees well with the mean-field results. The slightly differing results in Figs. \ref{Fig:SW_FiniteT} (c) and (g) can be explained by noting that, in addition to the phase transition visible in the figures, the DMFT solution also exhibits an upper critical $U$, which is relatively low for very low filling fractions.
 
Below the critical interaction, the superfluid vanishes as expected and it grows rapidly as the interaction strength exceeds the critical value and then decreases after reaching a maximum. 
The total superfluid weight shows a  nonmonotonic behavior for the flat as well as the dispersive bands. However, as shown in Fig. \ref{Fig:SW_FiniteT}, the geometric contribution is  important for the flat band while for the dispersive band the superfluid weight comes mainly from the conventional contribution.

To conclude this section, our  mean-field theory is in good agreement with the state of the art DMFT and ED. The advantage of the mean-field theory is that it provides important understanding of the superfluid weight  in terms of the conventional  and geometric contributions.  
We find that the geometric contribution is large and dominant or comparable to the conventional one for the flat band, while for the dispersive band the conventional contribution dominates. Therefore the concept of geometric contribution is important for a proper understanding of the superfluid properties of a (quasi-)flat band. 

\begin{figure*}
 \includegraphics[width=2\columnwidth]{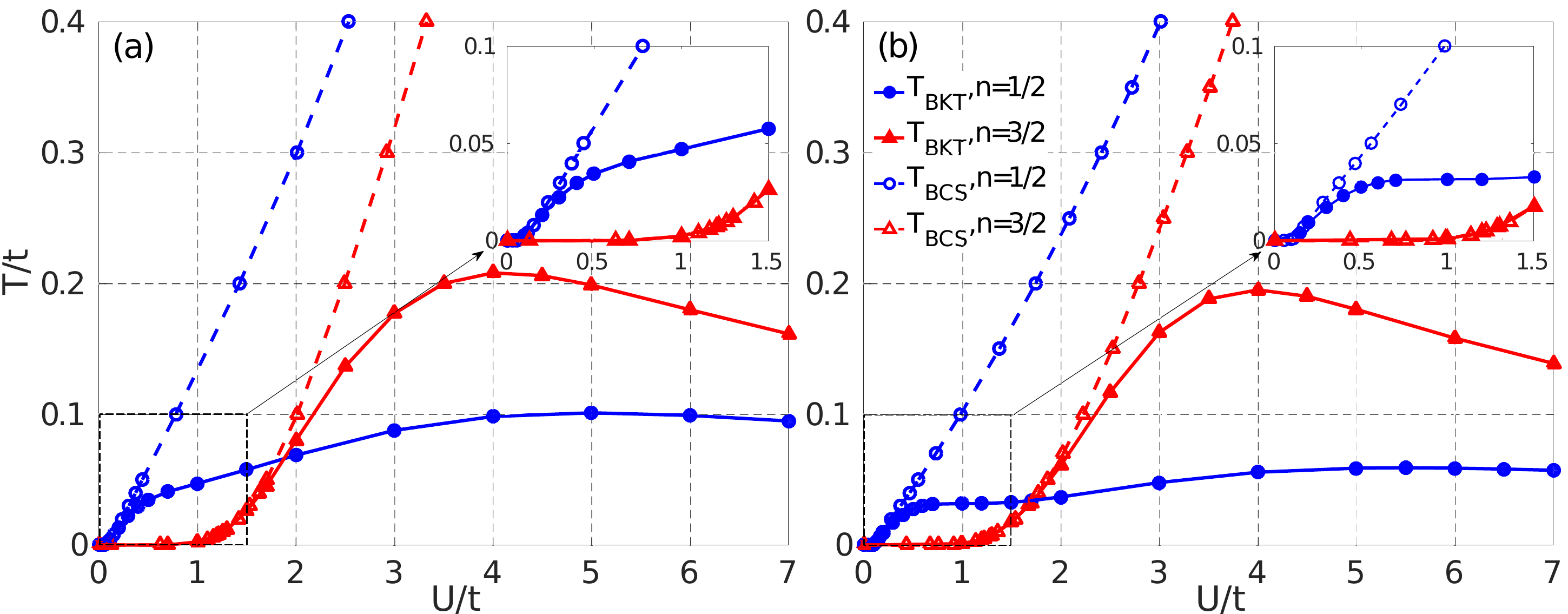}
\caption{The mean-field transition temperature $T_{\mathrm{BCS}}$ and BKT temperature $T_{\mathrm{BKT}}$ for the KMH (a) and HH (b) models.  In the weak coupling limit, $T_{\mathrm{BCS}}\propto U$ for the flat band ($n< 1$, blue) while it is exponentially small for the dispersive band ($n> 1$, red). In the strong coupling limit, $T_{\mathrm{BCS}} \propto U$. The BKT temperature is nonmonotonic. In the weak coupling limit it behaves  the same as $T_{\mathrm{BCS}}$ and in the strong coupling limit it decreases with increasing $U$. The superfluid transition temperature is determined by $T_\mathrm{BKT}$ and it is clearly shown that the flat band has a much higher transition temperature in the weak coupling limit.}\label{Fig:Tc}
\end{figure*}
\subsection{Mean-field transition temperature and Berezinskii-Kosterlitz-Thouless temperature}
Finally, we discuss the superfluid transition temperature in these models. The dashed lines in 
Fig. \ref{Fig:Tc} show the BCS mean-field transition temperature $T_{\mathrm{BCS}}$ as a function of $U$ for the flat and dispersive bands. In the weak coupling limit, $T_{\mathrm{BCS}}$ for the flat band is proportional to the interaction strength while for the dispersive band it is exponentially small,  which are consistent with previous theoretical predictions \cite{Volovik_Flatband1, Volovik_Flatband2, FlatBandTc}. In the strong coupling limit, the binding energy of Cooper pairs scales as $U$, and therefore the mean-field critical temperature also scales as $U$ in the strong coupling limit.

The superfluid weight gives the phase coherence energy scale and, in two dimensions, determines the Berezinskii-Kosterlitz-Thouless (BKT) temperature \cite{BKT_B,BKT_KT} via the universal relation $T_{\mathrm{BKT}}=\pi D^s(T_{\mathrm{BKT}})/8$. We use the mean-field $D^s$ to determine the BKT temperature, see the solid lines in  Fig. \ref{Fig:Tc}.  The BKT temperature , which is smaller than the mean-field transition temperature, gives the superfluid critical temperature. Different from $T_{\mathrm{BCS}}$, the BKT temperature 
$T_{\mathrm{BKT}}$ increases with $U$ for weak interactions until it reaches a maximum  and then decreases when $U$ increases further. 
 The nonmonotonic behavior of $T_{\mathrm{BKT}}$ is a reflection of the BCS-BEC crossover.

 For the KMH and HH models studied in this article, the dispersive band has a higher transition temperature than the flat band for intermediate and strong interactions, and $T_{\mathrm{BKT}}$ is maximized around  $U/t\approx 4$. This is not surprising since the dispersive band has larger superfluid weight than the flat band. However, in the weak coupling limit, the flat band has a higher critical temperature. In fact, one can see from the inserts in Fig. \ref{Fig:Tc} that, in the weak coupling limit, $T_{\mathrm{BKT}}$ behaves like $T_{\mathrm{BCS}}$: For the flat band it increases linearly with $U$, while for the dispersive band it is exponentially small. At first sight this may be surprising. Since the dispersive band has higher surperfluid density, one may expect that it also has higher transition temperature. This is indeed true for intermediate and strong interactions where the BCS temperature is high, but for weak interactions the BCS temperature of the dispersive band is much lower than  that of the flat band. Therefore the superfluid weight for the dispersive band decays much faster with increasing temperature, resulting in a lower BKT temperature.
This is visible also in Fig. \ref{Fig:SW_FiniteT} where the critical interactions for the flat bands are lower.

Our results confirm that flat bands indeed provide a way to  improve the superfluid transition temperature in the weak to intermediate coupling regime.
This is particularly important for real materials where the effective attractive interaction between electrons is expected to span this range of couplings.

\section{Conclusion and outlook}\label{Sec4}

In this work, we investigate the superfluid properties of generic attractive Hubbard models defined on a lattice with complex orbital structure. We focus on the effects due to the multiband (multiorbital) nature of the system that are not present in a single-band lattice model, i.e. a model defined on a simple Bravais lattice. Our work is based on linear response theory which provides a convenient framework for calculating the superfluid weight and  the foundation for addressing the superfluid properties of realistic systems. This approach is equivalent to the one used in Ref.~\cite{PT} based on thermodynamic potentials, but has several advantages. 

In the mean-field BCS approximation, we obtain the general result for the superfluid weight given in Eq.~\eqref{Eq:SW} expressed in terms of the BdG Hamiltonian $\mathcal{H}$ and the corresponding eigenstates $|\psi_i\rangle$ and eigenvalues $E_i$. Our essential finding is that even those quasiparticle states that are not adiabatically connected in the noninteracting limit to the isolated partially filled band of interest  can provide an important contribution to the superfluid weight in the isolated band limit, i.e. the limit where the band gap diverges. This contribution is crucial for understanding the transport properties of (quasi-)flat bands.

Linear response theory explains the cause behind this counterintuitive phenomenon in terms of the off-diagonal (interband) matrix elements of the current operator [Eq.~\eqref{eq:current_operator}] that are proportional to the band gap and therefore can be quite large in general. We also find that in the case of TRS breaking it is necessary to include band mixing in the quasiparticle states to obtain the correct result for the superfluid weight in the isolated band limit, while band mixing is absent in the case of TRS and uniform pairing.

The multiband effects on the superfluid properties that we find are interaction effects, since in the noninteracting limit all terms with off-diagonal matrix elements of the current operator disappear. Remarkably, the interband effects in the isolated band limit can be expressed solely in terms of the properties of the isolated band. Specifically, we find that they lead to a contibution proportional to the quantum metric, a band structure invariant obtained from the Bloch functions, both in the case with TRS and without TRS. We call this the geometric contribution to the superfluid weight.

With respect to Ref.~\cite{PT} we extend the general result in the TRS case to an isolated but not necessarily flat band, Eq. \eqref{Eq:geom}, and provide a novel relation between superfluid weight and Berry curvature, Eq. \eqref{Eq:geom_bound}, that is useful even in the case of a band with zero Chern number and is therefore of more general applicability. We emphasize that it is the nontrivial geometry rather than the nontrivial topology that affects the superfluid weight.

A physical interpretation of the connection between the superfluidity and geometry has started to emerge recently, at least in the TRS case~\cite{2016arXiv160800976T}. It can be traced back to the fact that the Hubbard interaction produces pair-hopping terms between overlapping Wannier functions, and these hopping processes provide the kinetic energy to the Cooper pairs even if the band is perfectly flat and the kinetic energy is zero for unpaired particles. The quantum metric enters precisely as a measure of the overlap between the Wannier functions. More work is need to understand the case without TRS.

The quantum metric also appears in the orbital magnetic susceptibility \cite{GYN2015,PRFM2016}, which, like the superfluid weight, is a response of an electron system to an external magnetic field. The role played by quantum geometry in these response functions is an interesting topic for further study.

As an application of our theory we study the superfluid weight in the attractive KMH and HH models, confirming our BCS mean-field results using state of the art DMFT and ED methods. We focus on a specific set of hopping parameters such that both the case of a quasi-flat and a highly dispersive band can be studied by tuning the filling in the same models. For the flat band, 
the geometric contribution to the superfluid weight is important at  both zero and finite temperature. Using our results for the superfluid weight we calculate the BKT temperature and find that  the flat band indeed has higher transition temperature in the weak coupling limit. Our results could also be of immediate experimental interest, as the Haldane-Hubbard model has recently been realized in cold atomic gases experiments~\cite{Jotzu2014,Flaschner1091}, and a realization of the Kane-Mele-Hubbard model has also been proposed~\cite{Jotzu2014}. In these experiments the tunable atom-atom interaction would provide an ideal platform for studying the interplay of Bloch band geometry and superfluidity.

A very interesting topic for further research is the relevance of ours findings in the context of solid state  systems. We have pointed out that the interplay of complex lattice geometry, band structure and interaction can produce qualitatively new effects on the superfluid properties that cannot be captured by single-band Hamiltonians such as a Hubbard model on a square lattice. As shown in Fig.~\ref{Fig:Tc} a flat band can significantly enhance the critical transition temperature in the range from weak to intermediate interactions,  $0\leq U \leq 1.5t$ in our case, with respect to a dispersive band. Thus flat bands or quasi-flat bands may be at the root of high-$T_{\rm c}$ superconductivity, since most unconventional superconductors are characterized by complex orbital structure and are in a interaction regime where the geometric term should be important. In contrast in conventional superconductors driven by weak electron-phonon coupling the geometric term, if present, is likely to be overshadowed for very small values of $U$ by the conventional one, both in the case of a dispersive band or in a quasi-flat band,  as seen in Figs.~\ref{Fig:SW_Zero} and~\ref{Fig:Ds_Finitesize}.

The superfluid weight, which is related to the magnetic penetration depth, is a powerful probe of the microscopic properties of carriers of the supercurrent and is currently being intensively investigated in high-$T_c$ superconductors \cite{I2016}. Note that the results of these recent experiments are interpreted in the framework of BCS theory but only accounting for the conventional contribution to the superfluid weight and neglecting the geometric contribution which may be large. Further work is necessary in order to assess the importance of the geometric term in unconventional high-$T_c$  superconductors.

\acknowledgments
We  acknowledge  helpful  discussions  with  Arun  Paramekanti. We thank Grigory Volovik for useful comments. We thank Timo Hyart for useful discussions and for bringing Ref.~\cite{Moon:1995} to our attention.
This work was supported by the Academy of Finland through its  Centers  of  Excellence  Programme  (2012-2017)  and under  Project  Nos.  263347,  284621, and  272490,  and by  the  European  Research  Council  (ERC-2013-AdG-340748-CODE). This project has received funding from the European Union's Horizon 2020 research and innovation programme under the Marie Sklodowska-Curie grant agreement No 702281 (FLATOPS). T.I.V. acknowledges the support from the V\"ais\"al\"a foundation.
Computing resources were provided by CSC - the Finnish IT Centre for Science and the Triton cluster at Aalto University.

\appendix
\section{Equivalence  of the superfluid weight defined through thermodynamic potential and linear response theory}\label{appendix:equivalence}
In this Appendix we show that the definition of the superfluid weight used in Ref.~\cite{PT} is equivalent to the one used in the present work. Following Ref.~\cite{swz}, we define the superfluid weight through the static Meissner effect.  The order of $q_x\to 0$ and $q_y\to 0$ does not affect the one-loop result \cite{swz}, so $D^s_{\mu\nu}=K_{\mu\nu}(\mathbf{q}\to 0,i\omega=0)$. In \cite{PT}, the superfluid weight is defined as the second order derivative of the free energy with respect to a vanishing constant phase $\mathbf{q}$ of the order parameter. Here  $\mathbf{q}$ is nothing but a constant vector potential $\mathbf{A}$. The free energy is  $F(\mathbf{A})=-\beta^{-1}\ln Z(\mathbf{A})$, with the partition function $Z(\mathbf{A})=\mathrm{Tr}\exp{\{-\beta H(\mathbf{A})\} }$. It is enough to expand $H(\mathbf{A})$ to the second order of $\mathbf{A}$, $H(\mathbf{A})\approx H+j^p_{\mu}A_{\mu}+T_{\mu\nu} A_\mu A_\nu/2$. Taking the second order derivative of the free energy with respect to $\mathbf{A}$, we get,
$D^s_{\mu\nu}=\langle T_{\mu\nu} \rangle-\int^{\beta}_0 \mathrm{d}\tau \langle j_{\mu}(\tau)j_{\nu}(0) \rangle=K_{\mu\nu}(\mathbf{q}\to 0, i\omega=0)$. 

We briefly discuss the relation between the superfluid weight and the phase stiffness.  Phase fluctuations are introduced to the order parameter as $\Delta(\mathbf{r})=\Delta e^{2i\phi(\mathbf{r})}$. A factor of 2 is introduced because the Cooper pair carries twice the charge of the fermion. 
To get an effective theory for the phase fluctuations, we perform a local $U(1)$ transformation $c^\dag(\mathbf{r})\to c^\dag(\mathbf{r}) e^{-i\phi(\mathbf{r})}$. The phase then enters the kinetic energy, i.e., $t_{ij}\to t_{ij} e^{-i[\phi(\mathbf{r}_i)-\phi(\mathbf{r}_j)]}$, and it is clear that the phase fluctuations can be absorbed into the gauge field.  For long wavelength fluctuations we can perform the gradient expansion
$\phi(\mathbf{r}_i)-\phi(\mathbf{r}_j)\approx(\mathbf{r}_i-\mathbf{r}_j)\partial_\mathbf{r}\phi(\mathbf{r})$. Integrating over fermions, the effective action for the phase fluctuations in the long wavelength limit is
\begin{eqnarray}\label{XY}
S_{\mathrm{eff}}[\theta]=\int\mathrm{d}\mathbf{r}\int \mathrm{d}\tau \frac{1}{8}D^s_{\mu\nu}\partial_\mu\theta\partial_\nu\theta,
\end{eqnarray}
where we have defined $\theta=2\phi$ such that the periodicity of the variable $\theta$ is  $2\pi$. Assuming that the superfluid weight is proportional to the identity (see Appendix \ref{appendix:C6}), Eq. (\ref{XY}) becomes a classical XY model. In two dimensions, the XY model has a finite temperature BKT transition and the critical temperature is given by the universal relation \cite{BKT_B,BKT_KT} $T_{\mathrm{BKT}}=\pi D^s(T_{\mathrm{BKT}})/8$.

\section{Derivation of the isolated band limit of the superfluid weight}\label{appendix:corrections}

\subsection{Derivation of Eq. (\ref{Eq:DS_TRS_geom_isolated})}
Suppose the band $\bar{m}$ is the isolated band we are interested in, that is to say, the chemical potential $\mu$ lies within $\varepsilon_{\bar{m}}$ and the other bands are far away from $\bar{m}$.  
Without loss of generality,  we take $\bar{m}$  to be the lowest band and write the dispersions of other bands as $\varepsilon_{n} = W_n+\varepsilon_{\bar{m}}$, where the band gaps $W_n$ are positive and $W_n>>|\varepsilon_{\bar{m}}-\mu|, |\Delta|$. 
 Now we can perform a large $W_n$ expansion to simplify the superfluid weight. 
 
 It is easy to verify that, for the conventional superfluid weight, the contribution from band $n\ne\bar{m}$ is of order $1/W^3_n$ and thus can be neglected. 
 
 The geometric contribution containing the isolated band is
 \begin{eqnarray}\label{Eq:expansion}
 && D^s_{\mathrm{geom},\mu\nu}= \nonumber\\
 && 2\Delta^2\sum_{\mathbf{k},m\ne \bar{m}}\left[\frac{\tanh{(\beta E_{\bar{m}}/2)}}{E_{\bar{m}}}-\frac{\tanh{(\beta E_m/2)}}{E_m}\right]\nonumber\\
  &&\times \frac{ W_m}{W_m+2 \varepsilon_{\bar{m}}-2\mu}(\langle  \partial_\mu  \bar{m}|  m \rangle  \langle m|\partial_\nu  \bar{m}\rangle+\mathrm{H.c.})\nonumber\\
 &&\approx 2\Delta^2 \sum_{\mathbf{k}}\frac{\tanh{(\beta E_{\bar{m}}/2)}}{E_{\bar{m}}}g^{\bar{m}}_{\mu\nu} \nonumber\\
 &&- 2 \Delta^2\sum_{\mathbf{k},m\ne \bar{m}}\frac{1}{W_m}\left[ 1+2\frac{\varepsilon_{\bar{m}}-\mu}{E_{\bar{m}}}\tanh{(\beta E_{\bar{m}}/2)}\right]\nonumber\\
 &&\times(\langle  \partial_\mu  \bar{m}|  m \rangle  \langle m|\partial_\nu  \bar{m}\rangle+\mathrm{H.c.}).
 \end{eqnarray}
 We have used the approximation $\tanh{(\beta E_m/2)}/E_m\approx 1/W_m$ and the correction to this approximation is of order $1/W^2_m$. 
 For the strictly flat band,  $\varepsilon_{\bar{m}}=\mu$, the  $1/W_m$ correction  in Eq. (\ref{Eq:expansion}) is negative.

The remaining terms are ($m,n\ne \bar{m}$)
\begin{eqnarray}\label{Eq:high_order}
&&\Delta^2\sum_{\mathbf{k},m\ne n}\left[\frac{\tanh{(\beta E_m/2)}}{E_m}-\frac{\tanh{(\beta E_n/2)}}{E_n}\right]\nonumber\\
&&\times \frac{\varepsilon_n-\varepsilon_m}{\varepsilon_m-\varepsilon_n-2\mu}(\langle  \partial_\mu  m|  n \rangle  \langle n|\partial_\nu  m\rangle+\mathrm{H.c.}).\nonumber\\
&\approx&\Delta^2\sum_{\mathbf{k}, m\ne n}\left[\frac{1}{W_n}-\frac{1}{W_m}\right]\frac{ (W_m-W_n)}{W_m+W_n+2 \varepsilon_{\bar{n}}-2\mu}\nonumber\\
&&\times (\langle  \partial_\mu  n|  m \rangle  \langle m|\partial_\nu  n\rangle+\mathrm{H.c.}),
\end{eqnarray}
which are at least of order $1/W_n$.  
If all the gaps are of the same order, i.e., $W_n\sim W$ and $|W_m-W_n|<<W$, then Eq. (\ref{Eq:high_order}) is of order $1/W^3$. Therefore, in this case  Eq. (\ref{Eq:expansion}) is exact up to order $1/W_m$, which is negative for the strictly flat band. Thus the lowest order result, Eq. (\ref{Eq:DS_TRS_geom_isolated}) in the main text, is actually an upper bound of the geometric superfluid weight.

\subsection{  Derivation of Eq. (\ref{Eq:SW_geom_isolated_HH})}
\label{app:derivation_HH}

Here we  present a derivation of the  geometric superfluid weight in the HH model and show that it is also related to the quantum metric in the isolated band limit. 

The BCS mean-field Hamiltonian for the HH model is $H_{\mathrm{MF}}=\sum_{\mathbf{k}}\Psi^\dag_{\mathbf{k}}\mathcal{H}(\mathbf{k})\Psi_{\mathbf{k}}$. The Nambu field is $\Psi_{\mathbf{k}}=(c_{A \mathbf{k}\uparrow},c_{B \mathbf{k}\uparrow}, c^{\dag}_{A -\mathbf{k}\downarrow},c^{\dag}_{B -\mathbf{k}\downarrow})^{T}$ and the BdG  Hamiltonian reads
\begin{eqnarray}
 \mathcal{H}(\mathbf{k})=\left[  \begin{array}{cc}
  \mathcal{H}_{\uparrow}(\mathbf{k})-\mu & \bm \Delta \\ 
 \bm \Delta & -\mathcal{H}^{\ast}_{\uparrow}(-\mathbf{k})+\mu 
 \end{array} \right],
 \end{eqnarray} 
 where $\mathcal{H}_{\uparrow}(\mathbf{k})=h_0(\mathbf{k}) I+\mathbf{h}(\mathbf{k})\cdot\bm{ \sigma}$ is the Bloch Hamiltonian with the eigenvalues $\varepsilon_{d}(\mathbf{k})=h_0(\mathbf{k})- |\mathbf{h}(\mathbf{k})|$ and $\varepsilon_{u}(\mathbf{k})=h_0(\mathbf{k})+ |\mathbf{h}(\mathbf{k})|$. The corresponding eigenvectors  can be constructed as
 \begin{eqnarray}
 |d_{\mathbf{k}}\rangle=\frac{P_-|A\rangle}{\sqrt{(1-\hat{h}_z)/2}},~~ |u_{\mathbf{k}}\rangle=\frac{P_+|A\rangle}{\sqrt{(1+\hat{h}_z)/2}},
 \end{eqnarray}
  with $P_{\pm}=[1\pm \hat{\mathbf{h}}(\mathbf{k})\bm{\sigma}]/2$ is the projection operator and $|\alpha = A\rangle$ is a reference state chosen arbitrarily.  In the presence of inversion symmetry one has $\varepsilon_{d/u}(\mathbf{k})=\varepsilon_{d/u}(-\mathbf{k})$ and $\bm \Delta=\Delta I$. It is convenient to choose  $|A\rangle=(1,0)^T$, and then $-|d_{-\mathbf{k}}\rangle=R|d_{\mathbf{k}}\rangle$ and $|u_{-\mathbf{k}}\rangle=R|u_{\mathbf{k}}\rangle$ with $R=e^{i\arg {(h_x-ih_y)}}\sigma^x$ is a representation of the inversion symmetry. The BdG Hamiltonian in the bases $|d_{\mathbf{k}}\rangle|+\rangle$, $-|d^\ast_{-\mathbf{k}}\rangle|-\rangle$, $|u_{\mathbf{k}}\rangle|+\rangle$ and $|u^\ast_{-\mathbf{k}}\rangle|-\rangle$ can be written as 
  \begin{eqnarray}\label{Eq:HH_Ham}
  \mathcal{H}(\mathbf{k})&=&(h_0-\mu)I\otimes\tau^z-|\mathbf{h}| s^z\otimes\tau^z\nonumber\\
  &&-\Delta_{\mathrm{intra}}s^z\otimes\tau^x-\Delta_{\mathrm{inter}} s^x\otimes\tau^x, 
  \end{eqnarray}
with $\Delta_{\mathrm{intra}}=\Delta\sqrt{1-\hat{h}^2_z}$ is the intraband pairing and $\Delta_{\mathrm{inter}}=\Delta \hat{h}_z$ is the interband pairing. 
 The Pauli matrix $s^i$ acts in the two-dimensional space spanned by $|d_{\mathbf{k}}\rangle$ ($-|d^\ast_{-\mathbf{k}}\rangle$) and $|u_{\mathbf{k}}\rangle$ ($|u^\ast_{-\mathbf{k}}\rangle$) and $\tau^i$ acts in particle-hole space. 
Note that  $\gamma=iI\otimes\tau^y$ anticommutes with  $\mathcal{H}$ while it commutes with $\mathcal{H}^2= (h_0-\mu)^2+\Delta^2+|\mathbf{h}|^2+2 |\mathbf{h}| E_0 P$, where $E_0=\sqrt{(h_0-\mu)^2+\Delta^2 \hat{h}^2_z}$,  and 
\begin{eqnarray}
P=-[(h_0-\mu)s^z\otimes I+\Delta\hat{h}_z s^y\otimes\tau^y]/E_0,
\end{eqnarray}
 whose eigenvalues are $\pm 1$. 
Now it is clear that the eigenvalues of the BdG Hamiltonian are $\pm E_{d}$ and $\pm E_{u}$, where  
\begin{eqnarray}
&&E_d=\sqrt{(h_0-\mu)^2+\Delta^2+|\mathbf{h}|^2-2 |\mathbf{h}| E_0},
\end{eqnarray}
and
\begin{eqnarray}
&&E_u=\sqrt{(h_0-\mu)^2+\Delta^2+|\mathbf{h}|^2+2 |\mathbf{h}| E_0}.
\end{eqnarray}
Using the projection operators
\begin{eqnarray}
 && P^+_d=\frac{1}{4}\bigg(1+\frac{\mathcal{H}(\mathbf{k})}{E_-}\bigg)\bigg(1-P\bigg),
 \end{eqnarray}
 and
 \begin{eqnarray}
 P^+_u=\frac{1}{4}\bigg(1+\frac{\mathcal{H}(\mathbf{k})}{E_+}\bigg)\bigg(1+P\bigg),
\end{eqnarray}
 the eigenvectors corresponding to $E_d$ and $E_u$ can be constructed as
 \begin{eqnarray}
   &&|\psi^+_d\rangle =\frac{1}{N_d}P^+_d|A\rangle |+\rangle, ~|\psi^+_u\rangle =\frac{1}{N_u}P^+_u|A\rangle |+\rangle, 
 \end{eqnarray}
where the normalization factors are
$N_d=\sqrt{\langle  +|\langle A | P^+_d|A \rangle |+\rangle}$ and  $N_u=\sqrt{\langle  +| \langle A | P^+_u|A\rangle |+\rangle}$.
 Explicitly,  
\begin{eqnarray}
|\psi^+_s\rangle=\sum_{t=u,d} \bigg(u_{st} |t_{\mathbf{k}}\rangle|+\rangle + v_{st} |t^\ast_{-\mathbf{k}}\rangle|-\rangle\bigg ).
\end{eqnarray}
 The eigenvectors corresponding to the negative eigenvalues are
$|\psi^-_d\rangle=\gamma |\psi^+_d\rangle$ and $|\psi^-_u\rangle=\gamma |\psi^+_u\rangle$. 

The superfluid weight can be calculated directly after obtaining the BdG wave functions. We are interested in the geometric contribution at zero temperature, given by
  \begin{eqnarray}\label{Eq:SW_geom_HH}
  && D^s_{\mathrm{geom},\mu\nu}=\nonumber \sum_{\substack{\mathbf{k}\\s=u,d}}C^s_1\frac{8|\mathbf{h}|^2}{E_s}g_{\mu\nu}- \sum_{\substack{\mathbf{k}\\s=u,d}}C^s_2\frac{16|\mathbf{h}|^2}{E_s}\frac{\partial_\mu \hat{h}_z \partial_\nu \hat{h}_z}{1-\hat{h}^2_z}
 \nonumber\\
  &&+\sum_{\mathbf{k}}\frac{32|\mathbf{h}|^2}{E_d+E_u} \bigg[C_3 g_{\mu\nu}
  -C_4\frac{\partial_{\mu}\hat{h}_z\partial_{\nu}\hat{h}_z}{1-\hat{h}^2_z}\bigg].
  \end{eqnarray}
 The coefficients are $C^s_1=(u_{sd}v_{su}+u_{su}v_{sd})^2$, $C^s_2=u_{sd}v_{sd}u_{su}v_{su}$,  $C_3=(u_{dd}v_{uu}+u_{du}v_{ud})(u_{uu}v_{dd}+u_{ud}v_{du})$ and $C_4=(u_{dd}v_{uu}u_{uu}v_{dd}+u_{du}v_{ud}u_{ud}v_{du})$. 
We observe that $u_{lh}$ and $u_{hl}$ are smaller than the other coefficients because the bands  $|u\rangle|+\rangle$ and $|d\rangle|+\rangle$ are decoupled up to first order in $\Delta$.  As a result, the terms containing $u_{du}$ or $u_{ud}$ can be dropped in the isolated band limit and Eq. (\ref{Eq:SW_geom_HH}) can be simplified as
  \begin{eqnarray}\label{Eq:SW_geom_HH_isolated}
  	&& D^s_{\mathrm{geom},\mu\nu}\approx\nonumber \\
  &&\Delta^2\sum_{\mathbf{k}}\frac{|\mathbf{h}| (1-\hat{h}^2_z)(E_0+h_0)^2}{2E^3_0E_d}\left[ g_{\mu\nu}-\frac{\partial_{\mu}\hat{h}_z\partial_{\nu}\hat{h}_z}{1-\hat{h}^2_z}\right]\nonumber\\
  &&+\Delta^2\sum_{\mathbf{k}}\frac{ h^2_z(E_d+E_0-|\mathbf{h}|)^2}{2E^2_0E^3_d}g_{\mu\nu}.
  \end{eqnarray}
  
\begin{figure} 
	\includegraphics[width=\columnwidth]{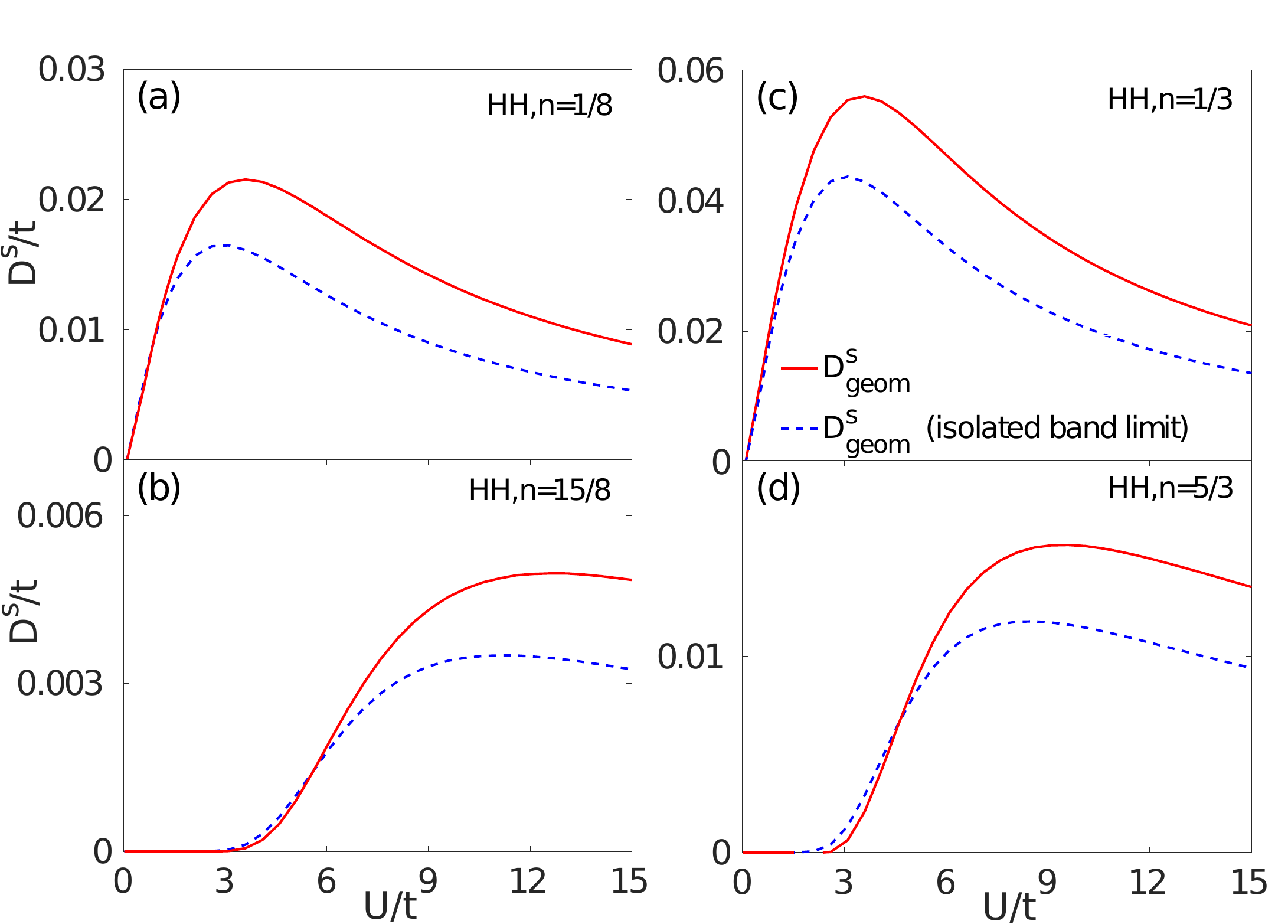}
	\caption{Geometric contribution of the superfluid weight in the Haldane-Hubbard model.
	The red solid curves are the exact results (in the mean-field sense) obtained using Eq. (\ref{Eq:SW_geom_HH}).	
	 The blue dash curves are obtained using  Eq. (\ref{Eq:SW_geom_HH_isolated2}), which is valid in the isolated band limit. Red and blue curves are in good agreement for weak couplings.}\label{Fig:HH_geom}
\end{figure}

The second term in the the right hand side comes from the interband pairing  and is  related to the quantum metric. This kind of contribution is absent in systems with TRS and uniform pairing since there is no band mixing in the quasiparticle states. In order to recover this term it is necessary to calculate the quasiparticle states $\left|\psi^\pm_{s}\right\rangle$ adiabatically connected to the isolated band up to first order in $\Delta$ before taking the isolated band limit.
The first term in Eq. (\ref{Eq:SW_geom_HH_isolated})  comes from intraband pairing, and has the same origin as in the time reversal symmetric systems. However, TRS breaking induces an extra term $\propto-\partial_{\mu}\hat{h}_z\partial_{\nu}\hat{h}_z/(1-\hat{h}^2_z)$ that cancels the contribution of the quantum metric. Because of the cancellation, this term is small in TRS breaking systems, and therefore  Eq. (\ref{Eq:SW_geom_HH_isolated}) can be further simplified as
  \begin{eqnarray}\label{Eq:SW_geom_HH_isolated2}
  	D^s_{\mathrm{geom},\mu\nu}
  &\approx&\Delta^2\sum_{\mathbf{k}}\frac{ h^2_z(E_d+E_0-|\mathbf{h}|)^2}{2E^2_0E^3_d}g_{\mu\nu}.
  \end{eqnarray}
In Fig. \ref{Fig:HH_geom} we
compare the approximation, Eq. (\ref{Eq:SW_geom_HH_isolated2}) to the exact mean-field result,  Eq. (\ref{Eq:SW_geom_HH}). They are in qualitative agreement for weak couplings where the isolated band approximation is good.  For strong couplings, Eq. (\ref{Eq:SW_geom_HH_isolated2}) also gives qualitatively the correct behavior. 

\section{Superfluid weight in superconducting graphene}\label{appendix:Dirac} 

In this appendix we apply  Eqs. (\ref{Eq:DS_TRS_conv}) and  (\ref{Eq:DS_TRS_2bands}) to superconducting graphene. The low energy properties of graphene are well described by the two Dirac points in the Brillouin zone. The two Dirac points are related by the inversion symmetry,  so it is enough to consider one and then multiply the result by a factor of 2. Near one Dirac point, the low energy effective Hamiltonian can be written as  $\mathcal{H}_{\uparrow}(\mathbf{k})=v_fk_x\sigma^x+v_fk_y\sigma^y-\mu I$, where $v_f$ is the Fermi velocity. The energy spectrum is $\varepsilon_{\pm}=\pm v_f k-\mu$ and the corresponding BdG excitation is $E_{\pm}=\sqrt{\varepsilon^2_{\pm}+\Delta^2}$. The quantum metric is $g_{\mu\nu}=(k^2\delta_{\mu\nu}-k_\mu k_\nu)/(2k^4)$. Substituting these into Eqs. (\ref{Eq:DS_TRS_conv}) and  (\ref{Eq:DS_TRS_2bands}) and using the fact that $D^s_{xx}=D^s_{yy}\equiv D^s$ and $D^s_{xy}=0$, we find the superfluid weight at zero temperature,
\begin{eqnarray}
D^s_{\mathrm{conv}}&=& 2\frac{1}{2}
\sum_{s=\pm}\int\frac{k\mathrm{d}k\mathrm{d}\theta}{(2\pi)^2}\frac{\Delta^2}{E^3_s}[(\partial_x\varepsilon_s)^2+(\partial_y\varepsilon_s)^2]\nonumber\\
&=&\frac{1}{\pi}\sqrt{\mu^2+\Delta^2},
\end{eqnarray}
\begin{eqnarray}
D^s_{\mathrm{geom}}&=& 2\frac{1}{2}
\sum_{s=\pm}\int\frac{k\mathrm{d}k\mathrm{d}\theta}{(2\pi)^2}\frac{s\Delta^2v_f k}{\mu E_s}(g_{xx}+g_{yy})\nonumber\\
&=&\frac{1}{\pi}\frac{\Delta^2}{|\mu|}\ln\frac{|\mu|+\sqrt{\Delta^2+\mu^2}}{|\Delta|}.
\end{eqnarray}
The factor of 2 counts the two Dirac points.  The total superfluid weight is \begin{eqnarray}
D^s&=&D^s_{\mathrm{conv}}+D^s_{\mathrm{geom}}\nonumber\\ 
&=&\frac{1}{\pi}\bigg(\sqrt{\Delta^2+\mu^2}+\frac{\Delta^2}{|\mu|}\ln\frac{|\mu|+\sqrt{\Delta^2+\mu^2}}{|\Delta|}\bigg),~
\end{eqnarray}
which coincides with Eq. (32) in  Ref. \cite{SuperfluidGraphene}. Note that to compare  with our result, the phase factor $\mathbf{k}$ in Eq. (32) in  Ref. \cite{SuperfluidGraphene} should be rescaled by a factor of 2, see Eq. (7) in the cited article.

\section{Superfluid weight in a system with $C_6$ symmetry}\label{appendix:C6}

In this appendix we prove that for a system with $C_6$ symmetry, the superfluid weight in an orthogonal basis is diagonal and proportional to the identity. Suppose the superfluid weight in a non-orthogonal primitive basis $\{\mathbf{e}_1, \mathbf{e}_2\}$ (see Fig. \ref{Fig:Lattice}) is $D^s_{n}=\bigg(\begin{array}{cc}
D^s_{11} & D^s_{12} \\ 
D^s_{21} & D^s_{22}
\end{array}\bigg) $. Notice that the superfluid weight is a symmetric tensor, i.e.,  $D^s_{12} =D^s_{21}$. Because of the $C_6$ symmetry, $D^s_n$ is invariant under $\pi/3$ rotation. This can be viewed as a basis transformation and the basis after and before the $\pi/3$ rotation is related by a matrix  $A=\bigg(\begin{array}{cc}
0 &-1 \\ 
1 & 1
\end{array}\bigg)$. Invariance under rotation means  that $D^s_n=A^{T} D^s_n A$ \cite{PRB.90.134503}. This gives $D^s_{11}=D^s_{22}=2D^s_{12}$.
In the orthogonal basis $\{\mathbf{e}_x, \mathbf{e}_y \}$, the superfluid weight is $D^s_{o}=\bigg(\begin{array}{cc}
D^s_{xx} & D^s_{xy} \\ 
D^s_{yx} & D^s_{yy}
\end{array}\bigg) $. Similarly, $D^s_n$ and $D^s_o$ are related by a basis transformation, $D^s_n=B^{T} D^s_o B$ with $B=\bigg(\begin{array}{cc}
1 &\cos{\pi/3} \\ 
0 & \sin{\pi/3}
\end{array}\bigg) $. We then find $D^s_{xx}=D^s_{yy}=D^s_{11}$ and $D^s_{xy}=D^s_{yx}=0$.

\section{Order parameter and superfluid weight from DMFT}\label{appendix:DMFT}

\begin{figure} [h]
\includegraphics[width=\columnwidth] {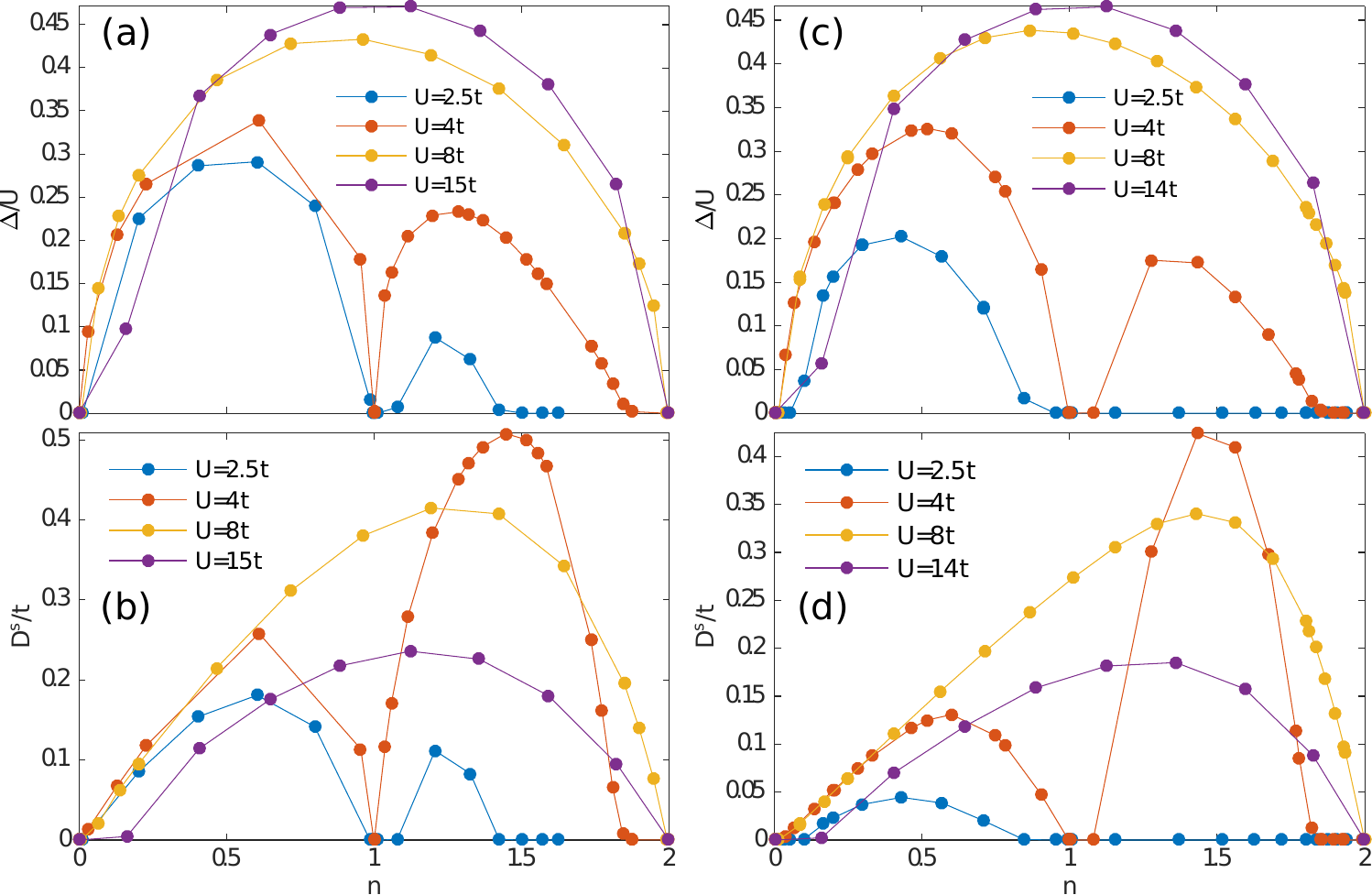}%
\caption{DMFT results for the order parameter $\Delta$ and the superfluid weight $D^s$ as a function of the filling for different values of $U$ for $T=0.1t$. Figures (a) and (b) are for the Kane-Mele model and figures (c) and (d) for the Haldane-Hubbard model.}\label{dmft_kuva}
\end{figure}

In this appendix we provide some details and further results from our DMFT \cite{RevModPhys.68.13,RevModPhys.77.1027} calculations. In this work we have used single-site DMFT to determine the superconducting order parameter $\Delta$ and the superfluid weight $D^s$. Because the unit cell of the hexagonal lattice includes two lattice sites, we get two single-site impurity problems that are solved independently. For weak and intermediate interactions ($U \lesssim 8$) we use a continuous-time interaction-expansion (CT-INT) impurity solver \cite{PhysRevB.72.035122,RevModPhys.83.349} and for larger $U$ we resort to exact diagonalization \cite{RevModPhys.68.13,PhysRevLett.72.1545} in solving the impurity problem.

We evaluate the superfluid density in a straightforward manner by adding a small constant (in time and space) vector potential $A_{\nu}$ to the model. This modifies the hopping amplitudes by multiplying them with a Peierls phase factor $t_{ij} \rightarrow e^{-i\mathbf{A}\cdot (\mathbf{r}_{j}-\mathbf{r}_{i})} t_{ij}$. We then calculate the current as a function of $A_{\nu}$ and determine the superfluid weight from the linear response formula 
$\langle j_{\mu}\rangle=D_{\mu\nu} A_{\nu}$ for small $A_{\nu}$.

Note that this procedure may seem contradictory, as a constant vector potential is gauge equivalent to a vanishing one, and thus should produce no current. In an exact calculation this is indeed true, provided that the vector potential is consistent with the periodic boundary conditions of the problem, i.e. $\mathbf{A} \cdot \mathbf{L}=2 \pi n$, where $\mathbf{L}$ is a period of the lattice.
For DMFT or mean-field theory, however, the single-unit-cell calculation implicitly imposes the constraint that the order parameter field $\Delta=\left \langle c_{i\uparrow} c_{i\downarrow} \right \rangle$ is uniform in space, which effectively breaks the gauge symmetry. 
Perhaps the easiest way to understand this is to perform a gauge transformation
\begin{equation}
c'_{i\sigma}=\exp(i \mathbf{A} \cdot \mathbf{r}_i) c_{i\sigma}.
\end{equation}
In the primed variables the Hamiltonian does not have any Peierls phases, but the order parameter gains a position dependent phase twist,
\begin{equation}
\Delta_i '= \left \langle c_{i \uparrow}' c_{i \downarrow}' \right \rangle
=\exp( 2 i \mathbf{A} \cdot \mathbf{r}_i ) \Delta.
\end{equation}
Thus we can see that we are in fact calculating the current response of the system to a phase twist of the order parameter. This is equivalent to the phase stiffness definition of the superfluid weight discussed in Appendix \ref{appendix:equivalence}, as the current is the first derivative of the free energy. In this way it is also easy to understand why the BCS mean-field calculation produces the same result for longitudinal and transverse gauge fields, as discussed in reference \cite{swz} for example.

This procedure should also be applicable to cluster DMFT without the need to calculate two-body correlation functions. However, it can only be applied in the phase where the gauge symmetry of the model is broken by a finite superconducting order parameter $\Delta$. In the symmetric case it is necessary to perform a more careful analysis of the self-energy \cite{PhysRevB.80.161105}.

The expectation value of the current operator, or indeed any single particle operator $O=\sum_{ij} M_{ij} c_i^\dagger c_j$, can be evaluated using the DMFT self-energy once the iteration has converged. In principle the expectation value can be expressed in terms of the Greens function $G_{ij}(\tau)=\left\langle c_i(\tau) c_j^\dagger(0) \right\rangle$ as
\begin{equation}
\left \langle O \right \rangle = \mathrm{Tr}(M)-\mathrm{Tr}(M G(\tau=0^+) ),
\end{equation}
where $i$ and $j$ index the orbitals of the whole lattice model. In practice we do not want to calculate the whole Greens function $G(\tau=0^+)$ in real space, and the trace has to be calculated in the Fourier transformed representation.
The Green's function in frequency and momentum space is given by
\begin{equation}
G_{\mathbf{k}}(i \omega_n) = \left( -i\omega_n + T_{\mathbf{k}} - \Sigma(i \omega_n) \right)^{-1},
\end{equation}
where $T_{\mathbf{k}}$ is the noninteracting Bloch Hamiltonian (given by equation \ref{Eq:BdgHamiltonian} with $\Delta=0$) and $\Sigma$ is the DMFT self-energy including the anomalous components. The expectation value of $O$ is then given by
\begin{equation}
\left \langle O \right \rangle = \frac{1}{N} \sum_{\mathbf{k}} \mathrm{Tr}(M_{\mathbf{k}}) - \frac{1}{\beta N} \sum_{n} \sum_{\mathbf{k}} \mathrm{Tr} \left( M_{\mathbf{k}} G_{\mathbf{k}}(i\omega_n) \right),
\end{equation}
where $\beta$ is the inverse temperature, $N$ is the number of $\mathbf{k}$-points and $M_{\mathbf{k}}$ is the k-space representation of $M$. To perform the frequency summation of the large frequency tail we perform a fitting procedure to find the lowest moments of the expansion of $\sum_{\mathbf{k}} \mathrm{Tr} \left( M_{\mathbf{k}} G_{\mathbf{k}}(i\omega_n) \right)$ in powers of $(i\omega_n)^{-1}$, and calculate the contribution from the tail analytically, as is commonly done with the Fourier transform of $G$ itself.

When dealing with almost flat bands it is not always easy to obtain a DMFT solution with a desired density. We alleviate this problem by tuning the chemical potential $\mu$ in the course of the iteration so that the sum of the chemical potential and the Hartree energy, $\mu_h=\mu+E_h$ is always given by some predefined value. We find that tuning $\mu_h$ instead of $\mu$ directly makes it easier to attain a specific density. Of course, one has to check that the chemical potential $\mu$ actually converges. We stress that this procedure is not a modification of the DMFT equations, but just a modified iterative method for their solution.

In Fig.\ \ref{dmft_kuva} we plot the DMFT superfluid weight as a function of the filling for different interaction strenghts. For weak values of the interaction $U$ one can observe two domes corresponding to the two bands. For stronger $U$ the bands are mixed and the two-dome structure disappears.


\section{ED calculations of the Drude weight}\label{appendix:ED}

The Drude weight is the singular part of the real part of the optical conductivity, given by
\begin{eqnarray}\label{Eq:OC}
D_{\mu\nu} 
&=&\langle 0| T_{\mu\nu}|0\rangle+2\Re \langle 0 |j^p_\mu \frac{1}{E_0+i 0^+ -H}j^p_\nu |0\rangle.
\end{eqnarray}
We calculate 
the ground state energy $E_0$ and ground state wave function $|0\rangle$  using the Lanczos algorithm realized on graphics processing units (GPU) \cite{Siro20121884} and the Green's function in the second term of Eq. (\ref{Eq:OC}) is evaluated through the continued fraction expansion method \cite{PhysRevLett.59.2999}.

\begin{figure} 
\includegraphics[width=\columnwidth]{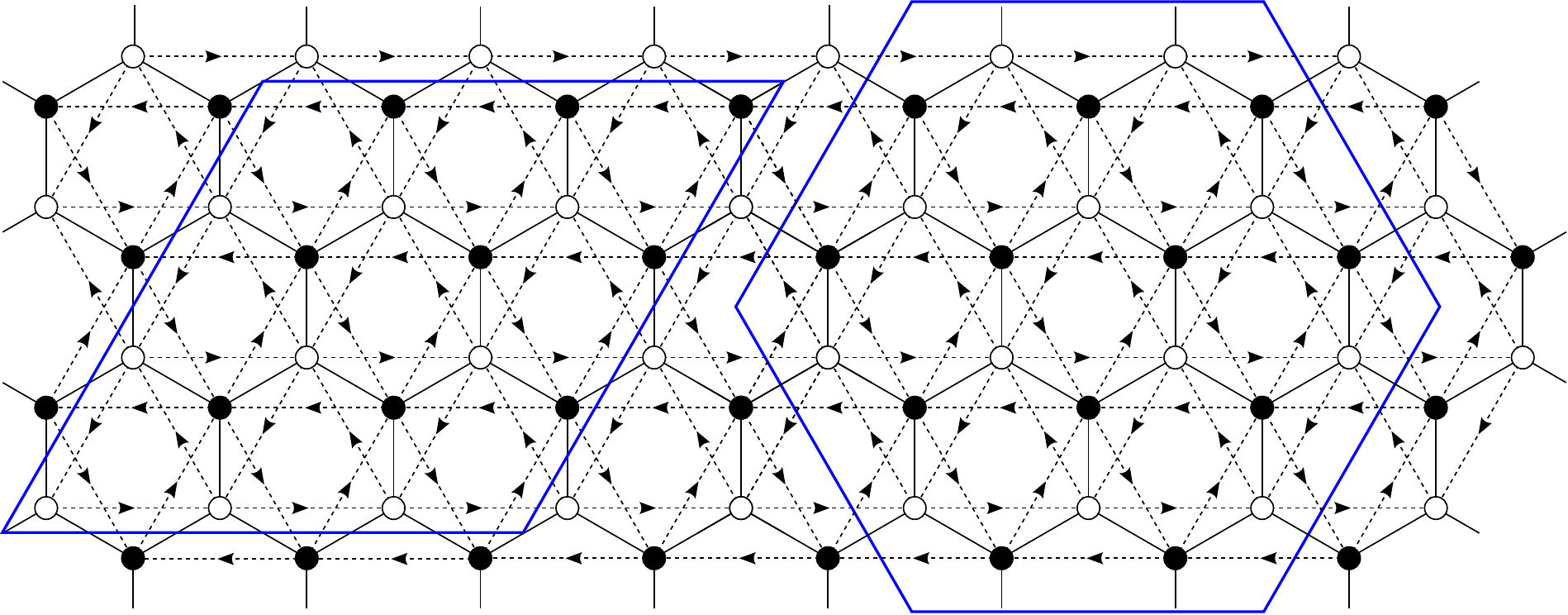}
	\caption{18- and 24-site  clusters used in ED calculations. }\label{Fig:Lattice_cluster}
\end{figure}
\begin{figure} 
	\includegraphics[width=\columnwidth]{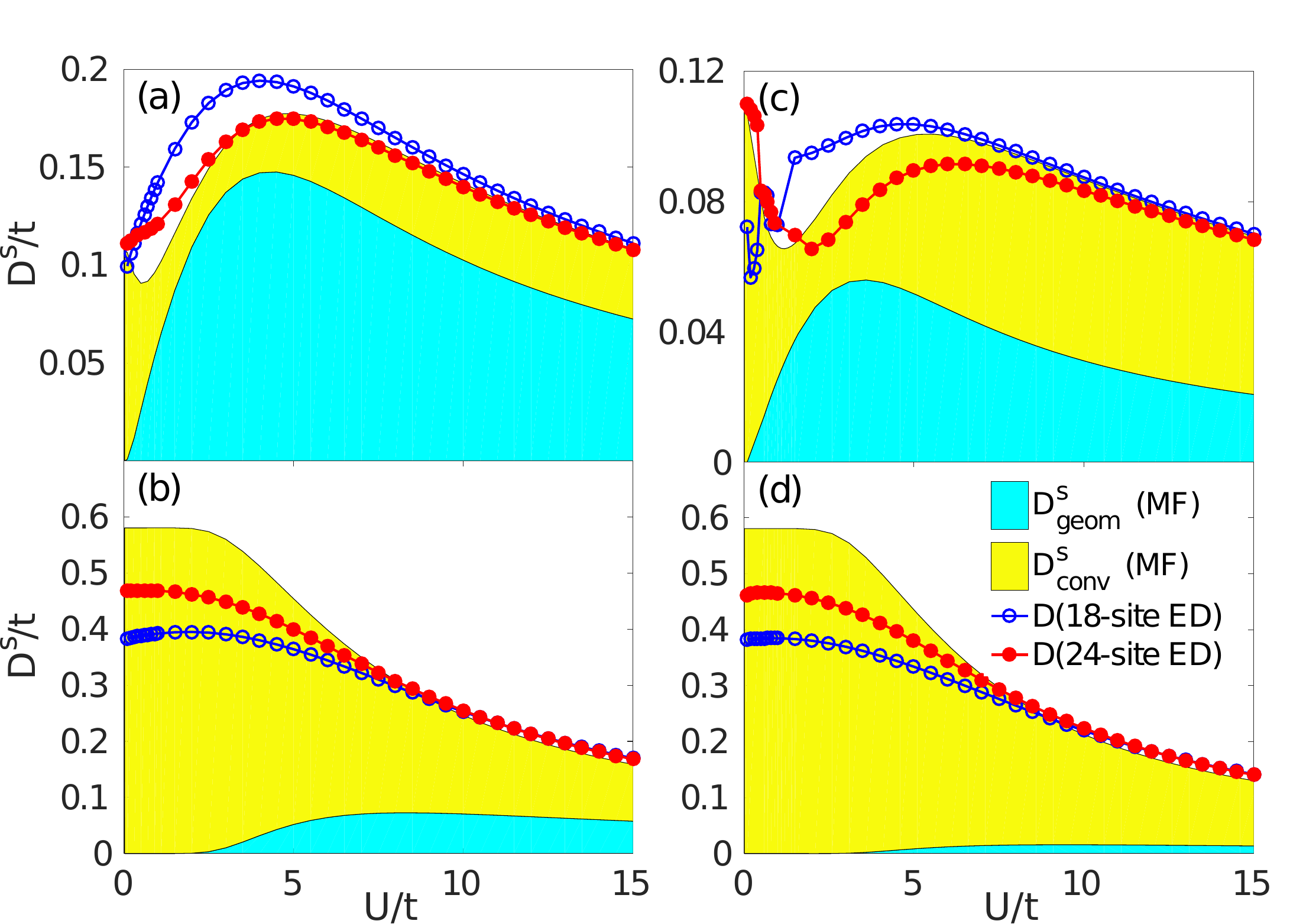}
	\caption{Zero temperature superfluid weight for the KMH  model [(a)-(b)] and the HH model [(c)-(d)] at filling  $n=1/3$ [(a) and (c), flat band], $5/3$ [(b) and (d), dispersive band]. ED is performed on  18- and 24-site  clusters. Finite size effects can be seen, especially for weak interactions.
	}\label{Fig:Ds_Finitesize}
\end{figure}

In the main text we compare the mean-field results of the superfluid weight against the Drude weight obtained from ED on a  32-site cluster. Since the cluster is large, we can only calculate at low and high fillings, corresponding to the bottom of the flat band and the top of the dispersive band. In this appendix we present our results for intermediate fillings, $n=1/3$ and $5/3$. We perform ED calculations on 18- and 24-site clusters that preserve the $C_6$ symmetry, see Fig. \ref{Fig:Lattice_cluster}. As shown in Fig. \ref{Fig:Ds_Finitesize}, for strong interactions, the ED results on both clusters are in good agreement with mean-field results. For weak interactions, finite size effects are visible  and increasing the cluster size improves the agreement significantly.

\bibliography{SW}

\begin{thebibliography}{71}%
\makeatletter
\providecommand \@ifxundefined [1]{%
 \@ifx{#1\undefined}
}%
\providecommand \@ifnum [1]{%
 \ifnum #1\expandafter \@firstoftwo
 \else \expandafter \@secondoftwo
 \fi
}%
\providecommand \@ifx [1]{%
 \ifx #1\expandafter \@firstoftwo
 \else \expandafter \@secondoftwo
 \fi
}%
\providecommand \natexlab [1]{#1}%
\providecommand \enquote  [1]{``#1''}%
\providecommand \bibnamefont  [1]{#1}%
\providecommand \bibfnamefont [1]{#1}%
\providecommand \citenamefont [1]{#1}%
\providecommand \href@noop [0]{\@secondoftwo}%
\providecommand \href [0]{\begingroup \@sanitize@url \@href}%
\providecommand \@href[1]{\@@startlink{#1}\@@href}%
\providecommand \@@href[1]{\endgroup#1\@@endlink}%
\providecommand \@sanitize@url [0]{\catcode `\\12\catcode `\$12\catcode
  `\&12\catcode `\#12\catcode `\^12\catcode `\_12\catcode `\%12\relax}%
\providecommand \@@startlink[1]{}%
\providecommand \@@endlink[0]{}%
\providecommand \url  [0]{\begingroup\@sanitize@url \@url }%
\providecommand \@url [1]{\endgroup\@href {#1}{\urlprefix }}%
\providecommand \urlprefix  [0]{URL }%
\providecommand \Eprint [0]{\href }%
\providecommand \doibase [0]{http://dx.doi.org/}%
\providecommand \selectlanguage [0]{\@gobble}%
\providecommand \bibinfo  [0]{\@secondoftwo}%
\providecommand \bibfield  [0]{\@secondoftwo}%
\providecommand \translation [1]{[#1]}%
\providecommand \BibitemOpen [0]{}%
\providecommand \bibitemStop [0]{}%
\providecommand \bibitemNoStop [0]{.\EOS\space}%
\providecommand \EOS [0]{\spacefactor3000\relax}%
\providecommand \BibitemShut  [1]{\csname bibitem#1\endcsname}%
\let\auto@bib@innerbib\@empty
\bibitem [{\citenamefont {Berry}(1989)}]{Berry:1989ai}%
  \BibitemOpen
  \bibfield  {author} {\bibinfo {author} {\bibfnamefont {M.~V.}\ \bibnamefont
  {Berry}},\ }\bibfield  {title} {\enquote {\bibinfo {title} {The quantum
  phase, five years after},}\ }in\ \href
  {https://www.phas.ubc.ca/~berciu/PHILIP/TEACHING/PHYS501/NOTES/FILES/Berry90_QPhase-5yrsAfter--Book88.pdf}
  {\emph {\bibinfo {booktitle} {Geometric Phases in Physics}}},\ \bibinfo
  {editor} {edited by\ \bibinfo {editor} {\bibfnamefont {A.}~\bibnamefont
  {Shapere}}\ and\ \bibinfo {editor} {\bibfnamefont {F.}~\bibnamefont
  {Wilczeck}}}\ (\bibinfo  {publisher} {World Scientific},\ \bibinfo {year}
  {1989})\ pp.\ \bibinfo {pages} {7--28}\BibitemShut {NoStop}%
\bibitem [{\citenamefont {Provost}\ and\ \citenamefont
  {Vallee}(1980)}]{Provost1980}%
  \BibitemOpen
  \bibfield  {author} {\bibinfo {author} {\bibfnamefont {J.~P.}\ \bibnamefont
  {Provost}}\ and\ \bibinfo {author} {\bibfnamefont {G.}~\bibnamefont
  {Vallee}},\ }\bibfield  {title} {\enquote {\bibinfo {title} {Riemannian
  structure on manifolds of quantum states},}\ }\href {\doibase
  10.1007/BF02193559} {\bibfield  {journal} {\bibinfo  {journal} {Commun. Math.
  Phys.}\ }\textbf {\bibinfo {volume} {76}},\ \bibinfo {pages} {289--301}
  (\bibinfo {year} {1980})}\BibitemShut {NoStop}%
\bibitem [{\citenamefont {Berry}(1984)}]{Berry84}%
  \BibitemOpen
  \bibfield  {author} {\bibinfo {author} {\bibfnamefont {M.~V.}\ \bibnamefont
  {Berry}},\ }\bibfield  {title} {\enquote {\bibinfo {title} {Quantal phase
  factors accompanying adiabatic changes},}\ }\href {\doibase
  10.1098/rspa.1984.0023} {\bibfield  {journal} {\bibinfo  {journal} {Proc. R.
  Soc. A: Mathematical, Physical and Engineering Sciences}\ }\textbf {\bibinfo
  {volume} {392}},\ \bibinfo {pages} {45--57} (\bibinfo {year}
  {1984})}\BibitemShut {NoStop}%
\bibitem [{\citenamefont {Xiao}\ \emph {et~al.}(2010)\citenamefont {Xiao},
  \citenamefont {Chang},\ and\ \citenamefont {Niu}}]{XCN2010}%
  \BibitemOpen
  \bibfield  {author} {\bibinfo {author} {\bibfnamefont {D.}~\bibnamefont
  {Xiao}}, \bibinfo {author} {\bibfnamefont {M.~C.}\ \bibnamefont {Chang}}, \
  and\ \bibinfo {author} {\bibfnamefont {Q.}~\bibnamefont {Niu}},\ }\bibfield
  {title} {\enquote {\bibinfo {title} {Berry phase effects on electronic
  properties},}\ }\href {\doibase 10.1103/RevModPhys.82.1959} {\bibfield
  {journal} {\bibinfo  {journal} {Rev. Mod. Phys.}\ }\textbf {\bibinfo {volume}
  {82}},\ \bibinfo {pages} {1959--2007} (\bibinfo {year} {2010})}\BibitemShut
  {NoStop}%
\bibitem [{\citenamefont {Resta}(1994)}]{resta1994}%
  \BibitemOpen
  \bibfield  {author} {\bibinfo {author} {\bibfnamefont {R.}~\bibnamefont
  {Resta}},\ }\bibfield  {title} {\enquote {\bibinfo {title} {Macroscopic
  polarization in crystalline dielectrics: the geometric phase approach},}\
  }\href {\doibase 10.1103/RevModPhys.66.899} {\bibfield  {journal} {\bibinfo
  {journal} {Rev. Mod. Phys.}\ }\textbf {\bibinfo {volume} {66}},\ \bibinfo
  {pages} {899--915} (\bibinfo {year} {1994})}\BibitemShut {NoStop}%
\bibitem [{\citenamefont {Resta}(2010)}]{resta2010}%
  \BibitemOpen
  \bibfield  {author} {\bibinfo {author} {\bibfnamefont {R.}~\bibnamefont
  {Resta}},\ }\bibfield  {title} {\enquote {\bibinfo {title} {Electrical
  polarization and orbital magnetization: the modern theories},}\ }\href
  {http://stacks.iop.org/0953-8984/22/i=12/a=123201} {\bibfield  {journal}
  {\bibinfo  {journal} {J. Phys.: Condens. Matter}\ }\textbf {\bibinfo {volume}
  {22}},\ \bibinfo {pages} {123201} (\bibinfo {year} {2010})}\BibitemShut
  {NoStop}%
\bibitem [{\citenamefont {Bengtsson}\ and\ \citenamefont
  {\.{Z}yczkowski}(2006)}]{GQS}%
  \BibitemOpen
  \bibfield  {author} {\bibinfo {author} {\bibfnamefont {I.}~\bibnamefont
  {Bengtsson}}\ and\ \bibinfo {author} {\bibfnamefont {K.}~\bibnamefont
  {\.{Z}yczkowski}},\ }\href {http://dx.doi.org/10.1017/CBO9780511535048}
  {\emph {\bibinfo {title} {Geometry of Quantum States}}}\ (\bibinfo
  {publisher} {Cambridge University Press},\ \bibinfo {year}
  {2006})\BibitemShut {NoStop}%
\bibitem [{\citenamefont {Gu}(2010)}]{QPT_rev}%
  \BibitemOpen
  \bibfield  {author} {\bibinfo {author} {\bibfnamefont {S.~J.}\ \bibnamefont
  {Gu}},\ }\bibfield  {title} {\enquote {\bibinfo {title} {Fidelity approach to
  quantum phase transitions},}\ }\href {\doibase 10.1142/S0217979210056335}
  {\bibfield  {journal} {\bibinfo  {journal} {Int. J. Mod. Phys. B}\ }\textbf
  {\bibinfo {volume} {24}},\ \bibinfo {pages} {4371--4458} (\bibinfo {year}
  {2010})}\BibitemShut {NoStop}%
\bibitem [{\citenamefont {Sun}\ \emph {et~al.}(2011)\citenamefont {Sun},
  \citenamefont {Gu}, \citenamefont {Katsura},\ and\ \citenamefont
  {Das~Sarma}}]{SarmaFlatBand}%
  \BibitemOpen
  \bibfield  {author} {\bibinfo {author} {\bibfnamefont {K.}~\bibnamefont
  {Sun}}, \bibinfo {author} {\bibfnamefont {Z.~C.}\ \bibnamefont {Gu}},
  \bibinfo {author} {\bibfnamefont {H.}~\bibnamefont {Katsura}}, \ and\
  \bibinfo {author} {\bibfnamefont {S.}~\bibnamefont {Das~Sarma}},\ }\bibfield
  {title} {\enquote {\bibinfo {title} {Nearly flatbands with nontrivial
  topology},}\ }\href {\doibase 10.1103/PhysRevLett.106.236803} {\bibfield
  {journal} {\bibinfo  {journal} {Phys. Rev. Lett.}\ }\textbf {\bibinfo
  {volume} {106}},\ \bibinfo {pages} {236803} (\bibinfo {year}
  {2011})}\BibitemShut {NoStop}%
\bibitem [{\citenamefont {Tang}\ \emph {et~al.}(2011)\citenamefont {Tang},
  \citenamefont {Mei},\ and\ \citenamefont {Wen}}]{WenFlatBand}%
  \BibitemOpen
  \bibfield  {author} {\bibinfo {author} {\bibfnamefont {E.}~\bibnamefont
  {Tang}}, \bibinfo {author} {\bibfnamefont {J.~W.}\ \bibnamefont {Mei}}, \
  and\ \bibinfo {author} {\bibfnamefont {X.~G.}\ \bibnamefont {Wen}},\
  }\bibfield  {title} {\enquote {\bibinfo {title} {High-temperature fractional
  quantum {H}all states},}\ }\href {\doibase 10.1103/PhysRevLett.106.236802}
  {\bibfield  {journal} {\bibinfo  {journal} {Phys. Rev. Lett.}\ }\textbf
  {\bibinfo {volume} {106}},\ \bibinfo {pages} {236802} (\bibinfo {year}
  {2011})}\BibitemShut {NoStop}%
\bibitem [{\citenamefont {Neupert}\ \emph {et~al.}(2011)\citenamefont
  {Neupert}, \citenamefont {Santos}, \citenamefont {Chamon},\ and\
  \citenamefont {Mudry}}]{fqh2011}%
  \BibitemOpen
  \bibfield  {author} {\bibinfo {author} {\bibfnamefont {T.}~\bibnamefont
  {Neupert}}, \bibinfo {author} {\bibfnamefont {L.}~\bibnamefont {Santos}},
  \bibinfo {author} {\bibfnamefont {C.}~\bibnamefont {Chamon}}, \ and\ \bibinfo
  {author} {\bibfnamefont {C.}~\bibnamefont {Mudry}},\ }\bibfield  {title}
  {\enquote {\bibinfo {title} {Fractional quantum {H}all states at zero
  magnetic field},}\ }\href {\doibase 10.1103/PhysRevLett.106.236804}
  {\bibfield  {journal} {\bibinfo  {journal} {Phys. Rev. Lett.}\ }\textbf
  {\bibinfo {volume} {106}},\ \bibinfo {pages} {236804} (\bibinfo {year}
  {2011})}\BibitemShut {NoStop}%
\bibitem [{\citenamefont {Wang}\ \emph {et~al.}(2011)\citenamefont {Wang},
  \citenamefont {Gu}, \citenamefont {Gong},\ and\ \citenamefont
  {Sheng}}]{ShengFlatBand_boson}%
  \BibitemOpen
  \bibfield  {author} {\bibinfo {author} {\bibfnamefont {Y.~F.}\ \bibnamefont
  {Wang}}, \bibinfo {author} {\bibfnamefont {Z.~C.}\ \bibnamefont {Gu}},
  \bibinfo {author} {\bibfnamefont {C.~D.}\ \bibnamefont {Gong}}, \ and\
  \bibinfo {author} {\bibfnamefont {D.~N.}\ \bibnamefont {Sheng}},\ }\bibfield
  {title} {\enquote {\bibinfo {title} {Fractional quantum {H}all effect of
  hard-core bosons in topological flat bands},}\ }\href {\doibase
  10.1103/PhysRevLett.107.146803} {\bibfield  {journal} {\bibinfo  {journal}
  {Phys. Rev. Lett.}\ }\textbf {\bibinfo {volume} {107}},\ \bibinfo {pages}
  {146803} (\bibinfo {year} {2011})}\BibitemShut {NoStop}%
\bibitem [{\citenamefont {Bergholtz}\ and\ \citenamefont
  {Liu}(2013)}]{doi:10.1142/S021797921330017X}%
  \BibitemOpen
  \bibfield  {author} {\bibinfo {author} {\bibfnamefont {E.~J.}\ \bibnamefont
  {Bergholtz}}\ and\ \bibinfo {author} {\bibfnamefont {Z.}~\bibnamefont
  {Liu}},\ }\bibfield  {title} {\enquote {\bibinfo {title} {Topological flat
  band models and fractional {C}hern insulators},}\ }\href {\doibase
  10.1142/S021797921330017X} {\bibfield  {journal} {\bibinfo  {journal} {Int.
  J. Mod. Phys. B}\ }\textbf {\bibinfo {volume} {27}},\ \bibinfo {pages}
  {1330017} (\bibinfo {year} {2013})}\BibitemShut {NoStop}%
\bibitem [{\citenamefont {Roy}(2014)}]{roy2014}%
  \BibitemOpen
  \bibfield  {author} {\bibinfo {author} {\bibfnamefont {R.}~\bibnamefont
  {Roy}},\ }\bibfield  {title} {\enquote {\bibinfo {title} {Band geometry of
  fractional topological insulators},}\ }\href {\doibase
  10.1103/PhysRevB.90.165139} {\bibfield  {journal} {\bibinfo  {journal} {Phys.
  Rev. B}\ }\textbf {\bibinfo {volume} {90}},\ \bibinfo {pages} {165139}
  (\bibinfo {year} {2014})}\BibitemShut {NoStop}%
\bibitem [{\citenamefont {Jackson}\ \emph {et~al.}(2015)\citenamefont
  {Jackson}, \citenamefont {Moller},\ and\ \citenamefont {Roy}}]{Roy2015}%
  \BibitemOpen
  \bibfield  {author} {\bibinfo {author} {\bibfnamefont {T.~S.}\ \bibnamefont
  {Jackson}}, \bibinfo {author} {\bibfnamefont {G.}~\bibnamefont {Moller}}, \
  and\ \bibinfo {author} {\bibfnamefont {R.}~\bibnamefont {Roy}},\ }\bibfield
  {title} {\enquote {\bibinfo {title} {Geometric stability of topological
  lattice phases},}\ }\href {http://dx.doi.org/10.1038/ncomms9629} {\bibfield
  {journal} {\bibinfo  {journal} {Nat. Commun.}\ }\textbf {\bibinfo {volume}
  {6}} (\bibinfo {year} {2015})}\BibitemShut {NoStop}%
\bibitem [{\citenamefont {Kopnin}\ \emph {et~al.}(2011)\citenamefont {Kopnin},
  \citenamefont {Heikkil\"a},\ and\ \citenamefont
  {Volovik}}]{Volovik_Flatband1}%
  \BibitemOpen
  \bibfield  {author} {\bibinfo {author} {\bibfnamefont {N.~B.}\ \bibnamefont
  {Kopnin}}, \bibinfo {author} {\bibfnamefont {T.~T.}\ \bibnamefont
  {Heikkil\"a}}, \ and\ \bibinfo {author} {\bibfnamefont {G.~E.}\ \bibnamefont
  {Volovik}},\ }\bibfield  {title} {\enquote {\bibinfo {title}
  {High-temperature surface superconductivity in topological flat-band
  systems},}\ }\href {\doibase 10.1103/PhysRevB.83.220503} {\bibfield
  {journal} {\bibinfo  {journal} {Phys. Rev. B}\ }\textbf {\bibinfo {volume}
  {83}},\ \bibinfo {pages} {220503} (\bibinfo {year} {2011})}\BibitemShut
  {NoStop}%
\bibitem [{\citenamefont {Heikkil{\"a}}\ \emph {et~al.}(2011)\citenamefont
  {Heikkil{\"a}}, \citenamefont {Kopnin},\ and\ \citenamefont
  {Volovik}}]{Volovik_Flatband2}%
  \BibitemOpen
  \bibfield  {author} {\bibinfo {author} {\bibfnamefont {T.~T.}\ \bibnamefont
  {Heikkil{\"a}}}, \bibinfo {author} {\bibfnamefont {N.~B.}\ \bibnamefont
  {Kopnin}}, \ and\ \bibinfo {author} {\bibfnamefont {G.~E.}\ \bibnamefont
  {Volovik}},\ }\bibfield  {title} {\enquote {\bibinfo {title} {Flat bands in
  topological media},}\ }\href {\doibase 10.1134/S0021364011150045} {\bibfield
  {journal} {\bibinfo  {journal} {JETP Lett.}\ }\textbf {\bibinfo {volume}
  {94}},\ \bibinfo {pages} {233--239} (\bibinfo {year} {2011})}\BibitemShut
  {NoStop}%
\bibitem [{\citenamefont {{Noda}}\ \emph {et~al.}(2015)\citenamefont {{Noda}},
  \citenamefont {{Inaba}},\ and\ \citenamefont {{Yamashita}}}]{FlatBandTc}%
  \BibitemOpen
  \bibfield  {author} {\bibinfo {author} {\bibfnamefont {K.}~\bibnamefont
  {{Noda}}}, \bibinfo {author} {\bibfnamefont {K.}~\bibnamefont {{Inaba}}}, \
  and\ \bibinfo {author} {\bibfnamefont {M.}~\bibnamefont {{Yamashita}}},\
  }\bibfield  {title} {\enquote {\bibinfo {title} {{BCS superconducting
  transitions in lattice fermions}},}\ }\href {http://arxiv.org/abs/1512.07858}
  {\bibfield  {journal} {\bibinfo  {journal} {ArXiv e-prints}\ } (\bibinfo
  {year} {2015})},\ \Eprint {http://arxiv.org/abs/1512.07858} {arXiv:1512.07858
  [cond-mat.supr-con]} \BibitemShut {NoStop}%
\bibitem [{\citenamefont {Kauppila}\ \emph {et~al.}(2016)\citenamefont
  {Kauppila}, \citenamefont {Aikebaier},\ and\ \citenamefont
  {Heikkil\"a}}]{PhysRevB.93.214505}%
  \BibitemOpen
  \bibfield  {author} {\bibinfo {author} {\bibfnamefont {V.~J.}\ \bibnamefont
  {Kauppila}}, \bibinfo {author} {\bibfnamefont {F.}~\bibnamefont {Aikebaier}},
  \ and\ \bibinfo {author} {\bibfnamefont {T.~T.}\ \bibnamefont {Heikkil\"a}},\
  }\bibfield  {title} {\enquote {\bibinfo {title} {Flat-band superconductivity
  in strained {D}irac materials},}\ }\href {\doibase
  10.1103/PhysRevB.93.214505} {\bibfield  {journal} {\bibinfo  {journal} {Phys.
  Rev. B}\ }\textbf {\bibinfo {volume} {93}},\ \bibinfo {pages} {214505}
  (\bibinfo {year} {2016})}\BibitemShut {NoStop}%
\bibitem [{\citenamefont {Peotta}\ and\ \citenamefont
  {{T{\"o}rm{\"a}}}(2015)}]{PT}%
  \BibitemOpen
  \bibfield  {author} {\bibinfo {author} {\bibfnamefont {S.}~\bibnamefont
  {Peotta}}\ and\ \bibinfo {author} {\bibfnamefont {P.}~\bibnamefont
  {{T{\"o}rm{\"a}}}},\ }\bibfield  {title} {\enquote {\bibinfo {title}
  {Superfluidity in topologically nontrivial flat bands},}\ }\href
  {http://dx.doi.org/10.1038/ncomms9944} {\bibfield  {journal} {\bibinfo
  {journal} {Nat. Commun.}\ }\textbf {\bibinfo {volume} {6}},\ \bibinfo {pages}
  {8944} (\bibinfo {year} {2015})}\BibitemShut {NoStop}%
\bibitem [{\citenamefont {Julku}\ \emph {et~al.}(2016)\citenamefont {Julku},
  \citenamefont {Peotta}, \citenamefont {Vanhala}, \citenamefont {Kim},\ and\
  \citenamefont {T\"orm\"a}}]{LiebLattice}%
  \BibitemOpen
  \bibfield  {author} {\bibinfo {author} {\bibfnamefont {A.}~\bibnamefont
  {Julku}}, \bibinfo {author} {\bibfnamefont {S.}~\bibnamefont {Peotta}},
  \bibinfo {author} {\bibfnamefont {T.~I.}\ \bibnamefont {Vanhala}}, \bibinfo
  {author} {\bibfnamefont {D.~H.}\ \bibnamefont {Kim}}, \ and\ \bibinfo
  {author} {\bibfnamefont {P.}~\bibnamefont {T\"orm\"a}},\ }\bibfield  {title}
  {\enquote {\bibinfo {title} {Geometric origin of superfluidity in the
  {L}ieb-lattice flat band},}\ }\href {\doibase 10.1103/PhysRevLett.117.045303}
  {\bibfield  {journal} {\bibinfo  {journal} {Phys. Rev. Lett.}\ }\textbf
  {\bibinfo {volume} {117}},\ \bibinfo {pages} {045303} (\bibinfo {year}
  {2016})}\BibitemShut {NoStop}%
\bibitem [{\citenamefont {Gao}\ \emph {et~al.}(2015)\citenamefont {Gao},
  \citenamefont {Yang},\ and\ \citenamefont {Niu}}]{GYN2015}%
  \BibitemOpen
  \bibfield  {author} {\bibinfo {author} {\bibfnamefont {Y.}~\bibnamefont
  {Gao}}, \bibinfo {author} {\bibfnamefont {S.~A.}\ \bibnamefont {Yang}}, \
  and\ \bibinfo {author} {\bibfnamefont {Q.}~\bibnamefont {Niu}},\ }\bibfield
  {title} {\enquote {\bibinfo {title} {Geometrical effects in orbital magnetic
  susceptibility},}\ }\href {\doibase 10.1103/PhysRevB.91.214405} {\bibfield
  {journal} {\bibinfo  {journal} {Phys. Rev. B}\ }\textbf {\bibinfo {volume}
  {91}},\ \bibinfo {pages} {214405} (\bibinfo {year} {2015})}\BibitemShut
  {NoStop}%
\bibitem [{\citenamefont {Pi\'echon}\ \emph {et~al.}(2016)\citenamefont
  {Pi\'echon}, \citenamefont {Raoux}, \citenamefont {Fuchs},\ and\
  \citenamefont {Montambaux}}]{PRFM2016}%
  \BibitemOpen
  \bibfield  {author} {\bibinfo {author} {\bibfnamefont {Fr\'ed\'eric}\
  \bibnamefont {Pi\'echon}}, \bibinfo {author} {\bibfnamefont {Arnaud}\
  \bibnamefont {Raoux}}, \bibinfo {author} {\bibfnamefont {Jean-No\"el}\
  \bibnamefont {Fuchs}}, \ and\ \bibinfo {author} {\bibfnamefont {Gilles}\
  \bibnamefont {Montambaux}},\ }\bibfield  {title} {\enquote {\bibinfo {title}
  {Geometric orbital susceptibility: Quantum metric without berry curvature},}\
  }\href {\doibase 10.1103/PhysRevB.94.134423} {\bibfield  {journal} {\bibinfo
  {journal} {Phys. Rev. B}\ }\textbf {\bibinfo {volume} {94}},\ \bibinfo
  {pages} {134423} (\bibinfo {year} {2016})}\BibitemShut {NoStop}%
\bibitem [{\citenamefont {Emery}\ and\ \citenamefont
  {Kivelson}(1995)}]{Emery1995}%
  \BibitemOpen
  \bibfield  {author} {\bibinfo {author} {\bibfnamefont {V.~J.}\ \bibnamefont
  {Emery}}\ and\ \bibinfo {author} {\bibfnamefont {S.~A.}\ \bibnamefont
  {Kivelson}},\ }\bibfield  {title} {\enquote {\bibinfo {title} {Importance of
  phase fluctuations in superconductors with small superfluid density},}\
  }\href {\doibase 10.1038/374434a0} {\bibfield  {journal} {\bibinfo  {journal}
  {Nature (London)}\ }\textbf {\bibinfo {volume} {374}},\ \bibinfo {pages}
  {434--437} (\bibinfo {year} {1995})}\BibitemShut {NoStop}%
\bibitem [{\citenamefont {Baym}(1968)}]{BaymStAndrews}%
  \BibitemOpen
  \bibfield  {author} {\bibinfo {author} {\bibfnamefont {G.~D.}\ \bibnamefont
  {Baym}},\ }\href@noop {} {\emph {\bibinfo {title} {Mathematical Methods in
  Solid State and Superfluid Theory}}},\ edited by\ \bibinfo {editor}
  {\bibfnamefont {G.~H.~Derrick}\ \bibnamefont {R.~C.~Clark}}\ (\bibinfo
  {publisher} {Springer (New York)},\ \bibinfo {year} {1968})\BibitemShut
  {NoStop}%
\bibitem [{\citenamefont {Stringari}\ and\ \citenamefont
  {Pitaevskii}(2016)}]{StringariBook}%
  \BibitemOpen
  \bibfield  {author} {\bibinfo {author} {\bibfnamefont {S.}~\bibnamefont
  {Stringari}}\ and\ \bibinfo {author} {\bibfnamefont {L.}~\bibnamefont
  {Pitaevskii}},\ }\href@noop {} {\emph {\bibinfo {title} {{B}ose-{E}instein
  Condensation and Superfluidity}}}\ (\bibinfo  {publisher} {Oxford University
  Press},\ \bibinfo {year} {2016})\BibitemShut {NoStop}%
\bibitem [{\citenamefont {Jotzu}\ \emph {et~al.}(2014)\citenamefont {Jotzu},
  \citenamefont {Messer}, \citenamefont {Desbuquois}, \citenamefont {Lebrat},
  \citenamefont {Uehlinger}, \citenamefont {Greif},\ and\ \citenamefont
  {Esslinger}}]{Jotzu2014}%
  \BibitemOpen
  \bibfield  {author} {\bibinfo {author} {\bibfnamefont {G.}~\bibnamefont
  {Jotzu}}, \bibinfo {author} {\bibfnamefont {M.}~\bibnamefont {Messer}},
  \bibinfo {author} {\bibfnamefont {R.}~\bibnamefont {Desbuquois}}, \bibinfo
  {author} {\bibfnamefont {M.}~\bibnamefont {Lebrat}}, \bibinfo {author}
  {\bibfnamefont {T.}~\bibnamefont {Uehlinger}}, \bibinfo {author}
  {\bibfnamefont {D.}~\bibnamefont {Greif}}, \ and\ \bibinfo {author}
  {\bibfnamefont {T.}~\bibnamefont {Esslinger}},\ }\bibfield  {title} {\enquote
  {\bibinfo {title} {Experimental realization of the topological {H}aldane
  model with ultracold fermions},}\ }\href
  {http://dx.doi.org/10.1038/nature13915} {\bibfield  {journal} {\bibinfo
  {journal} {Nature (London)}\ }\textbf {\bibinfo {volume} {515}},\ \bibinfo
  {pages} {237--240} (\bibinfo {year} {2014})}\BibitemShut {NoStop}%
\bibitem [{\citenamefont {Fl{\"a}schner}\ \emph {et~al.}(2016)\citenamefont
  {Fl{\"a}schner}, \citenamefont {Rem}, \citenamefont {Tarnowski},
  \citenamefont {Vogel}, \citenamefont {L{\"u}hmann}, \citenamefont
  {Sengstock},\ and\ \citenamefont {Weitenberg}}]{Flaschner1091}%
  \BibitemOpen
  \bibfield  {author} {\bibinfo {author} {\bibfnamefont {N.}~\bibnamefont
  {Fl{\"a}schner}}, \bibinfo {author} {\bibfnamefont {B.~S.}\ \bibnamefont
  {Rem}}, \bibinfo {author} {\bibfnamefont {M.}~\bibnamefont {Tarnowski}},
  \bibinfo {author} {\bibfnamefont {D.}~\bibnamefont {Vogel}}, \bibinfo
  {author} {\bibfnamefont {D.~S.}\ \bibnamefont {L{\"u}hmann}}, \bibinfo
  {author} {\bibfnamefont {K.}~\bibnamefont {Sengstock}}, \ and\ \bibinfo
  {author} {\bibfnamefont {C.}~\bibnamefont {Weitenberg}},\ }\bibfield  {title}
  {\enquote {\bibinfo {title} {Experimental reconstruction of the {B}erry
  curvature in a {F}loquet {B}loch band},}\ }\href {\doibase
  10.1126/science.aad4568} {\bibfield  {journal} {\bibinfo  {journal}
  {Science}\ }\textbf {\bibinfo {volume} {352}},\ \bibinfo {pages} {1091--1094}
  (\bibinfo {year} {2016})}\BibitemShut {NoStop}%
\bibitem [{\citenamefont {Haldane}(1988)}]{Haldane1988}%
  \BibitemOpen
  \bibfield  {author} {\bibinfo {author} {\bibfnamefont {F.~D.~M.}\
  \bibnamefont {Haldane}},\ }\bibfield  {title} {\enquote {\bibinfo {title}
  {Model for a quantum {H}all effect without {L}andau levels: Condensed-matter
  realization of the "parity anomaly"},}\ }\href {\doibase
  10.1103/PhysRevLett.61.2015} {\bibfield  {journal} {\bibinfo  {journal}
  {Phys. Rev. Lett.}\ }\textbf {\bibinfo {volume} {61}},\ \bibinfo {pages}
  {2015--2018} (\bibinfo {year} {1988})}\BibitemShut {NoStop}%
\bibitem [{\citenamefont {Kane}\ and\ \citenamefont {Mele}(2005)}]{KM2005_1}%
  \BibitemOpen
  \bibfield  {author} {\bibinfo {author} {\bibfnamefont {C.~L.}\ \bibnamefont
  {Kane}}\ and\ \bibinfo {author} {\bibfnamefont {E.~J.}\ \bibnamefont
  {Mele}},\ }\bibfield  {title} {\enquote {\bibinfo {title} {${Z}_{2}$
  topological order and the quantum spin {H}all effect},}\ }\href {\doibase
  10.1103/PhysRevLett.95.146802} {\bibfield  {journal} {\bibinfo  {journal}
  {Phys. Rev. Lett.}\ }\textbf {\bibinfo {volume} {95}},\ \bibinfo {pages}
  {146802} (\bibinfo {year} {2005})}\BibitemShut {NoStop}%
\bibitem [{\citenamefont {{Tovmasyan}}\ \emph {et~al.}(2016)\citenamefont
  {{Tovmasyan}}, \citenamefont {{Peotta}}, \citenamefont {{T{\"o}rm{\"a}}},\
  and\ \citenamefont {{Huber}}}]{2016arXiv160800976T}%
  \BibitemOpen
  \bibfield  {author} {\bibinfo {author} {\bibfnamefont {M.}~\bibnamefont
  {{Tovmasyan}}}, \bibinfo {author} {\bibfnamefont {S.}~\bibnamefont
  {{Peotta}}}, \bibinfo {author} {\bibfnamefont {P.}~\bibnamefont
  {{T{\"o}rm{\"a}}}}, \ and\ \bibinfo {author} {\bibfnamefont {S.~D.}\
  \bibnamefont {{Huber}}},\ }\bibfield  {title} {\enquote {\bibinfo {title}
  {{Effective theory and emergent $SU(2)$ symmetry in the flat bands of
  attractive {H}ubbard models}},}\ }\href@noop {} {\bibfield  {journal}
  {\bibinfo  {journal} {ArXiv e-prints}\ } (\bibinfo {year} {2016})},\ \Eprint
  {http://arxiv.org/abs/1608.00976} {arXiv:1608.00976 [cond-mat.str-el]}
  \BibitemShut {NoStop}%
\bibitem [{\citenamefont {Scalapino}\ \emph {et~al.}(1993)\citenamefont
  {Scalapino}, \citenamefont {White},\ and\ \citenamefont {Zhang}}]{swz}%
  \BibitemOpen
  \bibfield  {author} {\bibinfo {author} {\bibfnamefont {D.~J.}\ \bibnamefont
  {Scalapino}}, \bibinfo {author} {\bibfnamefont {S.~R.}\ \bibnamefont
  {White}}, \ and\ \bibinfo {author} {\bibfnamefont {S.~C.}\ \bibnamefont
  {Zhang}},\ }\bibfield  {title} {\enquote {\bibinfo {title} {Insulator, metal,
  or superconductor: The criteria},}\ }\href {\doibase
  10.1103/PhysRevB.47.7995} {\bibfield  {journal} {\bibinfo  {journal} {Phys.
  Rev. B}\ }\textbf {\bibinfo {volume} {47}},\ \bibinfo {pages} {7995--8007}
  (\bibinfo {year} {1993})}\BibitemShut {NoStop}%
\bibitem [{\citenamefont {Moon}\ \emph {et~al.}(1995)\citenamefont {Moon},
  \citenamefont {Mori}, \citenamefont {Yang}, \citenamefont {Girvin},
  \citenamefont {MacDonald}, \citenamefont {Zheng}, \citenamefont {Yoshioka},\
  and\ \citenamefont {Zhang}}]{Moon:1995}%
  \BibitemOpen
  \bibfield  {author} {\bibinfo {author} {\bibfnamefont {K.}~\bibnamefont
  {Moon}}, \bibinfo {author} {\bibfnamefont {H.}~\bibnamefont {Mori}}, \bibinfo
  {author} {\bibfnamefont {Kun}\ \bibnamefont {Yang}}, \bibinfo {author}
  {\bibfnamefont {S.~M.}\ \bibnamefont {Girvin}}, \bibinfo {author}
  {\bibfnamefont {A.~H.}\ \bibnamefont {MacDonald}}, \bibinfo {author}
  {\bibfnamefont {L.}~\bibnamefont {Zheng}}, \bibinfo {author} {\bibfnamefont
  {D.}~\bibnamefont {Yoshioka}}, \ and\ \bibinfo {author} {\bibfnamefont
  {Shou-Cheng}\ \bibnamefont {Zhang}},\ }\bibfield  {title} {\enquote {\bibinfo
  {title} {Spontaneous interlayer coherence in double-layer quantum hall
  systems: Charged vortices and kosterlitz-thouless phase transitions},}\
  }\href {\doibase 10.1103/PhysRevB.51.5138} {\bibfield  {journal} {\bibinfo
  {journal} {Phys. Rev. B}\ }\textbf {\bibinfo {volume} {51}},\ \bibinfo
  {pages} {5138--5170} (\bibinfo {year} {1995})}\BibitemShut {NoStop}%
\bibitem [{\citenamefont {Kopnin}\ and\ \citenamefont
  {Sonin}(2010)}]{SuperfluidGraphene}%
  \BibitemOpen
  \bibfield  {author} {\bibinfo {author} {\bibfnamefont {N.~B.}\ \bibnamefont
  {Kopnin}}\ and\ \bibinfo {author} {\bibfnamefont {E.~B.}\ \bibnamefont
  {Sonin}},\ }\bibfield  {title} {\enquote {\bibinfo {title} {Supercurrent in
  superconducting graphene},}\ }\href {\doibase 10.1103/PhysRevB.82.014516}
  {\bibfield  {journal} {\bibinfo  {journal} {Phys. Rev. B}\ }\textbf {\bibinfo
  {volume} {82}},\ \bibinfo {pages} {014516} (\bibinfo {year}
  {2010})}\BibitemShut {NoStop}%
\bibitem [{\citenamefont {{Mizoguchi}}\ and\ \citenamefont
  {{Ogata}}(2015)}]{2015JPSJ84h4704M}%
  \BibitemOpen
  \bibfield  {author} {\bibinfo {author} {\bibfnamefont {T.}~\bibnamefont
  {{Mizoguchi}}}\ and\ \bibinfo {author} {\bibfnamefont {M.}~\bibnamefont
  {{Ogata}}},\ }\bibfield  {title} {\enquote {\bibinfo {title} {Meissner effect
  of {D}irac electrons in superconducting state due to inter-band effect},}\
  }\href {\doibase 10.7566/JPSJ.84.084704} {\bibfield  {journal} {\bibinfo
  {journal} {J. Phys. Soc. Jpn.}\ }\textbf {\bibinfo {volume} {84}},\ \bibinfo
  {eid} {084704} (\bibinfo {year} {2015})}\BibitemShut {NoStop}%
\bibitem [{\citenamefont {Rachel}\ and\ \citenamefont
  {Le~Hur}(2010)}]{PhysRevB.82.075106}%
  \BibitemOpen
  \bibfield  {author} {\bibinfo {author} {\bibfnamefont {Stephan}\ \bibnamefont
  {Rachel}}\ and\ \bibinfo {author} {\bibfnamefont {Karyn}\ \bibnamefont
  {Le~Hur}},\ }\bibfield  {title} {\enquote {\bibinfo {title} {Topological
  insulators and mott physics from the hubbard interaction},}\ }\href {\doibase
  10.1103/PhysRevB.82.075106} {\bibfield  {journal} {\bibinfo  {journal} {Phys.
  Rev. B}\ }\textbf {\bibinfo {volume} {82}},\ \bibinfo {pages} {075106}
  (\bibinfo {year} {2010})}\BibitemShut {NoStop}%
\bibitem [{\citenamefont {He}\ \emph {et~al.}(2011)\citenamefont {He},
  \citenamefont {Zong}, \citenamefont {Kou}, \citenamefont {Liang},\ and\
  \citenamefont {Feng}}]{Kou_HH2011}%
  \BibitemOpen
  \bibfield  {author} {\bibinfo {author} {\bibfnamefont {J.}~\bibnamefont
  {He}}, \bibinfo {author} {\bibfnamefont {Y.~H.}\ \bibnamefont {Zong}},
  \bibinfo {author} {\bibfnamefont {S.~P.}\ \bibnamefont {Kou}}, \bibinfo
  {author} {\bibfnamefont {Y.}~\bibnamefont {Liang}}, \ and\ \bibinfo {author}
  {\bibfnamefont {S.~P.}\ \bibnamefont {Feng}},\ }\bibfield  {title} {\enquote
  {\bibinfo {title} {Topological spin density waves in the {H}ubbard model on a
  honeycomb lattice},}\ }\href {\doibase 10.1103/PhysRevB.84.035127} {\bibfield
   {journal} {\bibinfo  {journal} {Phys. Rev. B}\ }\textbf {\bibinfo {volume}
  {84}},\ \bibinfo {pages} {035127} (\bibinfo {year} {2011})}\BibitemShut
  {NoStop}%
\bibitem [{\citenamefont {He}\ \emph {et~al.}(2012)\citenamefont {He},
  \citenamefont {Liang},\ and\ \citenamefont {Kou}}]{Kou_HH2012}%
  \BibitemOpen
  \bibfield  {author} {\bibinfo {author} {\bibfnamefont {J.}~\bibnamefont
  {He}}, \bibinfo {author} {\bibfnamefont {Y.}~\bibnamefont {Liang}}, \ and\
  \bibinfo {author} {\bibfnamefont {S.~P.}\ \bibnamefont {Kou}},\ }\bibfield
  {title} {\enquote {\bibinfo {title} {Composite spin liquid in a correlated
  topological insulator: Spin liquid without spin-charge separation},}\ }\href
  {\doibase 10.1103/PhysRevB.85.205107} {\bibfield  {journal} {\bibinfo
  {journal} {Phys. Rev. B}\ }\textbf {\bibinfo {volume} {85}},\ \bibinfo
  {pages} {205107} (\bibinfo {year} {2012})}\BibitemShut {NoStop}%
\bibitem [{\citenamefont {Liang}\ \emph {et~al.}(2013)\citenamefont {Liang},
  \citenamefont {He}, \citenamefont {Wu}, \citenamefont {Zhu},\ and\
  \citenamefont {Kou}}]{HH_attractive_kou}%
  \BibitemOpen
  \bibfield  {author} {\bibinfo {author} {\bibfnamefont {Y.}~\bibnamefont
  {Liang}}, \bibinfo {author} {\bibfnamefont {J.}~\bibnamefont {He}}, \bibinfo
  {author} {\bibfnamefont {Y.~J.}\ \bibnamefont {Wu}}, \bibinfo {author}
  {\bibfnamefont {Y.~X.}\ \bibnamefont {Zhu}}, \ and\ \bibinfo {author}
  {\bibfnamefont {S.~P.}\ \bibnamefont {Kou}},\ }\bibfield  {title} {\enquote
  {\bibinfo {title} {Topological superconductors in correlated topological
  insulators on the honeycomb lattice},}\ }\href {\doibase
  10.1140/epjb/e2013-40521-5} {\bibfield  {journal} {\bibinfo  {journal} {Eur.
  Phys. J. B}\ }\textbf {\bibinfo {volume} {86}},\ \bibinfo {pages} {466}
  (\bibinfo {year} {2013})}\BibitemShut {NoStop}%
\bibitem [{\citenamefont {Zhu}\ \emph {et~al.}(2014)\citenamefont {Zhu},
  \citenamefont {He}, \citenamefont {Zang}, \citenamefont {Liang},\ and\
  \citenamefont {Kou}}]{Kou_2014}%
  \BibitemOpen
  \bibfield  {author} {\bibinfo {author} {\bibfnamefont {Y.~X.}\ \bibnamefont
  {Zhu}}, \bibinfo {author} {\bibfnamefont {J.}~\bibnamefont {He}}, \bibinfo
  {author} {\bibfnamefont {C.~L.}\ \bibnamefont {Zang}}, \bibinfo {author}
  {\bibfnamefont {Y.}~\bibnamefont {Liang}}, \ and\ \bibinfo {author}
  {\bibfnamefont {S.~P.}\ \bibnamefont {Kou}},\ }\bibfield  {title} {\enquote
  {\bibinfo {title} {Magnetic topological insulators at finite temperature},}\
  }\href {http://stacks.iop.org/0953-8984/26/i=17/a=175601} {\bibfield
  {journal} {\bibinfo  {journal} {J. Phys.: Condens. Matter}\ }\textbf
  {\bibinfo {volume} {26}},\ \bibinfo {pages} {175601} (\bibinfo {year}
  {2014})}\BibitemShut {NoStop}%
\bibitem [{\citenamefont {Wu}\ \emph {et~al.}(2015)\citenamefont {Wu},
  \citenamefont {Li},\ and\ \citenamefont {Kou}}]{HH_attractive_kou_2}%
  \BibitemOpen
  \bibfield  {author} {\bibinfo {author} {\bibfnamefont {Y.~J.}\ \bibnamefont
  {Wu}}, \bibinfo {author} {\bibfnamefont {N.}~\bibnamefont {Li}}, \ and\
  \bibinfo {author} {\bibfnamefont {S.~P.}\ \bibnamefont {Kou}},\ }\bibfield
  {title} {\enquote {\bibinfo {title} {Chiral topological superfluids in the
  attractive {H}aldane-{H}ubbard model with opposite {Z}eeman energy at two
  sublattice sites},}\ }\href {\doibase 10.1140/epjb/e2015-60412-y} {\bibfield
  {journal} {\bibinfo  {journal} {Eur. Phys. J. B}\ }\textbf {\bibinfo {volume}
  {88}},\ \bibinfo {pages} {255} (\bibinfo {year} {2015})}\BibitemShut
  {NoStop}%
\bibitem [{\citenamefont {Hung}\ \emph {et~al.}(2014)\citenamefont {Hung},
  \citenamefont {Chua}, \citenamefont {Wang},\ and\ \citenamefont
  {Fiete}}]{Fiete2014}%
  \BibitemOpen
  \bibfield  {author} {\bibinfo {author} {\bibfnamefont {H.~H.}\ \bibnamefont
  {Hung}}, \bibinfo {author} {\bibfnamefont {V.}~\bibnamefont {Chua}}, \bibinfo
  {author} {\bibfnamefont {L.}~\bibnamefont {Wang}}, \ and\ \bibinfo {author}
  {\bibfnamefont {G.~A.}\ \bibnamefont {Fiete}},\ }\bibfield  {title} {\enquote
  {\bibinfo {title} {Interaction effects on topological phase transitions via
  numerically exact quantum {M}onte {C}arlo calculations},}\ }\href {\doibase
  10.1103/PhysRevB.89.235104} {\bibfield  {journal} {\bibinfo  {journal} {Phys.
  Rev. B}\ }\textbf {\bibinfo {volume} {89}},\ \bibinfo {pages} {235104}
  (\bibinfo {year} {2014})}\BibitemShut {NoStop}%
\bibitem [{\citenamefont {Chen}\ \emph {et~al.}(2015)\citenamefont {Chen},
  \citenamefont {Hung}, \citenamefont {Su}, \citenamefont {Fiete},\ and\
  \citenamefont {Ting}}]{Ding_2015}%
  \BibitemOpen
  \bibfield  {author} {\bibinfo {author} {\bibfnamefont {Y.~H.}\ \bibnamefont
  {Chen}}, \bibinfo {author} {\bibfnamefont {H.~H.}\ \bibnamefont {Hung}},
  \bibinfo {author} {\bibfnamefont {G.}~\bibnamefont {Su}}, \bibinfo {author}
  {\bibfnamefont {G.~A.}\ \bibnamefont {Fiete}}, \ and\ \bibinfo {author}
  {\bibfnamefont {C.~S.}\ \bibnamefont {Ting}},\ }\bibfield  {title} {\enquote
  {\bibinfo {title} {Cellular dynamical mean-field theory study of an
  interacting topological honeycomb lattice model at finite temperature},}\
  }\href {\doibase 10.1103/PhysRevB.91.045122} {\bibfield  {journal} {\bibinfo
  {journal} {Phys. Rev. B}\ }\textbf {\bibinfo {volume} {91}},\ \bibinfo
  {pages} {045122} (\bibinfo {year} {2015})}\BibitemShut {NoStop}%
\bibitem [{\citenamefont {Hickey}\ \emph {et~al.}(2015)\citenamefont {Hickey},
  \citenamefont {Rath},\ and\ \citenamefont {Paramekanti}}]{PRB.91.134414}%
  \BibitemOpen
  \bibfield  {author} {\bibinfo {author} {\bibfnamefont {C.}~\bibnamefont
  {Hickey}}, \bibinfo {author} {\bibfnamefont {P.}~\bibnamefont {Rath}}, \ and\
  \bibinfo {author} {\bibfnamefont {A.}~\bibnamefont {Paramekanti}},\
  }\bibfield  {title} {\enquote {\bibinfo {title} {Competing chiral orders in
  the topological {H}aldane-{H}ubbard model of spin-$\frac{1}{2}$ fermions and
  bosons},}\ }\href {\doibase 10.1103/PhysRevB.91.134414} {\bibfield  {journal}
  {\bibinfo  {journal} {Phys. Rev. B}\ }\textbf {\bibinfo {volume} {91}},\
  \bibinfo {pages} {134414} (\bibinfo {year} {2015})}\BibitemShut {NoStop}%
\bibitem [{\citenamefont {Zheng}\ \emph {et~al.}(2015)\citenamefont {Zheng},
  \citenamefont {Shen}, \citenamefont {Wang},\ and\ \citenamefont
  {Zhai}}]{PRB.91.161107}%
  \BibitemOpen
  \bibfield  {author} {\bibinfo {author} {\bibfnamefont {W.}~\bibnamefont
  {Zheng}}, \bibinfo {author} {\bibfnamefont {H.}~\bibnamefont {Shen}},
  \bibinfo {author} {\bibfnamefont {Z.}~\bibnamefont {Wang}}, \ and\ \bibinfo
  {author} {\bibfnamefont {H.}~\bibnamefont {Zhai}},\ }\bibfield  {title}
  {\enquote {\bibinfo {title} {Magnetic-order-driven topological transition in
  the {H}aldane-{H}ubbard model},}\ }\href {\doibase
  10.1103/PhysRevB.91.161107} {\bibfield  {journal} {\bibinfo  {journal} {Phys.
  Rev. B}\ }\textbf {\bibinfo {volume} {91}},\ \bibinfo {pages} {161107}
  (\bibinfo {year} {2015})}\BibitemShut {NoStop}%
\bibitem [{\citenamefont {Arun}\ \emph {et~al.}(2016)\citenamefont {Arun},
  \citenamefont {Sohal}, \citenamefont {Hickey},\ and\ \citenamefont
  {Paramekanti}}]{PRB.93.115110}%
  \BibitemOpen
  \bibfield  {author} {\bibinfo {author} {\bibfnamefont {V.~S.}\ \bibnamefont
  {Arun}}, \bibinfo {author} {\bibfnamefont {R.}~\bibnamefont {Sohal}},
  \bibinfo {author} {\bibfnamefont {C.}~\bibnamefont {Hickey}}, \ and\ \bibinfo
  {author} {\bibfnamefont {A.}~\bibnamefont {Paramekanti}},\ }\bibfield
  {title} {\enquote {\bibinfo {title} {Mean field study of the topological
  {H}aldane-{H}ubbard model of spin-$\frac{1}{2}$ fermions},}\ }\href {\doibase
  10.1103/PhysRevB.93.115110} {\bibfield  {journal} {\bibinfo  {journal} {Phys.
  Rev. B}\ }\textbf {\bibinfo {volume} {93}},\ \bibinfo {pages} {115110}
  (\bibinfo {year} {2016})}\BibitemShut {NoStop}%
\bibitem [{\citenamefont {Wu}\ \emph {et~al.}(2016)\citenamefont {Wu},
  \citenamefont {Faye}, \citenamefont {S\'en\'echal},\ and\ \citenamefont
  {Maciejko}}]{PRB.93.075131}%
  \BibitemOpen
  \bibfield  {author} {\bibinfo {author} {\bibfnamefont {J.}~\bibnamefont
  {Wu}}, \bibinfo {author} {\bibfnamefont {J.~P.~L.}\ \bibnamefont {Faye}},
  \bibinfo {author} {\bibfnamefont {D.}~\bibnamefont {S\'en\'echal}}, \ and\
  \bibinfo {author} {\bibfnamefont {J.}~\bibnamefont {Maciejko}},\ }\bibfield
  {title} {\enquote {\bibinfo {title} {Quantum cluster approach to the spinful
  {H}aldane-{H}ubbard model},}\ }\href {\doibase 10.1103/PhysRevB.93.075131}
  {\bibfield  {journal} {\bibinfo  {journal} {Phys. Rev. B}\ }\textbf {\bibinfo
  {volume} {93}},\ \bibinfo {pages} {075131} (\bibinfo {year}
  {2016})}\BibitemShut {NoStop}%
\bibitem [{\citenamefont {Hickey}\ \emph {et~al.}(2016)\citenamefont {Hickey},
  \citenamefont {Cincio}, \citenamefont {Papi\ifmmode~\acute{c}\else
  \'{c}\fi{}},\ and\ \citenamefont {Paramekanti}}]{PRL.116.137202}%
  \BibitemOpen
  \bibfield  {author} {\bibinfo {author} {\bibfnamefont {C.}~\bibnamefont
  {Hickey}}, \bibinfo {author} {\bibfnamefont {L.}~\bibnamefont {Cincio}},
  \bibinfo {author} {\bibfnamefont {Z.}~\bibnamefont
  {Papi\ifmmode~\acute{c}\else \'{c}\fi{}}}, \ and\ \bibinfo {author}
  {\bibfnamefont {A.}~\bibnamefont {Paramekanti}},\ }\bibfield  {title}
  {\enquote {\bibinfo {title} {{H}aldane-{H}ubbard {M}ott insulator: From
  tetrahedral spin crystal to chiral spin liquid},}\ }\href {\doibase
  10.1103/PhysRevLett.116.137202} {\bibfield  {journal} {\bibinfo  {journal}
  {Phys. Rev. Lett.}\ }\textbf {\bibinfo {volume} {116}},\ \bibinfo {pages}
  {137202} (\bibinfo {year} {2016})}\BibitemShut {NoStop}%
\bibitem [{\citenamefont {{Gu}}\ \emph {et~al.}(2015)\citenamefont {{Gu}},
  \citenamefont {{Li}},\ and\ \citenamefont {{Li}}}]{151205118}%
  \BibitemOpen
  \bibfield  {author} {\bibinfo {author} {\bibfnamefont {Z.~L.}\ \bibnamefont
  {{Gu}}}, \bibinfo {author} {\bibfnamefont {K.}~\bibnamefont {{Li}}}, \ and\
  \bibinfo {author} {\bibfnamefont {J.~X.}\ \bibnamefont {{Li}}},\ }\bibfield
  {title} {\enquote {\bibinfo {title} {Topological phase transitions and
  topological {M}ott insulator in {H}aldane-{H}ubbard model},}\ }\href
  {http://arxiv.org/abs/1512.05118} {\bibfield  {journal} {\bibinfo  {journal}
  {ArXiv e-prints}\ } (\bibinfo {year} {2015})},\ \Eprint
  {http://arxiv.org/abs/1512.05118} {arXiv:1512.05118 [cond-mat.str-el]}
  \BibitemShut {NoStop}%
\bibitem [{\citenamefont {{Zhang}}\ \emph {et~al.}(2015)\citenamefont
  {{Zhang}}, \citenamefont {{Xu}},\ and\ \citenamefont {{Zhang}}}]{151103833Z}%
  \BibitemOpen
  \bibfield  {author} {\bibinfo {author} {\bibfnamefont {Y.~C.}\ \bibnamefont
  {{Zhang}}}, \bibinfo {author} {\bibfnamefont {Z.}~\bibnamefont {{Xu}}}, \
  and\ \bibinfo {author} {\bibfnamefont {S.}~\bibnamefont {{Zhang}}},\
  }\bibfield  {title} {\enquote {\bibinfo {title} {{Topological superfluids and
  {BEC}-{BCS} crossover in attractive {H}aldane-{H}ubbard model}},}\ }\href
  {http://arxiv.org/abs/1511.03833} {\bibfield  {journal} {\bibinfo  {journal}
  {ArXiv e-prints}\ } (\bibinfo {year} {2015})},\ \Eprint
  {http://arxiv.org/abs/1511.03833} {arXiv:1511.03833 [cond-mat.quant-gas]}
  \BibitemShut {NoStop}%
\bibitem [{\citenamefont {Vanhala}\ \emph {et~al.}(2016)\citenamefont
  {Vanhala}, \citenamefont {Siro}, \citenamefont {Liang}, \citenamefont
  {Troyer}, \citenamefont {Harju},\ and\ \citenamefont {T\"orm\"a}}]{C1}%
  \BibitemOpen
  \bibfield  {author} {\bibinfo {author} {\bibfnamefont {T.~I.}\ \bibnamefont
  {Vanhala}}, \bibinfo {author} {\bibfnamefont {T.}~\bibnamefont {Siro}},
  \bibinfo {author} {\bibfnamefont {L.}~\bibnamefont {Liang}}, \bibinfo
  {author} {\bibfnamefont {M.}~\bibnamefont {Troyer}}, \bibinfo {author}
  {\bibfnamefont {A.}~\bibnamefont {Harju}}, \ and\ \bibinfo {author}
  {\bibfnamefont {P.}~\bibnamefont {T\"orm\"a}},\ }\bibfield  {title} {\enquote
  {\bibinfo {title} {Topological phase transitions in the repulsively
  interacting {H}aldane-{H}ubbard model},}\ }\href {\doibase
  10.1103/PhysRevLett.116.225305} {\bibfield  {journal} {\bibinfo  {journal}
  {Phys. Rev. Lett.}\ }\textbf {\bibinfo {volume} {116}},\ \bibinfo {pages}
  {225305} (\bibinfo {year} {2016})}\BibitemShut {NoStop}%
\bibitem [{\citenamefont {Imri\ifmmode~\check{s}\else \v{s}\fi{}ka}\ \emph
  {et~al.}(2016)\citenamefont {Imri\ifmmode~\check{s}\else \v{s}\fi{}ka},
  \citenamefont {Wang},\ and\ \citenamefont {Troyer}}]{PhysRevB.94.035109}%
  \BibitemOpen
  \bibfield  {author} {\bibinfo {author} {\bibfnamefont {J.}~\bibnamefont
  {Imri\ifmmode~\check{s}\else \v{s}\fi{}ka}}, \bibinfo {author} {\bibfnamefont
  {L.}~\bibnamefont {Wang}}, \ and\ \bibinfo {author} {\bibfnamefont
  {M.}~\bibnamefont {Troyer}},\ }\bibfield  {title} {\enquote {\bibinfo {title}
  {First-order topological phase transition of the {H}aldane-{H}ubbard
  model},}\ }\href {\doibase 10.1103/PhysRevB.94.035109} {\bibfield  {journal}
  {\bibinfo  {journal} {Phys. Rev. B}\ }\textbf {\bibinfo {volume} {94}},\
  \bibinfo {pages} {035109} (\bibinfo {year} {2016})}\BibitemShut {NoStop}%
\bibitem [{\citenamefont {Shankar}(1994)}]{Shankar_RMP_1994}%
  \BibitemOpen
  \bibfield  {author} {\bibinfo {author} {\bibfnamefont {R.}~\bibnamefont
  {Shankar}},\ }\bibfield  {title} {\enquote {\bibinfo {title}
  {Renormalization-group approach to interacting fermions},}\ }\href {\doibase
  10.1103/RevModPhys.66.129} {\bibfield  {journal} {\bibinfo  {journal} {Rev.
  Mod. Phys.}\ }\textbf {\bibinfo {volume} {66}},\ \bibinfo {pages} {129--192}
  (\bibinfo {year} {1994})}\BibitemShut {NoStop}%
\bibitem [{\citenamefont {Leggett}(1980)}]{Leggett1980}%
  \BibitemOpen
  \bibfield  {author} {\bibinfo {author} {\bibfnamefont {A.~J.}\ \bibnamefont
  {Leggett}},\ }\enquote {\bibinfo {title} {Diatomic molecules and {C}ooper
  pairs},}\ in\ \href {\doibase 10.1007/BFb0120125} {\emph {\bibinfo
  {booktitle} {Modern Trends in the Theory of Condensed Matter}}},\ \bibinfo
  {editor} {edited by\ \bibinfo {editor} {\bibfnamefont {A.}~\bibnamefont
  {P{\k{e}}kalski}}\ and\ \bibinfo {editor} {\bibfnamefont {J.~A.}\
  \bibnamefont {Przystawa}}}\ (\bibinfo  {publisher} {Springer Berlin
  Heidelberg},\ \bibinfo {address} {Berlin, Heidelberg},\ \bibinfo {year}
  {1980})\ pp.\ \bibinfo {pages} {13--27}\BibitemShut {NoStop}%
\bibitem [{\citenamefont {Nozi{\`e}res}\ and\ \citenamefont
  {Schmitt-Rink}(1985)}]{BCSBEC1985}%
  \BibitemOpen
  \bibfield  {author} {\bibinfo {author} {\bibfnamefont {P.}~\bibnamefont
  {Nozi{\`e}res}}\ and\ \bibinfo {author} {\bibfnamefont {S.}~\bibnamefont
  {Schmitt-Rink}},\ }\bibfield  {title} {\enquote {\bibinfo {title} {Bose
  condensation in an attractive fermion gas: From weak to strong coupling
  superconductivity},}\ }\href {\doibase 10.1007/BF00683774} {\bibfield
  {journal} {\bibinfo  {journal} {J. Low Temp. Phys.}\ }\textbf {\bibinfo
  {volume} {59}},\ \bibinfo {pages} {195--211} (\bibinfo {year}
  {1985})}\BibitemShut {NoStop}%
\bibitem [{\citenamefont {Micnas}\ \emph {et~al.}(1990)\citenamefont {Micnas},
  \citenamefont {Ranninger},\ and\ \citenamefont
  {Robaszkiewicz}}]{RevModPhys.62.113}%
  \BibitemOpen
  \bibfield  {author} {\bibinfo {author} {\bibfnamefont {R.}~\bibnamefont
  {Micnas}}, \bibinfo {author} {\bibfnamefont {J.}~\bibnamefont {Ranninger}}, \
  and\ \bibinfo {author} {\bibfnamefont {S.}~\bibnamefont {Robaszkiewicz}},\
  }\bibfield  {title} {\enquote {\bibinfo {title} {Superconductivity in
  narrow-band systems with local nonretarded attractive interactions},}\ }\href
  {\doibase 10.1103/RevModPhys.62.113} {\bibfield  {journal} {\bibinfo
  {journal} {Rev. Mod. Phys.}\ }\textbf {\bibinfo {volume} {62}},\ \bibinfo
  {pages} {113--171} (\bibinfo {year} {1990})}\BibitemShut {NoStop}%
\bibitem [{\citenamefont {Ho}\ \emph {et~al.}(2009)\citenamefont {Ho},
  \citenamefont {Cazalilla},\ and\ \citenamefont
  {Giamarchi}}]{PhysRevA.79.033620}%
  \BibitemOpen
  \bibfield  {author} {\bibinfo {author} {\bibfnamefont {A.~F.}\ \bibnamefont
  {Ho}}, \bibinfo {author} {\bibfnamefont {M.~A.}\ \bibnamefont {Cazalilla}}, \
  and\ \bibinfo {author} {\bibfnamefont {T.}~\bibnamefont {Giamarchi}},\
  }\bibfield  {title} {\enquote {\bibinfo {title} {Quantum simulation of the
  {H}ubbard model: The attractive route},}\ }\href {\doibase
  10.1103/PhysRevA.79.033620} {\bibfield  {journal} {\bibinfo  {journal} {Phys.
  Rev. A}\ }\textbf {\bibinfo {volume} {79}},\ \bibinfo {pages} {033620}
  (\bibinfo {year} {2009})}\BibitemShut {NoStop}%
\bibitem [{\citenamefont {Anderson}(1951)}]{PhysRev.83.1260}%
  \BibitemOpen
  \bibfield  {author} {\bibinfo {author} {\bibfnamefont {P.~W.}\ \bibnamefont
  {Anderson}},\ }\bibfield  {title} {\enquote {\bibinfo {title} {Limits on the
  energy of the antiferromagnetic ground state},}\ }\href {\doibase
  10.1103/PhysRev.83.1260} {\bibfield  {journal} {\bibinfo  {journal} {Phys.
  Rev.}\ }\textbf {\bibinfo {volume} {83}},\ \bibinfo {pages} {1260--1260}
  (\bibinfo {year} {1951})}\BibitemShut {NoStop}%
\bibitem [{\citenamefont {Fouet}\ \emph {et~al.}(2001)\citenamefont {Fouet},
  \citenamefont {Sindzingre},\ and\ \citenamefont {Lhuillier}}]{Fouet2001}%
  \BibitemOpen
  \bibfield  {author} {\bibinfo {author} {\bibfnamefont {J.~B.}\ \bibnamefont
  {Fouet}}, \bibinfo {author} {\bibfnamefont {P.}~\bibnamefont {Sindzingre}}, \
  and\ \bibinfo {author} {\bibfnamefont {C.}~\bibnamefont {Lhuillier}},\
  }\bibfield  {title} {\enquote {\bibinfo {title} {An investigation of the
  quantum \textsc{J}$_1$-\textsc{J}$_2$-\textsc{J}$_3$ model on the honeycomb
  lattice},}\ }\href {\doibase 10.1007/s100510170273} {\bibfield  {journal}
  {\bibinfo  {journal} {Eur. Phys. J. B}\ }\textbf {\bibinfo {volume} {20}},\
  \bibinfo {pages} {241--254} (\bibinfo {year} {2001})}\BibitemShut {NoStop}%
\bibitem [{\citenamefont {Berezinskii}(1971)}]{BKT_B}%
  \BibitemOpen
  \bibfield  {author} {\bibinfo {author} {\bibfnamefont {V.~L.}\ \bibnamefont
  {Berezinskii}},\ }\bibfield  {title} {\enquote {\bibinfo {title} {Destruction
  of long-range order in one-dimensional and two-dimensional systems having a
  continuous symmetry group {I}. classical systems},}\ }\href
  {http://www.jetp.ac.ru/cgi-bin/e/index/e/32/3/p493?a=list} {\bibfield
  {journal} {\bibinfo  {journal} {Sov. Phys. JETP}\ }\textbf {\bibinfo {volume}
  {32}},\ \bibinfo {pages} {493} (\bibinfo {year} {1971})}\BibitemShut
  {NoStop}%
\bibitem [{\citenamefont {Kosterlitz}\ and\ \citenamefont
  {Thouless}(1973)}]{BKT_KT}%
  \BibitemOpen
  \bibfield  {author} {\bibinfo {author} {\bibfnamefont {J.~M.}\ \bibnamefont
  {Kosterlitz}}\ and\ \bibinfo {author} {\bibfnamefont {D.~J.}\ \bibnamefont
  {Thouless}},\ }\bibfield  {title} {\enquote {\bibinfo {title} {Ordering,
  metastability and phase transitions in two-dimensional systems},}\ }\href
  {http://stacks.iop.org/0022-3719/6/i=7/a=010} {\bibfield  {journal} {\bibinfo
   {journal} {J. Phys. C: Solid State Physics}\ }\textbf {\bibinfo {volume}
  {6}},\ \bibinfo {pages} {1181} (\bibinfo {year} {1973})}\BibitemShut
  {NoStop}%
\bibitem [{\citenamefont {Bo\ifmmode~\check{z}\else \v{z}\fi{}ovi\'c}\ \emph
  {et~al.}(2016)\citenamefont {Bo\ifmmode~\check{z}\else \v{z}\fi{}ovi\'c},
  \citenamefont {He}, \citenamefont {Wu},\ and\ \citenamefont
  {Bollinger}}]{I2016}%
  \BibitemOpen
  \bibfield  {author} {\bibinfo {author} {\bibfnamefont {I.}~\bibnamefont
  {Bo\ifmmode~\check{z}\else \v{z}\fi{}ovi\'c}}, \bibinfo {author}
  {\bibfnamefont {X.}~\bibnamefont {He}}, \bibinfo {author} {\bibfnamefont
  {J.}~\bibnamefont {Wu}}, \ and\ \bibinfo {author} {\bibfnamefont {A.~T.}\
  \bibnamefont {Bollinger}},\ }\bibfield  {title} {\enquote {\bibinfo {title}
  {Dependence of the critical temperature in overdoped copper oxides on
  superfluid density},}\ }\href {http://dx.doi.org/10.1038/nature19061}
  {\bibfield  {journal} {\bibinfo  {journal} {Nature (London)}\ }\textbf
  {\bibinfo {volume} {536}},\ \bibinfo {pages} {309--311} (\bibinfo {year}
  {2016})}\BibitemShut {NoStop}%
\bibitem [{\citenamefont {Rousseau}(2014)}]{PRB.90.134503}%
  \BibitemOpen
  \bibfield  {author} {\bibinfo {author} {\bibfnamefont {V.~G.}\ \bibnamefont
  {Rousseau}},\ }\bibfield  {title} {\enquote {\bibinfo {title} {Superfluid
  density in continuous and discrete spaces: Avoiding misconceptions},}\ }\href
  {\doibase 10.1103/PhysRevB.90.134503} {\bibfield  {journal} {\bibinfo
  {journal} {Phys. Rev. B}\ }\textbf {\bibinfo {volume} {90}},\ \bibinfo
  {pages} {134503} (\bibinfo {year} {2014})}\BibitemShut {NoStop}%
\bibitem [{\citenamefont {Georges}\ \emph {et~al.}(1996)\citenamefont
  {Georges}, \citenamefont {Kotliar}, \citenamefont {Krauth},\ and\
  \citenamefont {Rozenberg}}]{RevModPhys.68.13}%
  \BibitemOpen
  \bibfield  {author} {\bibinfo {author} {\bibfnamefont {A.}~\bibnamefont
  {Georges}}, \bibinfo {author} {\bibfnamefont {G.}~\bibnamefont {Kotliar}},
  \bibinfo {author} {\bibfnamefont {W.}~\bibnamefont {Krauth}}, \ and\ \bibinfo
  {author} {\bibfnamefont {M.~J.}\ \bibnamefont {Rozenberg}},\ }\bibfield
  {title} {\enquote {\bibinfo {title} {Dynamical mean-field theory of strongly
  correlated fermion systems and the limit of infinite dimensions},}\ }\href
  {\doibase 10.1103/RevModPhys.68.13} {\bibfield  {journal} {\bibinfo
  {journal} {Rev. Mod. Phys.}\ }\textbf {\bibinfo {volume} {68}},\ \bibinfo
  {pages} {13--125} (\bibinfo {year} {1996})}\BibitemShut {NoStop}%
\bibitem [{\citenamefont {Maier}\ \emph {et~al.}(2005)\citenamefont {Maier},
  \citenamefont {Jarrell}, \citenamefont {Pruschke},\ and\ \citenamefont
  {Hettler}}]{RevModPhys.77.1027}%
  \BibitemOpen
  \bibfield  {author} {\bibinfo {author} {\bibfnamefont {T.}~\bibnamefont
  {Maier}}, \bibinfo {author} {\bibfnamefont {M.}~\bibnamefont {Jarrell}},
  \bibinfo {author} {\bibfnamefont {T.}~\bibnamefont {Pruschke}}, \ and\
  \bibinfo {author} {\bibfnamefont {M.~H.}\ \bibnamefont {Hettler}},\
  }\bibfield  {title} {\enquote {\bibinfo {title} {Quantum cluster theories},}\
  }\href {\doibase 10.1103/RevModPhys.77.1027} {\bibfield  {journal} {\bibinfo
  {journal} {Rev. Mod. Phys.}\ }\textbf {\bibinfo {volume} {77}},\ \bibinfo
  {pages} {1027--1080} (\bibinfo {year} {2005})}\BibitemShut {NoStop}%
\bibitem [{\citenamefont {Rubtsov}\ \emph {et~al.}(2005)\citenamefont
  {Rubtsov}, \citenamefont {Savkin},\ and\ \citenamefont
  {Lichtenstein}}]{PhysRevB.72.035122}%
  \BibitemOpen
  \bibfield  {author} {\bibinfo {author} {\bibfnamefont {A.~N.}\ \bibnamefont
  {Rubtsov}}, \bibinfo {author} {\bibfnamefont {V.~V.}\ \bibnamefont {Savkin}},
  \ and\ \bibinfo {author} {\bibfnamefont {A.~I.}\ \bibnamefont
  {Lichtenstein}},\ }\bibfield  {title} {\enquote {\bibinfo {title}
  {Continuous-time quantum {M}onte {C}arlo method for fermions},}\ }\href
  {\doibase 10.1103/PhysRevB.72.035122} {\bibfield  {journal} {\bibinfo
  {journal} {Phys. Rev. B}\ }\textbf {\bibinfo {volume} {72}},\ \bibinfo
  {pages} {035122} (\bibinfo {year} {2005})}\BibitemShut {NoStop}%
\bibitem [{\citenamefont {Gull}\ \emph {et~al.}(2011)\citenamefont {Gull},
  \citenamefont {Millis}, \citenamefont {Lichtenstein}, \citenamefont
  {Rubtsov}, \citenamefont {Troyer},\ and\ \citenamefont
  {Werner}}]{RevModPhys.83.349}%
  \BibitemOpen
  \bibfield  {author} {\bibinfo {author} {\bibfnamefont {E.}~\bibnamefont
  {Gull}}, \bibinfo {author} {\bibfnamefont {A.~J.}\ \bibnamefont {Millis}},
  \bibinfo {author} {\bibfnamefont {A.~I.}\ \bibnamefont {Lichtenstein}},
  \bibinfo {author} {\bibfnamefont {A.~N.}\ \bibnamefont {Rubtsov}}, \bibinfo
  {author} {\bibfnamefont {M.}~\bibnamefont {Troyer}}, \ and\ \bibinfo {author}
  {\bibfnamefont {P.}~\bibnamefont {Werner}},\ }\bibfield  {title} {\enquote
  {\bibinfo {title} {Continuous-time {M}onte {C}arlo methods for quantum
  impurity models},}\ }\href {\doibase 10.1103/RevModPhys.83.349} {\bibfield
  {journal} {\bibinfo  {journal} {Rev. Mod. Phys.}\ }\textbf {\bibinfo {volume}
  {83}},\ \bibinfo {pages} {349--404} (\bibinfo {year} {2011})}\BibitemShut
  {NoStop}%
\bibitem [{\citenamefont {Caffarel}\ and\ \citenamefont
  {Krauth}(1994)}]{PhysRevLett.72.1545}%
  \BibitemOpen
  \bibfield  {author} {\bibinfo {author} {\bibfnamefont {M.}~\bibnamefont
  {Caffarel}}\ and\ \bibinfo {author} {\bibfnamefont {W.}~\bibnamefont
  {Krauth}},\ }\bibfield  {title} {\enquote {\bibinfo {title} {Exact
  diagonalization approach to correlated fermions in infinite dimensions:
  {M}ott transition and superconductivity},}\ }\href {\doibase
  10.1103/PhysRevLett.72.1545} {\bibfield  {journal} {\bibinfo  {journal}
  {Phys. Rev. Lett.}\ }\textbf {\bibinfo {volume} {72}},\ \bibinfo {pages}
  {1545--1548} (\bibinfo {year} {1994})}\BibitemShut {NoStop}%
\bibitem [{\citenamefont {Lin}\ \emph {et~al.}(2009)\citenamefont {Lin},
  \citenamefont {Gull},\ and\ \citenamefont {Millis}}]{PhysRevB.80.161105}%
  \BibitemOpen
  \bibfield  {author} {\bibinfo {author} {\bibfnamefont {N.}~\bibnamefont
  {Lin}}, \bibinfo {author} {\bibfnamefont {E.}~\bibnamefont {Gull}}, \ and\
  \bibinfo {author} {\bibfnamefont {A.~J.}\ \bibnamefont {Millis}},\ }\bibfield
   {title} {\enquote {\bibinfo {title} {Optical conductivity from cluster
  dynamical mean-field theory: Formalism and application to high-temperature
  superconductors},}\ }\href {\doibase 10.1103/PhysRevB.80.161105} {\bibfield
  {journal} {\bibinfo  {journal} {Phys. Rev. B}\ }\textbf {\bibinfo {volume}
  {80}},\ \bibinfo {pages} {161105} (\bibinfo {year} {2009})}\BibitemShut
  {NoStop}%
\bibitem [{\citenamefont {Siro}\ and\ \citenamefont
  {Harju}(2012)}]{Siro20121884}%
  \BibitemOpen
  \bibfield  {author} {\bibinfo {author} {\bibfnamefont {T.}~\bibnamefont
  {Siro}}\ and\ \bibinfo {author} {\bibfnamefont {A.}~\bibnamefont {Harju}},\
  }\bibfield  {title} {\enquote {\bibinfo {title} {Exact diagonalization of the
  {H}ubbard model on graphics processing units},}\ }\href {\doibase
  http://dx.doi.org/10.1016/j.cpc.2012.04.006} {\bibfield  {journal} {\bibinfo
  {journal} {Comput. Phys. Commun.}\ }\textbf {\bibinfo {volume} {183}},\
  \bibinfo {pages} {1884 -- 1889} (\bibinfo {year} {2012})}\BibitemShut
  {NoStop}%
\bibitem [{\citenamefont {Gagliano}\ and\ \citenamefont
  {Balseiro}(1987)}]{PhysRevLett.59.2999}%
  \BibitemOpen
  \bibfield  {author} {\bibinfo {author} {\bibfnamefont {E.~R.}\ \bibnamefont
  {Gagliano}}\ and\ \bibinfo {author} {\bibfnamefont {C.~A.}\ \bibnamefont
  {Balseiro}},\ }\bibfield  {title} {\enquote {\bibinfo {title} {Dynamical
  properties of quantum many-body systems at zero temperature},}\ }\href
  {\doibase 10.1103/PhysRevLett.59.2999} {\bibfield  {journal} {\bibinfo
  {journal} {Phys. Rev. Lett.}\ }\textbf {\bibinfo {volume} {59}},\ \bibinfo
  {pages} {2999--3002} (\bibinfo {year} {1987})}\BibitemShut {NoStop}%
\end{thebibliography}%

\end{document}